\documentclass[twocolumn, floatfix]{revtex4}

\usepackage{graphicx}
\usepackage{dcolumn}
\usepackage{amsmath}

\begin{document}

\title{Correlated-Electron Systems and High-Temperature Superconductivity
}

\author{Takashi Yanagisawa$^a$, Mitake Miyazaki$^b$ and Kunihiko Yamaji$^a$}

\affiliation{$^a$Electronics and Photonics Research Group, 
National Institute of Advanced Industrial Science and Technology (AIST),
Tsukuba Central 2, 1-1-1 Umezono, Tsukuba, Ibaraki 305-8568, Japan\\
$^b$Hakodate National College of Technology, 14-1 Tokura, Hakodate 042-8501,
Japan}


\begin{abstract}
\vspace{0.3cm}
\begin{center}
ABSTRACT
\end{center}

We present recent theoretical results on superconductivity in correlated-electron
systems, especially in the two-dimensional Hubbard model and the three-band d-p model.
The mechanism of superconductivity in high-temperature 
superconductors has been extensively studied  on the basis of various electronic
models and also electron-phonon models. 
In this study we investigate the properties of superconductivity in 
correlated-electron systems by using numerical methods such as the variational
Monte Carlo method and the quantum Monte Carlo method.

The Hubbard model is one of basic models for strongly correlated electron systems,
and is regarded as the model of cuprate high temperature superconductors.
The d-p model is more realistic model for cuprates.
The superconducting condensation energy obtained by adopting the Gutzwiller ansatz
is in reasonable agreement with the condensation energy estimated for
YBa$_2$Cu$_3$O$_7$. 
We show the phase diagram of the ground state using this method.
We have further investigated the stability
of striped and checkerboard states in the under-doped region. 
Holes doped in a half-filled square lattice lead to an incommensurate spin and
charge density
wave. The relationship of the hole density $x$ and incommensurability $\delta$, 
$\delta\sim x$, is satisfied in
the lower doping region, as indicated by the variational Monte Carlo calculations
for the two-dimensional Hubbard model. 
A checkerboard-like charge-density modulation with a roughly $4\times 4$ period
has also been observed by scanning tunneling microscopy experiments in 
Bi2212 and Na-CCOC compounds. 
We have performed a variational Monte Carlo simulation on a two-dimensional 
$t$-$t'$-$t''$-$U$ Hubbard model
with a Bi-2212 type band structure and found that the $4\times 4$ period checkerboard 
spin modulation, that is characterized by multi Q vectors, is indeed stabilized. 

We have further performed an investigation by using a quantum Monte 
Carlo method which is a
numerical method that can be used to simulate the behavior of correlated 
electron systems. 
We present a new algorithm of the quantum Monte Carlo diagonalization that
is a method
for the evaluation of expectation value without the negative sign problem.
We compute pair correlation functions and show that pair correlation is
indeed enhanced with hole doping. 
\vspace{1.3cm}
\end{abstract}

\maketitle



\section{Introduction}
\begin{figure}[htbp]
\includegraphics[width=8cm]{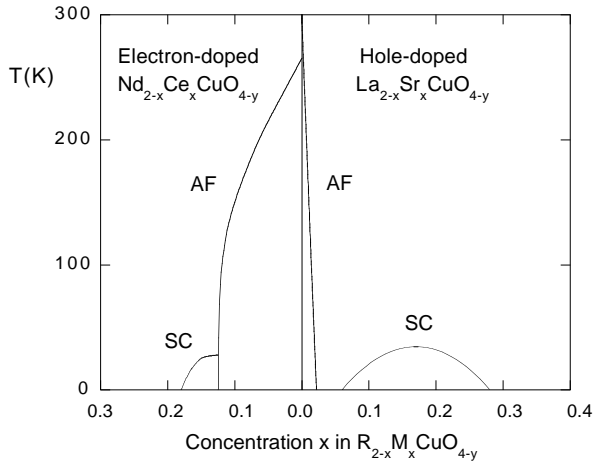}
\caption{
Phase diagram delineating the regions of superconductivity and antiferromagnetic
ordering of the Cu$^{2+}$ ions for the hole-doped La$_{2-x}$Sr$_x$CuO$_4$ and
electron-doped Nd$_{2-x}$Ce$_x$CuO$_{4-y}$ systems.
}
\label{fig1}
\end{figure}

The effect of the strong correlation between electrons is important for many
quantum critical phenomena, such as unconventional superconductivity
(SC) and the metal-insulator transition.
Typical correlated electron systems are high-temperature
superconductors\cite{bed86,dag94,and97,mor00,ben03},
heavy fermions\cite{ste84,lee86,ott87,map00} and organic
conductors\cite{ish98}.
The phase diagram for the typical high-T$_c$ cuprates is shown in 
Fig.\ref{fig1}.\cite{map00}
It has a  characteristics that the region of antiferromagnetic order
exists at low carrier concentrations and the superconducting phase is adjacent to
the antiferromagnetism. 

\begin{figure}[htbp]
\includegraphics[width=8cm]{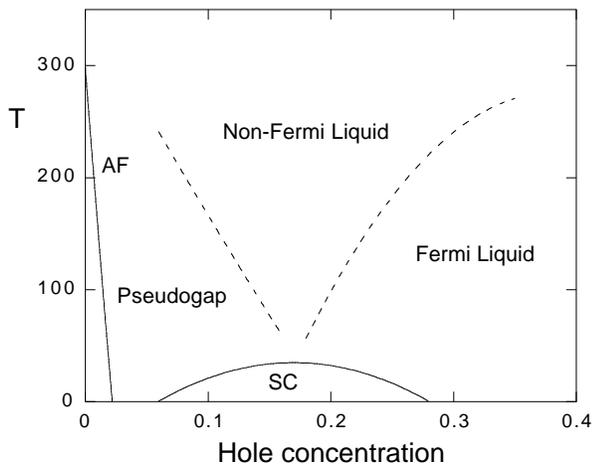}
\caption{
Phase diagram showing the regions of non-Fermi liquid and pseudogap metal
for the hole-doped case.  The boundaries indicated in the figure are not
confirmed yet.
}
\label{fig2}
\end{figure}

In the low-carrier region shown in Fig.\ref{fig2} there is the anomalous metallic 
region where the susceptibility and $1/T_1$ show a peak above T$_c$
suggesting an existence of the pseudogap.  To clarify
an origin of the anomalous metallic behaviors is also a subject attracting 
many physicists as a challenging problem.

It has been established that the Cooper pairs of high-T$_c$ cuprates have the
$d$-wave symmetry in the hole-doped materials.\cite{tsu94,wol95}
Several evidences of $d$-wave pairing symmetry were provided for the
electron-doped cuprates Nd$_{2-x}$Ce$_x$CuO$_4$.\cite{tsu00,sat01,yan01c} 
Thus it is expected that the superconductivity of electronic origin is a
candidate for the high-T$_c$ superconductivity.
We can also expect that the origin of $d$-wave superconductivity lies in the
on-site Coulomb interaction of the two-dimensional Hubbard model.

The antiferromagnetism should also be examined in correlated electron systems.
The undoped oxide compounds exhibit  rich structures of 
antiferromagnetic correlations over a wide range of temperature that are 
described by the two-dimensional quantum 
antiferromagnetism.\cite{shi87,lyo88,man89}
A small number of holes introduced by doping
are responsible for the disappearance of long-range antiferromagnetic
order.\cite{pre88,inu88,yan92,yan93,yan96,yan01}

Recent neutron scattering experiments have suggested an existence of 
incommensurate
ground states with modulation vectors given by $Q_s=(\pi\pm 2\pi\delta,\pi)$
and $Q_c=(\pm4\pi\delta,0)$ (or $Q_s=(\pi,\pi\pm2\pi\delta)$ and 
$Q_c=(0,\pm 4\pi\delta)$) where $Â\delta$ denotes the hole-doping 
ratio.\cite{tra95}
We can expect that the incommensurate correlations are induced by holes
doped into the Cu-O plane in the underdoped region.
A checkerboard-like charge-density modulation with a roughly $4\times 4$ period
has also been observed by scanning tunneling microscopy experiments in 
Bi2212 and Na-CCOC compounds.

\begin{figure}[htbp]
\includegraphics[width=8cm]{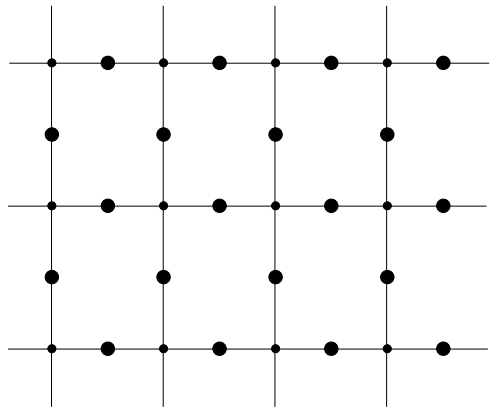}
\caption{
The lattice of the three-band Hubbard model on the CuO$_2$ plane.  Small circles 
denote Cu sites and large ones denote O sites.
}
\label{fig3}
\end{figure}

Recently the mechanism of
superconductivity in high-temperature superconductors 
has been extensively studied using various two-dimensional
(2D) models of electronic interactions.
Among them the 2D Hubbard model\cite{hub63} is the simplest and most
fundamental model.
This model has been studied intensively using numerical tools, such as the
Quantum Monte Carlo method
\cite{hir83,hir85,sor88,whi89,ima89,sor89,loh90,mor91,fur92,mor92,fah91,zha97,
zha97b,kas01,yan07,yan13},
and the variational Monte Carlo
method\cite{yan01,yok87,gro87,nak97,yam98,yan02,yan03,yan05,miy04}.
The two-leg ladder Hubbard model was also investigated with respect
to the mechanism of high-temperature
superconductivity\cite{yam94,yam94b,koi99,koi00,noa96,noa97,kur96,dau00,san05}.

Since the discovery of cuprate high-temperature superconductors, many
researchers have tried to explain the occurrence of superconductivity of
these materials in terms of the two-dimensional (2D) Hubbard model.
However,
it remains matter of considerable controversial as to whether the 2D Hubbard
model accounts for the properties of high-temperature cuprate superconductors.
This is because the membership of the the two-dimensional Hubbard model in the category
of strongly correlated systems is a considerable barrier to progress on this problem.
The quest for the existence of a superconducting transition in the 2D Hubbard
model is a long-standing problem in correlated-electron physics,
and has been the subject of intensive study\cite{fur92,mor92,yam98,zha97,bul02,aim07}.
In particular, the results of quantum Monte Carlo methods,
which are believed to be exact unbiased methods, have failed to show
the existence of superconductivity in this model\cite{zha97,aim07}.

In the weak coupling limit we can answer this question.
We can obtain the superconducting order parameter of the Hubbard model in the limit 
of small $U$, that is given by\cite{kon01,hlu99,koi01,koi06,yan08}

\begin{equation}
\Delta= {\rm exp}\left (-\frac{2}{xU^2}\right ) ,
\end{equation}
where $U$ is the strength of the on-site Coulomb interaction and the exponent $x$
is determined by solving the gap equation.  Thus the existence of the superconducting
gap is guaranteed by the weak coupling theory although $\Delta$ is extremely
small because of the exponential behavior given above.
$x$ indicates the strength of superconductivity.
In the intermediate or large coupling region, we must study it beyond the perturbation
theory. 

We investigate the ground state of the Hubbard model by employing
the variational Monte Carlo method.
In the region $6\leq U\leq 12$, the finite superconducting 
gap is obtained by using the quantum variational Monte Carlo method. 
The superconducting condensation energy obtained by adopting the Gutzwiller
ansatz is in reasonable agreement with the condensation energy derived for
YBa$_2$Cu$_3$O$_7$. 
We have further investigated the stability of striped 
and checkerboard states in the under-doped region. 
Holes doped in a half-filled square lattice lead to an incommensurate spin 
and charge density
wave. The relationship of the hole density x and incommensurability $\delta$, 
$\delta\sim x$, is satisfied in
the lower doping region.  This is consistent with the results by neutron
scattering measurements.
To examine the stability of a $4\times 4$ checkerboard state, we have 
performed a variational Monte Carlo simulation on a two-dimensional 
$t-t'-t''-U$ Hubbard model 
with a Bi-2212 type band structure. We found that the $4\times 4$ period 
checkerboard 
checkerboard spin modulation that is characterized by multi $Q$ vectors is 
stabilized. 

Further investigation has been performed by using the quantum Monte Carlo method
which is a numerical method that can be used to simulate the behavior of 
correlated 
electron systems. This method is believed to be an exact unbiased 
method. We compute pair correlation functions to examine a possibility
of superconductivity.

The Quantum Monte Carlo (QMC) method is a numerical method employed to
simulate the behavior of correlated electron systems. 
It is well known, however, that there are significant issues associated with
the application to the QMC.
First, the standard Metropolis(or heat bath) algorithm is associated with the 
negative sign problem.
Second, the convergence of the trial wave function is sometimes not monotonic,
and further, is sometimes slow.
In past studies, workers have investigated the possibility of eliminating
the negative sign 
problem\cite{fah91,zha97,kas01,yan07,yan13}.
We present the results obtained by a method, quantum Monte Carlo diagonalization,
without the negative sign problem.

\section{Hubbard Hamiltonian}
 
The Hubbard Hamiltonian is
\begin{equation}
H= -\sum_{ij}t_{ij}c_{i\sigma}^{\dag}c_{i\sigma}+U\sum_i n_{i\uparrow}n_{i\downarrow},
\end{equation}
where $c_{i\sigma}^{\dag}$ and $c_{i\sigma}$ denote the creation and annihilation
operators of electrons, respectively, and $n_{i\sigma}=c_{i\sigma}^{\dag}c_{i\sigma}$
is the number operator.  The second term represents the on-site Coulomb interaction
which acts when the two electrons occupy the same site.  The numbers of lattice sites and
electrons are denoted as $N$ and $N_e$, respectively.  The electron density is
$n_e=N_e/N$.  

In the non-interacting limit $U=0$, the Hamiltonian is easily diagonalized in terms
of the Fourier transformation.  In the ground state each energy level is occupied by electrons
up to the Fermi energy.  In the other limit $t_{ij}=0$, each site is occupied by the
up- or down-spin electron, or is empty.  The non-zero $t_{ij}$ induces the movement
of electrons that leads to a metallic state id $N_e\neq N$.
The ground state is probably insulating at half-filling $N_e=N$ if $U$ is sufficiently
large.

If $t_{ij}=t$ are non-zero only for the nearest-neighbor pairs,
the Hubbard Hamiltonian is transformed to the following effective Hamiltonian
for large $U/t$:\cite{har67}

\begin{eqnarray}
H&=& -t\sum_{\langle ij\rangle\sigma}a^{\dag}_{i\sigma}a_{j\sigma}
-\frac{t^2}{U}\sum_{\mu\mu'}[a^{\dag}_{j+\mu\uparrow}a^{\dag}_{\downarrow}
a_{j\downarrow}a_{j+\mu'\uparrow}\nonumber\\
&+& a^{\dag}_{j\uparrow}a^{\dag}_{j+\mu\downarrow}
a_{j+\mu'\downarrow}a_{j\uparrow}
+a^{\dag}_{j+\mu\uparrow}a^{\dag}_{j\downarrow}
a_{j+\mu'\downarrow}a_{j\uparrow}\nonumber\\
&+& a^{\dag}_{j\uparrow}a^{\dag}_{j+\mu\downarrow}
a_{j\downarrow}a_{j+\mu\uparrow}] ,
\end{eqnarray}
where $a_{i\sigma}=c_{i\sigma}(1-n_{i,-\sigma})$ and $j+\mu$ and $j+\mu'$ indicate
the nearest-neighbor sites in the $\mu$ and $\mu'$ directions, respectively.
The second term contains the three-site terms when $\mu\neq\mu'$.
If we neglect the three-site terms, this effective Hamiltonian is reduced to
the t-J model:
\begin{equation}
H= -t\sum_{\langle ij\rangle\sigma}(a^{\dag}_{i\sigma}a_{j\sigma}+{\rm h.c.})
+J\sum_{\langle ij\rangle}({\bf S}_i\cdot {\bf S}_j-\frac{1}{4}n_in_j),
\nonumber\\
\end{equation}
where $J=4t^2/U$.

The Hubbard model has a long history in describing the magnetism of materials
since the early works by Hubbard\cite{hub63}, Gutzwiller\cite{gut63} and
Kanamori.\cite{kan63}
One-dimensional Hubbard model has been well understood by means of the Bethe
ansatz\cite{bet31,yan67,lie68} and conformal field theory.\cite{sch90,fra90,kaw90}
The solutions established a novel concept of the Tomonaga-Luttinger liquid\cite{hal81}
which is described by the scalar bosons corresponding to charge and spin sectors,
respectively.
The correlated electrons in two- and three-dimensional space are still far from a
complete understanding in spite of the success for the one-dimensional Hubbard model.
A possibility of superconductivity is a hot topic as well as the magnetism and
metal-insulator transition for the two- and three-dimensional Hubbard model.

The three-band Hubbard model that contains $d$ and $p$ orbitals has also been 
investigated intensively with respect to high temperature superconductors.
\cite{koi01,yan01,eme87,tje89,ste89,hir89,sca91,dop90,dop92,asa96,kur96b,              
tak97,gue98,koi00b}
This model is also called the d-p model.
The 2D three-band Hubbard model is the more realistic and relevant model for
two-dimensional CuO$_2$ planes which are contained usually
in the crystal structures of high-$T_c$ superconductors.
The network of CuO$_2$ layer is shown in Fig.\ref{fig3}.
The parameters of the three-band Hubbard model are given by the Coulomb
repulsion $U_d$, energy levels of $p$ electrons $\epsilon_p$ and $d$
electron $\epsilon_d$, and transfer between $p$ orbitals given by $t_{pp}$.
Typical parameter values for the three-band ($d$-$p$) Hubbard model are
shown in Table I.
The Hamiltonian of the three-band Hubbard model is written as\cite{hir89,yan01,yan02,yan03}
\begin{eqnarray}
H_{dp}&=&\epsilon_d\sum_{i\sigma}d^{\dag}_{i\sigma}d_{i\sigma} +
U_d\sum_i d^{\dag}_{i\uparrow}d_{i\uparrow}d^{\dag}_{i\downarrow}d_{i\downarrow}\nonumber\
\\
&+& \epsilon_p\sum_{i\sigma}(p^{\dag}_{i+\hat{x}/2,\sigma}p_{i+\hat{x}/2,\sigma}
+p^{\dag}_{i+\hat{y}/2,\sigma}p_{i+\hat{y}/2,\sigma})\nonumber\\
&-&t_{dp}\sum_{i\sigma}[d^{\dag}_{i\sigma}(p_{i+\hat{x}/2,\sigma}
+p_{i+\hat{y}/2,\sigma}-p_{i-\hat{x}/2,\sigma}\nonumber\\
&-& p_{i-\hat{y}/2,\sigma})
+{\rm h.c.}]\nonumber\\
&-& t_{pp}\sum_{i\sigma}[ p^{\dag}_{i+\hat{y}/2,\sigma}p_{i+\hat{x}/2,\sigma}
-p^{\dag}_{i+\hat{y}/2,\sigma}p_{i-\hat{x}/2,\sigma}\nonumber\\
&-& p^{\dag}_{i-\hat{y}/2,\sigma}p_{i+\hat{x}/2,\sigma}
+p^{\dag}_{i-\hat{y}/2,\sigma}p_{i-\hat{x}/2,\sigma}\nonumber\\
&+& {\rm h.c.}].
\label{dpm}
\end{eqnarray}
$\hat{x}$ and $\hat{y}$ represent unit vectors along x and y directions,
respectively.
$p^{\dag}_{i\pm\hat{x}/2,\sigma}$
and $p_{i\pm\hat{x}/2,\sigma}$ denote the operators for the $p$ electrons at
site $R_i\pm\hat{x}/2$.  Similarly $p^{\dag}_{i\pm\hat{y}/2,\sigma}$ and
$p_{i\pm\hat{y}/2,\sigma}$ are  defined.
$U_d$ denotes the strength of Coulomb interaction between $d$ 
electrons.
For simplicity we neglect the Coulomb interaction among $p$ electrons in this paper.
Other notations are standard and energies are measured in $t_{dp}$ units.
The number of cells is denoted as $N$ for the three-band Hubbard model.
In the non-interacting case ($U_d=0$) the Hamiltonian in the
${\bf k}$-space is written as:
\begin{eqnarray}
H^0_{dp}&=& \epsilon_d\sum_{{\bf k}\sigma}d^{\dag}_{{\bf k}\sigma}d_{{\bf k}\sigma}
+\epsilon_p\sum_{{\bf k}\sigma}(p^{\dag}_{x{\bf k}\sigma}p_{x{\bf k}\sigma}
+p^{\dag}_{y{\bf k}\sigma}p_{y{\bf k}\sigma})\nonumber\\ 
&+&\sum_{{\bf k}\sigma}(2{\rm i}t_{dp}{\rm sin}(k_x/2)
d^{\dag}_{{\bf k}\sigma}p_{x{\bf k}\sigma}+{\rm h.c.})\nonumber\\
&+& \sum_{{\bf k}\sigma}(2{\rm i}t_{dp}{\rm sin}(k_y/2)
d^{\dag}_{{\bf k}\sigma}p_{y{\bf k}\sigma}+{\rm h.c.})\nonumber\\
&+&\sum_{{\bf k}\sigma}(-4t_{pp}{\rm sin}(k_x/2){\rm sin}(k_y/2))
(p^{\dag}_{x{\bf k}\sigma}p_{y{\bf k}\sigma}+{\rm h.c.}),\nonumber\\
\end{eqnarray}
where $d_{{\bf k}\sigma}$ ($d^{\dag}_{{\bf k}\sigma}$), 
$p_{x{\bf k}\sigma}$ ($p^{\dag}_{x{\bf k}\sigma}$) and 
$p_{y{\bf k}\sigma}$ ($p^{\dag}_{y{\bf k}\sigma}$)
are operators for $d$-, $p_x$- and $p_y$-electron of the momentum ${\bf k}$
and spin $\sigma$, respectively.

In the limit
$t_{dp}\ll U_d-(\epsilon_p-\epsilon_d)$, $t_{dp}\ll\epsilon_p-\epsilon_d$ ,
and $\epsilon_p-\epsilon_d < U_d$, the $d$-$p$ model is mapped to the
t-J model with
\begin{equation}
J= 4t_{eff}^2 \left( \frac{1}{U_d}+\frac{2}{2(\epsilon_p-\epsilon_d)+U_p} \right) ,
\end{equation}
where $t_{eff}\simeq t^2_{dp}/(\epsilon_p-\epsilon_d)$.
$J_K=4t_{eff}$ gives the antiferromagnetic coupling between the neighboring
$d$ and $p$ electrons.
In real materials $(\epsilon_p-\epsilon_d)/t_{dp}$ is not so large.
Thus it seems that the mapping to the t-J model is not necessarily justified.

\begin{table}
\caption{Typical parameter values for the three-band Hubbard model.
Energies are measured in eV.}
\begin{tabular}{ccccc}
\colrule
   & Ref.\cite{hyb90} & Ref.\cite{esk89}  & Ref.\cite{mcm90} & Ref.\cite{tje89}\\
\colrule
$\epsilon_p-\epsilon_d$   & 3.6  & $2.75-3.75$ & 3.5 & 2.5 \\
$t_{dp}$                  & 1.3  & 1.5         & 1.5 & 1.47 \\
$t_{pp}$                  & 0.65 & 0.65        & 0.6 & \\
$U_d$                     & 10.5 & 8.8         & 9.4 & 9.7 \\
$U_p$                     & 4.0  & 6.0         & 4.7 & 5.7 \\
$U_{dp}$                  & 1.2  & $<$1.0      & 0.8 & $<$1  \\
\colrule
\end{tabular}
\end{table}

\section{Variational Monte Carlo Studies}

In this Section we present studies on the two-dimensional Hubbard model by
using the variational Monte Carlo method.

\subsection{Variational Monte Carlo Method}

Let us start by describing the method based on the 2D Hubbard model.
The Hamiltonian is given by
\begin{eqnarray}
H &=& -t\sum_{\langle ij\rangle\sigma}(c^{\dag}_{i\sigma}c_{j\sigma}+{\rm h.c.})
-t'\sum_{\langle\langle j\ell\rangle\rangle\sigma}(c^{\dag}_{j\sigma}c_{\ell\sigma}
+{\rm h.c.})\nonumber\\
&+& U\sum_jn_{j\uparrow}n_{j\downarrow} ,
\end{eqnarray}
where $\langle ij\rangle$ denotes summation over all the nearest-neighbor bonds
and $\langle\langle j\ell\rangle\rangle$ means summation over the next
nearest-neighbor pairs.  $t$ is our energy unit. The dispersion is given by
\begin{equation}
\epsilon_{{\bf k}} = -2t({\rm cos}(k_x)+{\rm cos}(k_y))-4t'{\rm cos}(k_x)
{\rm cos}(k_y).
\end{equation}

Our trial wave function is the Gutzwiller-projected wave functions defined as
\begin{eqnarray}
\psi_n &=& P_{N_e}P_G\psi_0 ,\\
\psi_s &=& P_{N_e}P_G\psi_{BCS} ,
\end{eqnarray}
where
\begin{eqnarray}
\psi_0 &=& \prod_{|{\bf k}|\leq k_F,\sigma}c^{\dag}_{{\bf k}\sigma}|0\rangle ,
\\
\psi_{BCS} &=& \prod_{{\bf k}}(u_{{\bf k}}+v_{{\bf k}}c^{\dag}_{{\bf k}\uparrow}
c^{\dag}_{-{\bf k}\downarrow})|0\rangle .
\end{eqnarray}
$P_G$ is the Gutzwiller projection operator given by
\begin{equation}
P_G = \prod_j [1-(1-g)n_{j\uparrow}n_{j\downarrow}];
\end{equation}
$g$ is a variational parameter in the range from 0 to unity and $j$ labels a site
in the real space.  $P_{N_e}$ is a projection operator which extracts only
the sites with a fixed total electron number $N_e$.
Coefficients $u_{{\bf k}}$ and $v_{{\bf k}}$ in $\psi_{BCS}$ appear in the
ratio defined by
\begin{equation}
\frac{v_{{\bf k}}}{u_{{\bf k}}} = \frac{\Delta_{{\bf k}}}
{\xi_{{\bf k}}+(\xi^2_{{\bf k}}+\Delta^2_{{\bf k}})^{1/2}},
\end{equation}
where $\xi_{{\bf k}}=\epsilon_{{\bf k}}-\mu$ and $\Delta_{{\bf k}}$ is a
${\bf k}$-dependent gap function.
$\mu$ is a variational parameter working like the chemical potential.
$c_{{\bf k}\sigma}$ is the Fourier transform of $c_{j\sigma}$.
The wave functions $\psi_n$ and $\psi_s$ are expressed by the Slater determinants
for which the expectations values are evaluated using a Monte Carlo 
procedure.\cite{cep77,yok87,gro87}
$\psi_s$ is written as
\begin{eqnarray}
\psi_s &\propto& P_{N_e}P_G{\rm exp}[\sum_{{\bf k}}(v_{{\bf k}}/u_{{\bf k}})
c^{\dag}_{{\bf k}\uparrow}c^{\dag}_{-{\bf k}\downarrow}]|0\rangle\nonumber\\
&=& P_{N_e}P_G{\rm exp}[\sum_{j\ell}a(j,\ell)c^{\dag}_{j\uparrow}
c^{\dag}_{\ell\downarrow}]|0\rangle\nonumber\\
&\propto& P_G[\sum_{j\ell}a(j,\ell)c^{\dag}_{j\uparrow}
c^{\dag}_{\ell\downarrow}]^{N_e/2}|0\rangle,
\end{eqnarray}
where
\begin{equation}
a(j,\ell) = (1/N)\sum_{{\bf k}}(v_{{\bf k}}/u_{{\bf k}}){\rm exp}
[i{\bf k}\cdot ({\bf R}_{\ell}-{\bf R}_j)].
\end{equation}
Then $\psi_s$ is written using the Slater determinants as
\begin{eqnarray}
\psi_s&=&P_G\sum_{j_1\cdots j_{N_e/2}\ell_1\cdots \ell_{N_e/2}}\nonumber\\
&& A(j_1\cdots i_{N_e/2},\ell_1\cdots \ell_{N_e/2})\nonumber\\
&\times& c^{\dag}_{j_1\uparrow}c^{\dag}_{j_2\uparrow}\cdots 
c^{\dag}_{j_{N_e/2}\uparrow}c^{\dag}_{\ell_1\downarrow}c^{\dag}_{\ell_2\downarrow}
\cdots c^{\dag}_{\ell_{N_e/2}\downarrow}|0\rangle,\nonumber\\
\end{eqnarray}
where $A(j_1\cdots i_{N_e/2},\ell_1\cdots \ell_{N_e/2})$ is the Slater
determinant defined by
\begin{eqnarray}
A(j_1&\cdots&\ell_{N_e/2})=\nonumber\\
&&
\left | \begin{array}{cccc}
a(j_1,\ell_1) & a(j_1,\ell_2) & \cdots & a(j_1,\ell_{N_e/2})  \\
a(j_2,\ell_1) & a(j_2,\ell_2) & \cdots & a(j_2,\ell_{N_e/2})  \\
\cdots        & \cdots        & \cdots & \cdots                \\
a(j_{N_e/2},\ell_1) & a(j_{N_e/2},\ell_2) & \cdots & a(j_{N_e/2},\ell_{N_e/2})
\end{array} \right |.\nonumber\\
\label{sdet}
\end{eqnarray} 
In the process of Monte Carlo procedure the values of cofactors of the 
matrix in eq.(\ref{sdet}) are stored and corrected at each time when the
electron distribution is modified.
We optimized the ground state energy 
\begin{equation}
E_g = \langle H\rangle = \langle\psi_s |H|\psi_s\rangle/\langle\psi_s|\psi_s\rangle
\end{equation}
with respect to $g$, $\Delta_{{\bf k}}$ and $\mu$ for $\psi_s$ for $\psi_s$.
For $\psi_n$ the variational parameter is only $g$.  We can employ the 
correlated measurements 
method\cite{umr88} in the process of searching optimal parameter values
minimizing $E_g$.

A Monte Carlo algorithm developed in the auxiliary field quantum Monte Carlo
calculations can also be employed in evaluating the expectation values for the wave
functions shown above.\cite{bla81,yan98,yan99}
Note that the Gutzwiller projection operator is written as
\begin{equation}
P_G = {\rm exp}(-\alpha\sum_i n_{i\uparrow}n_{i\downarrow}) ,
\end{equation}
where $\alpha={\rm log}(1/g)$.
Then using the discrete Hubbard-Stratonovich transformation, the Gutzwiller
operator is the bilinear form:
\begin{eqnarray}
&&{\rm exp}(-\alpha\sum_i n_{i\uparrow}n_{i\downarrow})=
(1/2)^N\sum_{\{s_i\}}{\rm exp}[2a\nonumber\\
&\times&\sum_i s_i(n_{i\uparrow}-n_{i\downarrow})-\frac{\alpha}{2}\sum_i (n_{i\uparrow}+n_{i\downarrow})],
\end{eqnarray}
where ${\rm cosh}(2a)={\rm e}^{\alpha/2}$.
The Hubbard-Stratonovich auxiliary field $s_i$ takes the values of $\pm 1$.
The norm $\langle\psi_n|\psi_n\rangle$ is written as
\begin{eqnarray}
\langle\psi_n|\psi_n\rangle &=& {\rm const.} \sum_{\{u_i\}\{s_i\}}
\prod_{\sigma}
{\rm det}(\phi^{\sigma\dag}_0{\rm exp}(V^{\sigma}(u,\alpha))\nonumber\\
&\times& 
{\rm exp}(V^{\sigma}(s,\alpha))\phi^{\sigma}_0),
\end{eqnarray}
where $V^{\sigma}(s,\alpha)$ is a diagonal $N\times N$ matrix corresponding
to the potential
\begin{equation}
h^{\sigma}(s)= 2a\sigma\sum_i s_in_{i\sigma}-\frac{\alpha}{2}\sum_i
n_{i\sigma}.
\end{equation}
$V^{\sigma}(s,\alpha)$ is written as
\begin{equation}
V^{\sigma}(s,\alpha)= {\rm diag}(2a\sigma s_1-\alpha/2,\cdots,
2a\sigma s_N-\alpha/2,0,\cdots),
\end{equation}
where diag($a,\cdots$) denotes a diagonal matrix with elements given by the
arguments $a,\cdots$.
The elements of $(\phi^{\sigma}_0)_{ij}$ ($i=1,\cdots,N;j=1,\cdots,N_e/2$) 
are given by linear combinations of plane waves.  For example,
\begin{equation}
(\phi^{\sigma}_0)_{ij}={\rm exp}(i{\bf r}_i\cdot
{\bf k}_{j}) .
\end{equation}
Then we can apply the standard Monte Carlo sampling method to
evaluate the expectation values.\cite{bla81,yan98}
This method is used to consider an off-diagonal Jastrow correlation factor of
exp($-S$)-type.  The results for the improved wave functions are discussed
in Section III.J.

\begin{figure}[htbp]
\includegraphics[width=8cm]{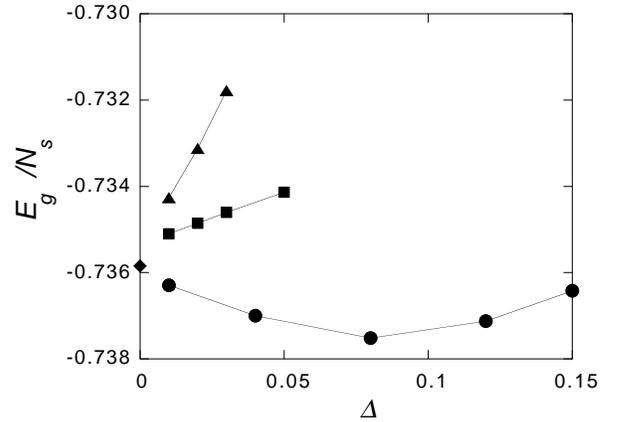}
\caption{
Ground state energy per site $E_g/N$ for the 2D Hubbard model 
is plotted against $\Delta$ for the case
of 84 electrons on the $10\times 10$ lattice with $U=8$ and $t'=0$.
Solid curves are for the $d$-wave gap function.
Squares and triangles are for the $s^*$- and $s$-wave gap functions, respectively.
The diamond shows the normal state value.\cite{yam98}
}
\label{fig13}
\end{figure}

\begin{figure}[htbp]
\includegraphics[width=\columnwidth]{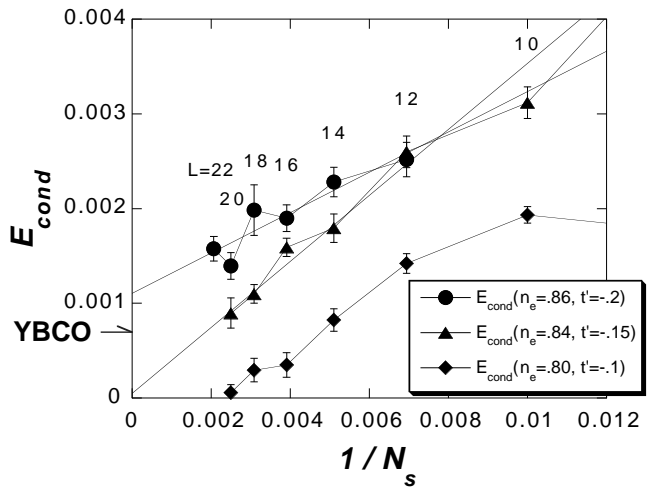}
\caption{
Energy gain per site in the SC state with reference to the normal state
for the 2D Hubbard model is plotted
as a function of $1/N$.  $L$ is the length of the edge of the square lattice.
YBCO attached to the vertical axis indicates the experimental value of the
SC condensation energy for YBa$_2$Cu$_3$O$_4$.\cite{yam00b}
}
\label{fig14}
\end{figure}

\begin{figure}[htbp]
\includegraphics[width=8cm]{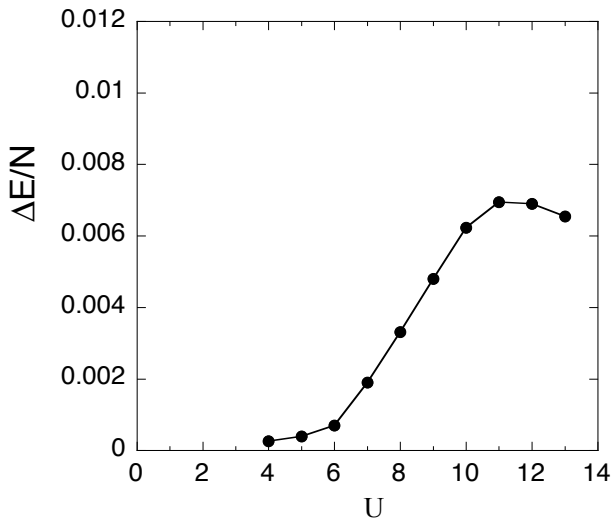}
\caption{
Energy gain per site in the SC state with reference to the normal state
for the 2D Hubbard model as a function of the Coulomb repulsion $U$.
The system is $10\times 10$ with the electron number $N_e=84$ and $t'=-0.3$.
}
\label{dE-U}
\end{figure}

\begin{figure}[htbp]
\includegraphics[width=7cm]{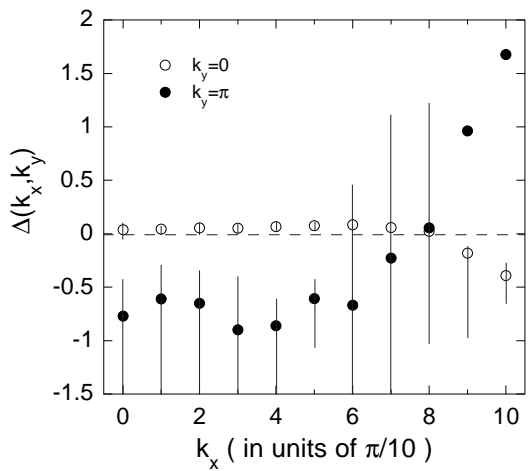}
\caption{
The values of components of $\Delta_k$ for the two-chain Hubbard model.
All the values of $k_x$ of the bonding band ($k_y=0$) and antibonding band
($k_y=\pi$) correspond to the energy minimum for $20\times 2$ lattice with
34 electrons.  The parameters in the Hamiltonian are $t_d=1.8$ and
$U_0=8$ and the variational parameters are $\mu=0.0182$ and $g=0.415$.\cite{koi00}
}
\label{fig15}
\end{figure}

\begin{figure}[htbp]
\includegraphics[width=\columnwidth]{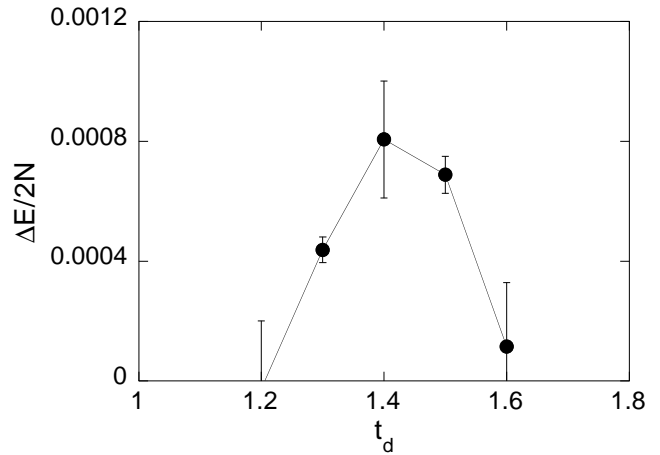}
\caption{
$t_d$ dependence of the SC condensation energy $\Delta E/2N$ for the two-chain
Hubbard model in the bulk limit.\cite{koi00}
}
\label{fig17}
\end{figure}

\begin{figure}[htbp]
\includegraphics[width=\columnwidth]{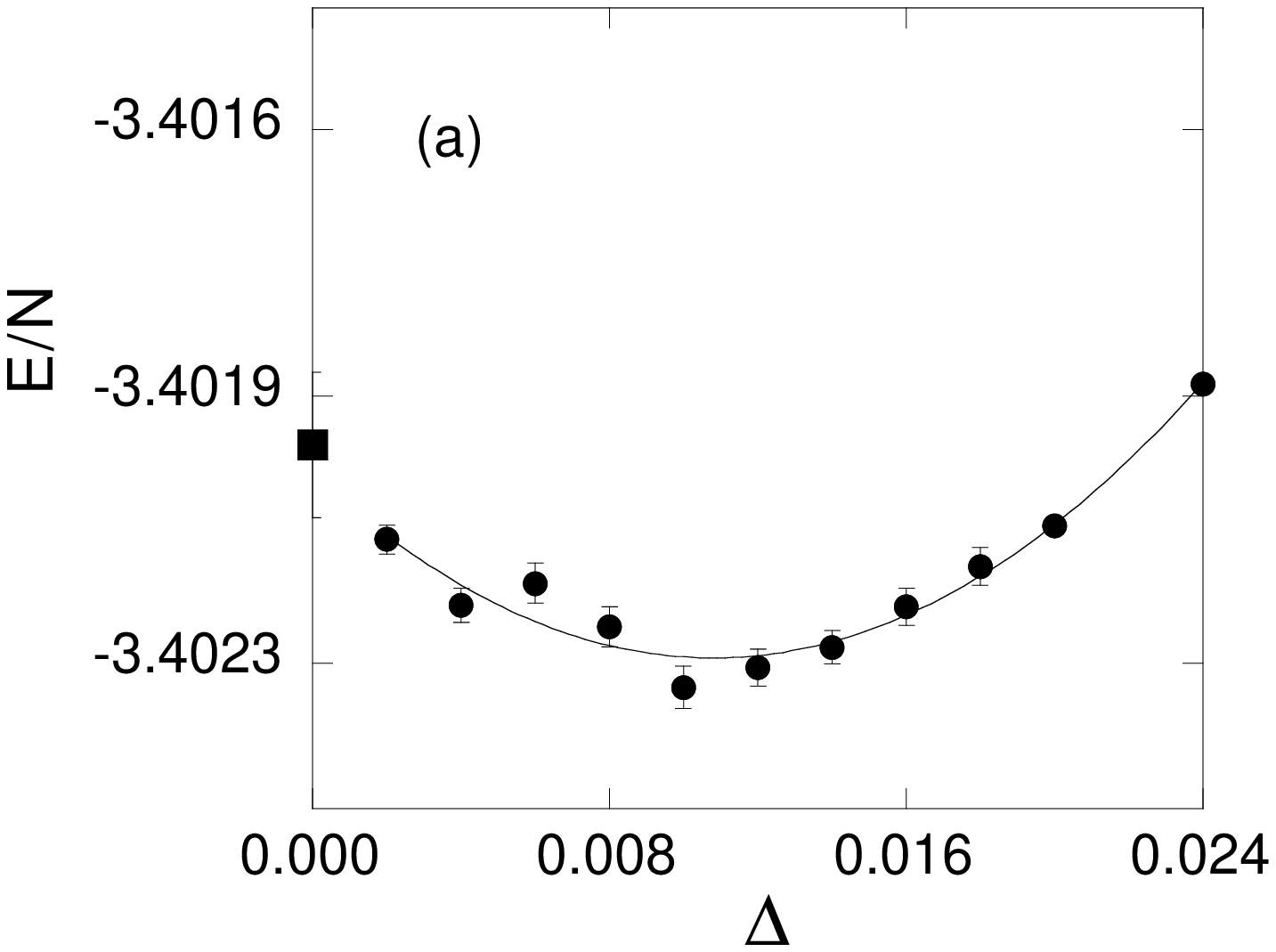}
\includegraphics[width=\columnwidth]{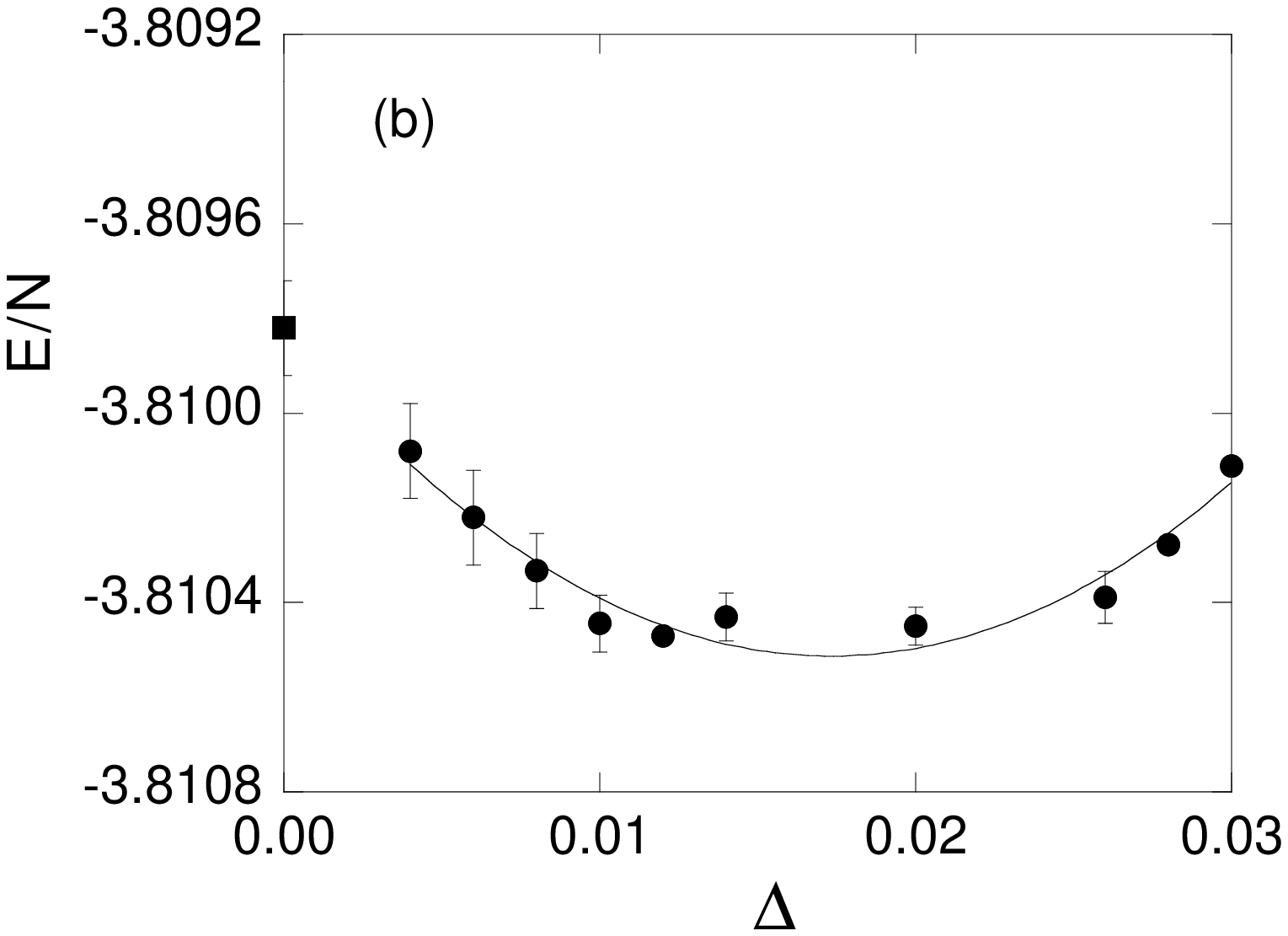}
\caption{
Ground-state energy per site as a function of $\Delta$ with the $d$-wave gap
function for the three-band Hubbard model.  The size of lattice is $6\times 6$.
Parameters are $U=8$, $t_{pp}=0$ and $\epsilon_p-\epsilon_d=2$ in units of 
$t_{dp}$.  The doping rate is $\delta=0.111$ for (a) and $\delta=0.333$ for
(b).  Squares denote the energies for the normal-state wave function.\cite{yan01}
}
\label{fig18}
\end{figure}

\begin{figure}[htbp]
\includegraphics[width=7.5cm]{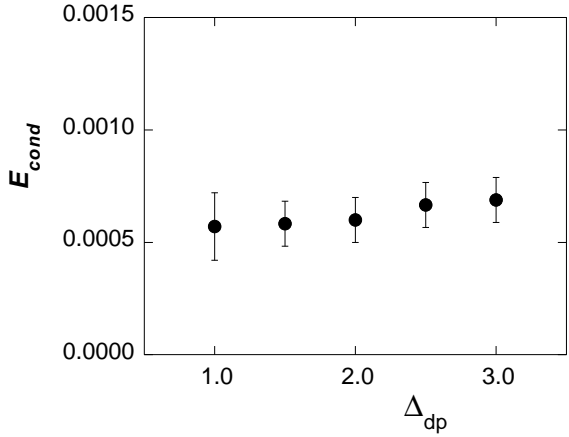}
\caption{
Energy gain per site in the SC state as a function of the level difference
$\Delta_{dp}=\epsilon_p-\epsilon_d$ for the three-band Hubbard model with
$U_d=8$ and $t_{pp}=0.2$.\cite{yan01}
The size of lattice is $6\times 6$ sites.   
}
\label{fig18c}
\end{figure}

\begin{figure}[htbp]
\includegraphics[width=\columnwidth]{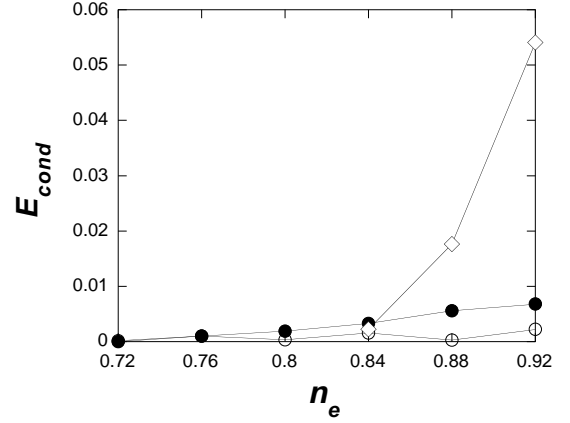}
\caption{
Energy gain per site in the SDW state (diamonds) against electron density for $t'=0$
and the energy gain in the SC state for $t'=0$ (open circles) and
$t'=-0.1$ (solid circles).  The model is the 2D Hubbard model on $10\times 10$
lattice.\cite{yam98}
}
\label{fig19}
\end{figure}

\begin{figure}[htbp]
\includegraphics[width=7cm]{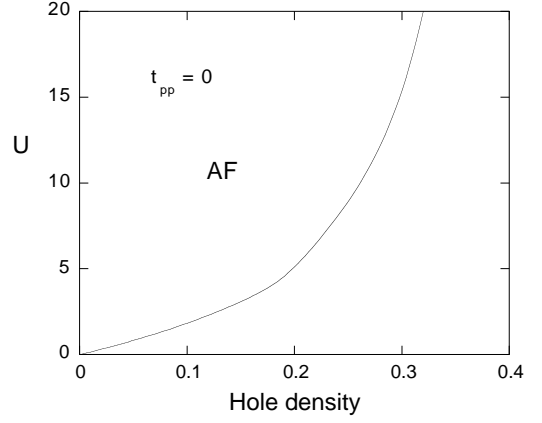}
\caption{
Antiferromagnetic region in the plane of $U$ and the hole density for
$t_{pp}=0.0$ and $\epsilon_p-\epsilon_d=2$. 
}
\label{fig20}
\end{figure}

\subsection{Superconducting Condensation Energy}

We study the cases of the $d$-, extended $s$- ($s^*$-) and $s$-wave gap
functions in the following:
\begin{eqnarray}
d~~~~\Delta_{{\bf k}}&=& \Delta ({\rm cos}(k_x)-{\rm cos}(k_y)),\\
s^*~~~\Delta_{{\bf k}}&=& \Delta ({\rm cos}(k_x)+{\rm cos}(k_y)),\\
s~~~~\Delta_{{\bf k}}&=& \Delta.
\end{eqnarray}

In Fig.\ref{fig13} calculated energies per site with $N_e=84$ on the $10\times 10$
lattice are shown for the case of $U=8$ and $t'=0$.\cite{yam98}
$E_g/N$ is plotted as a function of $\Delta$ for three types of gap
functions shown above.  We impose the periodic and the antiperiodic
boundary conditions for $x$- and $y$-direction, respectively.
This set of boundary conditions is chosen so that $\Delta_{{\bf k}}$
does not vanish for any ${\bf k}$-points occupied by electrons.
$E_g$ was obtained as the average of the results of several Monte Carlo
calculations each with $5\times 10^7$ steps.
$E_g/N$ has minimum at a finite value of $\Delta\simeq 0.08$ in the case
of the $d$-wave gap function.

The energy gain $\Delta E_g$ in the superconducting state is called the
SC condensation energy in this paper.  $\Delta E_g/N$ is plotted as a function
of $1/N$ in Fig.\ref{fig14}
in order to examine  the size dependence of the SC energy gain.\cite{yam00b}
Lattice sizes treated are from $8\times 8$ to $22\times 22$.
The electron density $n_e$ is in the range of $0.80\geq n_e\leq 0.86$.
Other parameters are $-0.20\leq t'\leq 0.0$ and $U/t=8$ in $t$ units.
Bulk limit $N\rightarrow\infty$ of SC condensation energy $E_{cond}$ was 
obtained by plotting as a
function of $1/N$.  The linear fitting line indicates very clearly that the
bulk limit remains finite when $-0.25\leq t'\leq -0.10$ and $n_e\geq 0.84$.
When $n_e=0.86$, $t'=-0.20$ and $U=8$, the bulk-limit $E_{cond}$ is
$E_{cond}=0.00117$/site $\simeq$ 0.60 meV/site, where $t=0.51$eV is 
used.\cite{fei96}  
Thus the superconductivity is a real bulk property, not a spurious size effect.
The value is remarkably close to experimental values 
$0.17\sim 0.26$ meV/site estimated from specific heat data\cite{lor93,and98} and
0.26 meV/site from the critical magnetic field $H_c$\cite{hao91} for optimally
doped YBa$_2$Cu$_3$O$_4$ (YBCO).
This good agreement strongly indicates that the 2D Hubbard model includes
essential ingredients for the superconductivity in the cuprates.

We just point out that the t-J model gives $E_{cond}=0.026t\simeq 13$ meV/site
at $n_e=0.84$ for $J=4t^2/U=0.5$ and $t'=0$.\cite{yok96}  This value is
50 times larger than the experimental values indicating a serious
quantitative problem with this model.  This means that the t-J model made
from the leading two terms in the expansion in terms of $t/U$ of the canonical
transformation of the Hubbard model should be treated with the higher-order
terms in order to give a realistic SC condensation energy.

Here we show the SC condensation energy as a function of $U$ in Fig.\ref{dE-U}.
The condensation energy $E_{cond}=\Delta E/N$ is increased as $U/t$ is increased
as far as $U/t\leq 12$.
In the strong coupling region $U>8t$, we obtain the large condensation energy.

\subsection{Fermi Surface and Condensation Energy}

Now let us consider the relationship between the Fermi surface structure and 
the strength of superconductivity.
The experimental SC condensation energy for (La,Sr)$_2$CuO$_4$ (LSCO) is estimated 
at 0.029meV/(Cu site)
or 0.00008 in units of $t$ which is much smaller than that for YBCO.\cite{mat04}
The band parameter values of LSCO were estimated as $t'=-0.12$ and $t''=0.08$.
\cite{toh00}
This set corresponds roughly to $E_{cond}\simeq 0.0010$.
The latter value is much larger than the above-mentioned experimental value
for LSCO.  However, the stripe-type SDW state coexists with 
superconductivity\cite{miy02,yan09} and the SC part of the whole $E_{cond}$ is
much reduced.  Therefore, such a coexistence allows us to qualitatively
understand the SC $E_{cond}$ in LSCO.

On the other hand, Tl2201 ($T_c=93$K) and Hg1201 ($T_c=98$K) band calculations
by Singh and Pickett\cite{sin92} give very much deformed Fermi surfaces
that can be fitted by large $|t'|$ such as $t'\sim -0.4$.
For Tl2201, an Angular Magnetoresistance Oscillations (AMRO) work\cite{hus03}
gives information of the Fermi surface, which allows to get $t'\sim -0.2$
and $t''\sim 0.165$. 
There is also an Angle-Resolved Photoemission Study (ARPES)\cite{pla05}, 
which provides similar values.
In the case of Hg1201, there is an ARPES
work\cite{lee06}, form which
we obtain by fitting $t'\sim -0.2$ and $t''\sim 0.175$.
For such a deformed Fermi surface, $E_{cond}$ in the bulk limit is reduced
considerably.\cite{yam08,yam11}
Therefore, the SC $E_{cond}$ calculated by VMC indicates that the Fermi surface 
of LSCO-type 
is more favorable for high $T_c$.
The lower $T_c$ in LSCO may be attributed to the coexistence with 
antiferromagnetism of stripe type.

\subsection{Ladder Hubbard Model}

The SC condensation energy in the bulk limit for the ladder Hubbard model has
also been evaluated using the variational Monte Carlo method.\cite{koi00}
The Hamiltonian is given by the 1D two-chain Hubbard 
model:\cite{yam94,yam94b,fab92,fab93,yan95,bal96,kur96b,noa96,noa97}
\begin{eqnarray}
H_{ladder} &=& -t_d\sum_{\ell\sigma}(c^{\dag}_{1\ell\sigma}c_{2\ell\sigma}+{\rm h.c.})
\nonumber\\
&-&t\sum_{j=1}^2\sum_{\ell\sigma}(c^{\dag}_{j\ell\sigma}c_{j,\ell+1,\sigma}
+{\rm h.c.})\nonumber\\
&+& U_0\sum_{j=1}^2\sum_{\ell}c^{\dag}_{j\ell\uparrow}c_{j\ell\uparrow}
c^{\dag}_{j\ell\downarrow}c_{j\ell\downarrow},
\end{eqnarray}
where $c^{\dag}_{j\ell\sigma}$ ($c_{j\ell\sigma}$) is the creation (annihilation)
operator of an electron with spin $\sigma$ at the $\ell$th site along the
$j$th chain ($j=1,2$).  $t$ is the intrachain nearest-neighbor transfer and
$t_d$ is the interchain nearest-neighbor transfer energy.  The energy is
measured in $t$ units.
The energy minimum was given when the components of the gap function $\Delta_k$
take finite values plotted in Fig.\ref{fig15} for the lattice of $20\times 2$ sites
with 34 electrons imposing the periodic boundary condition.\cite{koi00}
Each component of $\Delta_k$ was optimized for $U_0=8$ and $t_d=1.8$.
There are two characteristic features; one is that the components of the bonding
and antibonding bands have opposite signs each other and second is that the
absolute values of $\Delta_k$ of the antibonding band ($k_y=\pi$) is much larger 
than that of the bonding band ($k_y=0$).
In order to reduce the computation cpu time, $\Delta_k$ of each band was
forced to take a fixed value specific to each band, i.e. $\Delta_1$ for the
bonding band and $\Delta_2$ for the antibonding band.
This drastically reduces the number of the variational parameters but still
allows us to get a substantial value of the condensation energy.
$\Delta_1$ and $\Delta_2$ take opposite sign, which is similar to that of the
$d_{x^2-y^2}$ gap function.

The energy gain $\Delta F_{2c}$ remains
finite in the bulk limit when $1.2<t_d<1.6$.
The SC condensation energy per site in the bulk limit is plotted as a function
of $t_d$ in Fig.\ref{fig17}.\cite{koi00}
The SC region derived from the SC condensation energy in the bulk limit
is consistent with the results obtained from the density-matrix renormalization
group\cite{noa96,noa97} and the exact-diagonalization 
method.\cite{yam94,yam94b,yan95}
The maximum value of $\Delta E_{2c}$ is 0.0008 which is of the same order of
magnitude as the maximum condensation energy obtained for the 2D Hubbard 
model.\cite{yam98}

\subsection{Condensation Energy in the d-p Model}

The SC energy gain for the d-p model, namely, three-band Hubbard model in eq.(\ref{dpm})
has also been evaluated using the variational Monte Carlo method.
For the three-band model the wave functions are written as
\begin{eqnarray}
\psi_n&=& P_G\prod_{|k|\leq k_F,\sigma}\alpha_{k\sigma}^{\dag}|0\rangle ,\\
\psi_{SC}&=& P_GP_{N_e}\prod_{k}(u_k+v_k\alpha^{\dag}_{k\uparrow}
\alpha^{\dag}_{-k\downarrow})|0\rangle ,
\label{sc}
\end{eqnarray}
where $\alpha_{k\sigma}$ is the linear combination of $d_{k\sigma}$,
$p_{xk\sigma}$ and $p_{yk\sigma}$ constructed to express the operator for the
lowest band (in the hole picture) or the highest band (in the electron picture) 
of the non-interacting Hamiltonian.
The numerical calculations have been done in the hole picture.
The Gutzwiller parameter $g$, effective level difference 
$\tilde{\epsilon_p}-\tilde{\epsilon_d}$, chemical potential $\mu$ and superconducting
order parameter $\Delta$ are the variational parameters.

The similar results to the single-band Hubbard model were obtained as shown in
Fig.\ref{fig18} for $t_{pp}=0.0$, $U_d=8$ and $\epsilon_d-\epsilon_p=2$ in
$t_{dp}$ units where the calculations were performed in the hole picture.\cite{yan01}  
The SC condensation energy for the three-band model is
$E_{cond}\simeq 0.0005t_{dp}\simeq 0.75$ meV per site in the optimally doped 
region.  We set $t_{dp}=1.5$ eV as estimated in Table I.
There is a tendency that $E_{cond}$ increases as $\epsilon_d-\epsilon_p$
increases which is plotted in Fig.\ref{fig18c}.
This tendency is not, however, in accordance with NQR (nuclear quadrupole
resonance) study on cuprates.\cite{zhe95}
We think that the NQR experiments indicate an importance of the Coulomb
interaction on oxygen sites.  This will be discussed in section III.K.

\begin{figure}[htbp]
\includegraphics[width=7cm]{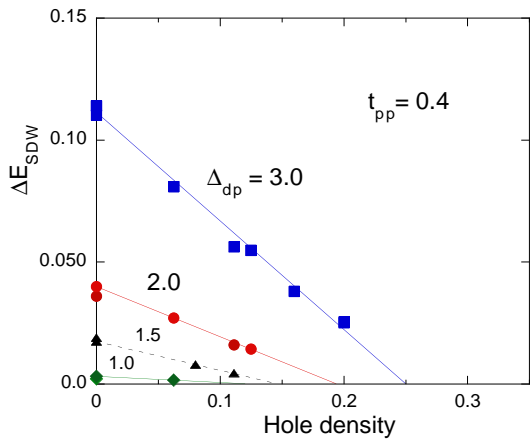}
\caption{
Energy gain per site $(E_{normal}-E)/N$ in the SDW state as a function of hole
density $\delta$ for the three-band Hubbard model.
Parameters are $t_{pp}=0.4$ and $U_d=8$ in $t_{dp}$ units.
From the top, $\Delta_{dp}\equiv\epsilon_p-\epsilon_d=3$, 2, 1.5 and 1.
The results are for $6\times 6$, $8\times 8$, $10\times 10$ and $16\times 12$
systems.  Antiperiodic and periodic boundary conditions are imposed in $x$-
and $y$-direction, respectively.
Monte Carlo statistical errors are smaller than the size of symbols.
Curves are a guide to the eye.\cite{yan01}
}
\label{fig21}
\end{figure}

\begin{figure}[htbp]
\includegraphics[width=8cm]{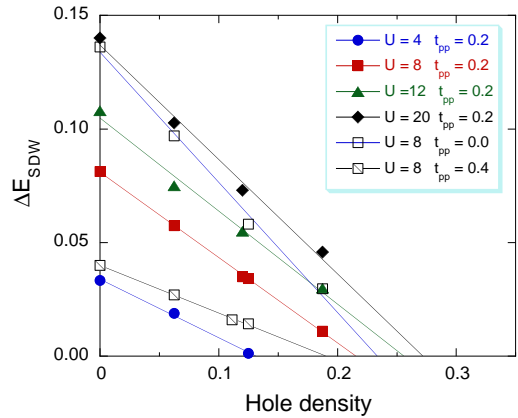}
\caption{
Uniform SDW energy gain per site with reference to the normal-state energy
as a function of the hole density for the three-band Hubbard model.  
Data are from $8\times 8$, $10\times 10$,
$12\times 12$ and $16\times 12$ systems for $\epsilon_p-\epsilon_d=2$.
For solid symbols $U_d=4$ (circles), $U_d=8$ (squares), $U_d=12$ (triangles) and
$U_d=20$ (diamonds) for $t_{pp}=0.2$.  For open symbols $U_d=8$ and $t_{pp}=0$,
and for open squares with slash $U_d=8$ and $t_{pp}=0.4$.
The lines are a guide to the eye.  The Monte Carlo statistical errors are
smaller than the size of symbols.\cite{yan02}
}
\label{fig22}
\end{figure}


\begin{figure}[htbp]
\includegraphics[width=9cm]{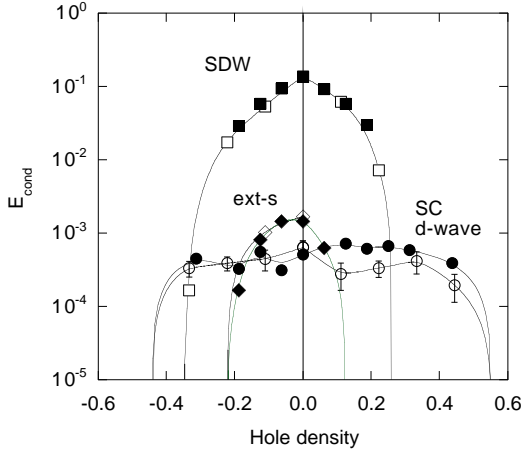}
\caption{
Condensation energy per site as a function of hole density for the three-band
Hubbard model where
$t_{pp}=0.0$, $\epsilon_p-\epsilon_d=2$ and $U_d=8$.
Circles and squares denote the energy gain per site with
reference to the normal-state energy for $d$-wave, ext-$s$ wave and SDW states, 
respectively.
For extremely small doping rate, the extended $s$-wave state is more favorable than
the $d$-wave state.
Solid symbols are for $8\times 8$ and open symbols are for $6\times 6$.
Curves are a guide to the eye.  
}
\label{phase-6x6}
\end{figure}

\begin{figure}[htbp]
\includegraphics[width=7.5cm]{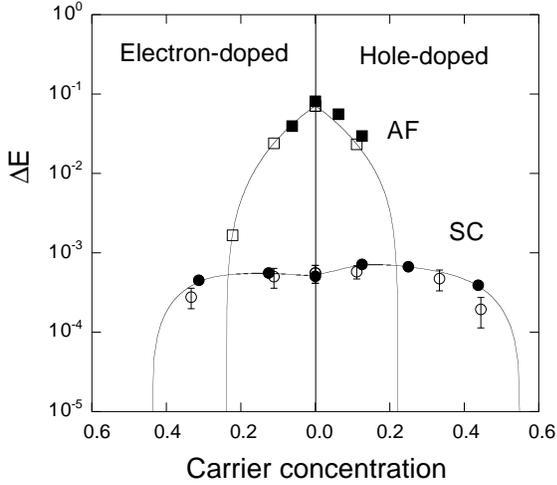}
\caption{
Condensation energy per site as a function of hole density for the three-band
Hubbard model where
$t_{pp}=0.2$, $\epsilon_p-\epsilon_d=2$ and $U_d=8$.
Circles and squares denote the energy gain per site with
reference to the normal-state energy for $d$-wave and SDW states, respectively.
Solid symbols are for $8\times 8$ and open symbols are for $6\times 6$.
Curves are a guide to the eye.  
}
\label{dEP}
\end{figure}

\begin{figure}[htbp]
\includegraphics[width=\columnwidth]{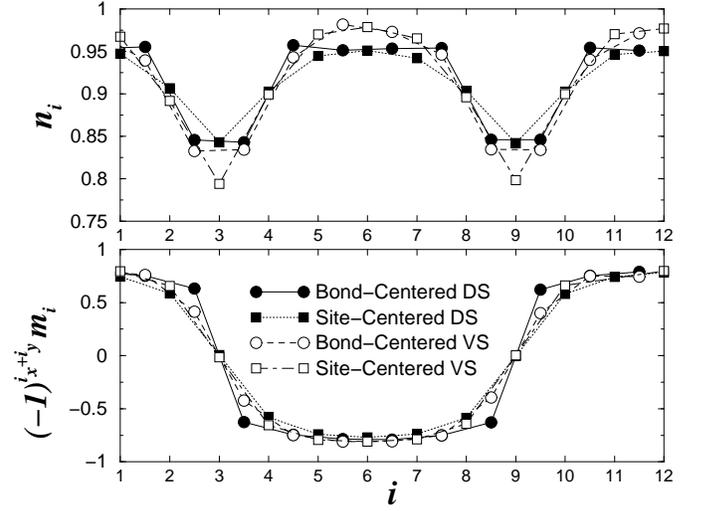}
\caption{
Charge and spin density as a function of the distance for a striped 
state.\cite{miy04}
}
\label{tranq}
\end{figure}

\begin{figure}[htbp]
\includegraphics[width=7cm]{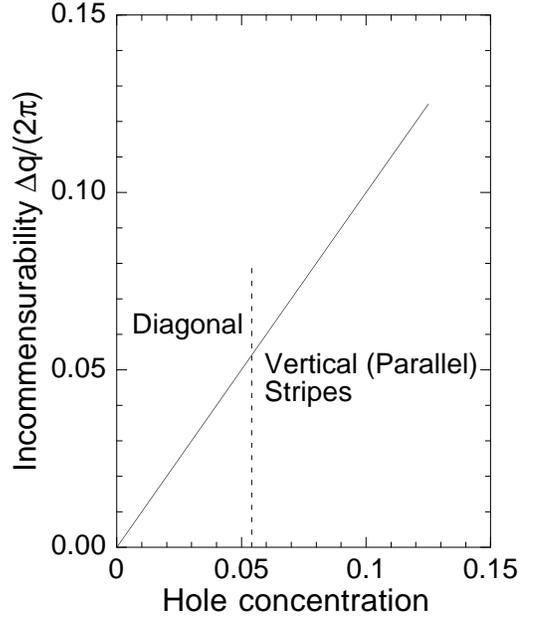}
\caption{
Schematic illustration of the incommensurability versus hole density.
}
\label{incoms}
\end{figure}

\begin{figure}[htbp]
\includegraphics[width=7cm]{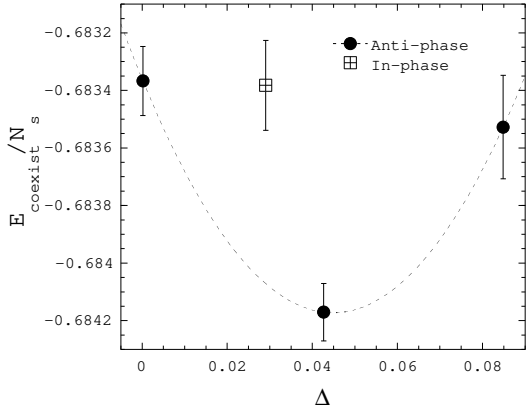}
\caption{
Coexistent state energy per site $E_{coexist}/N_s$ versus $\Delta$ for the
case of 84 electrons on $12\times 8$ sites with $U=8$ and $t'=-0.2$.
Here the vertical stripe state has 8-lattice periodicity for the hole
density $p=0.125$.  Only $E_{coexist}/N_s$ for the optimized gap is
plotted for the in-phase superconductivity.
}
\label{miyafig1}
\end{figure}

\begin{figure}[htbp]
\includegraphics[width=8cm]{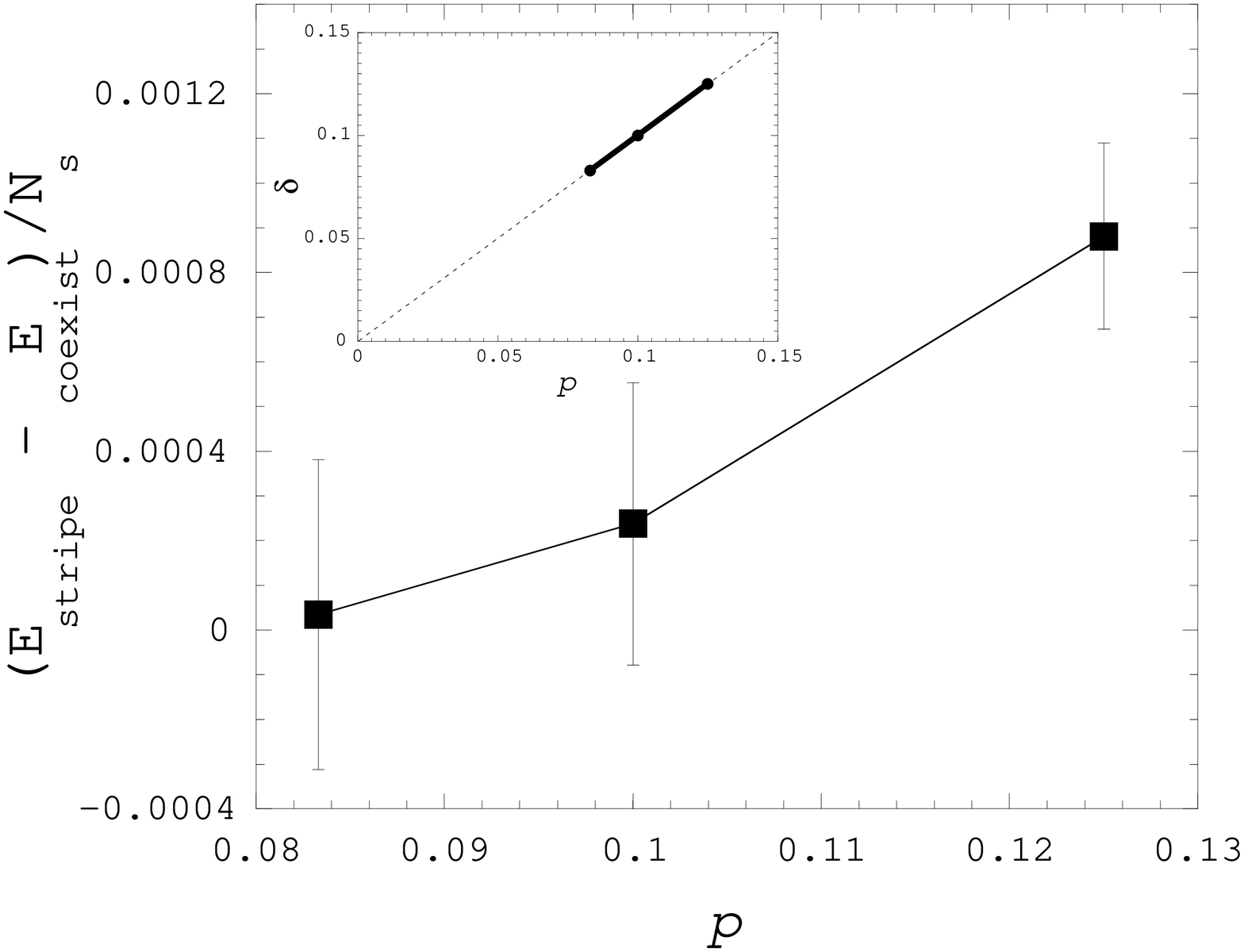}
\caption{
Superconducting condensation energy per site in the coexistence state as a 
function of the hole density $p=0.0833$, 0.10 and 0.125.
The model is the single-band Hubbard Hamiltonian with $t'=-0.20$.  
The stripe interval is preserved constant.
The inset shows the hole dependence of the incommensurability in
the coexistent state.\cite{miy02}
}
\label{coexi}
\end{figure}

\begin{figure}[htbp]
\includegraphics[width=7cm, angle=90]{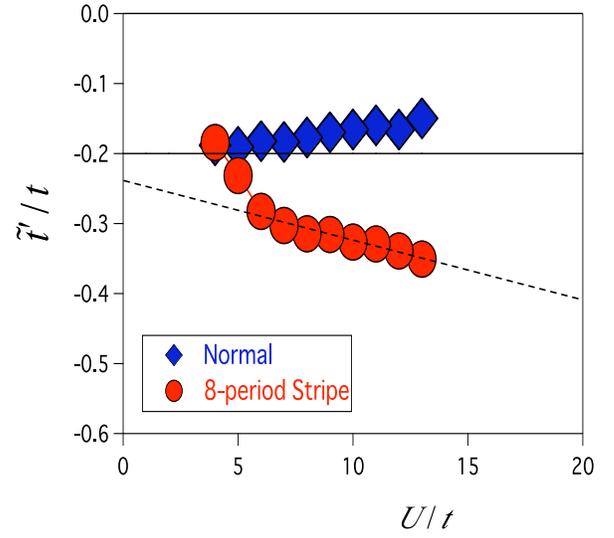}
\caption{
Optimized effective second neighbor transfer energy $\tilde{t}'/t$ as a function 
of $U/t$.  The system is a $16\times 16$ lattice with $t'/t=-0.2$ and the
electron density 0.875.
}
\label{stripe-t2}
\end{figure}

\begin{figure}[htbp]
\includegraphics[width=7cm]{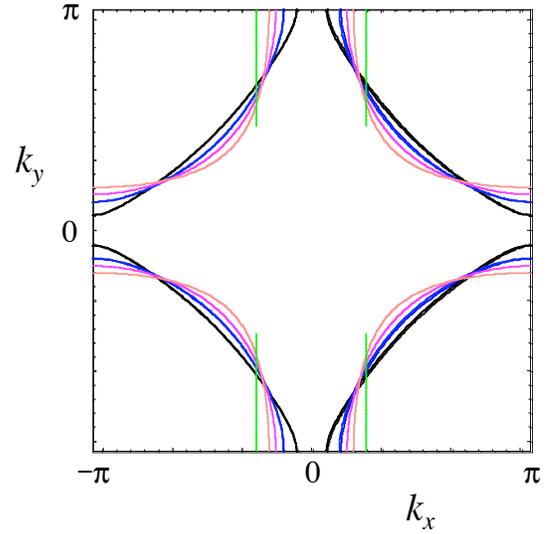}
\caption{
Renormalized quasi-Fermi surface for $\tilde{t}'/t=-0.3$, $-0.4$ and $-0.5$. 
The system is the same as that in Fig.\ref{stripe-t2}.
}
\label{stripe-FS}
\end{figure}

\begin{figure}[htbp]
\includegraphics[width=7cm, angle=270]{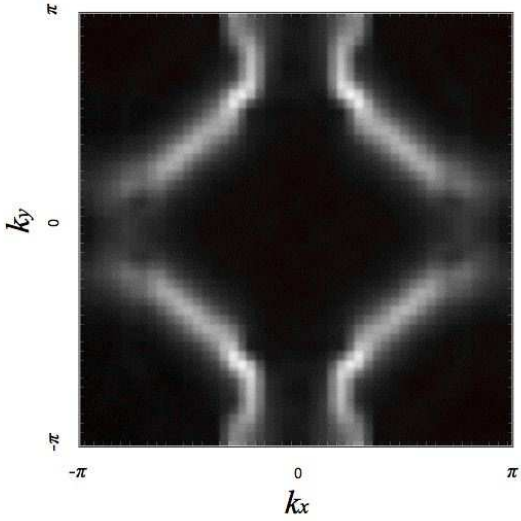}
\caption{
Contour plot of $|\nabla n_k|$ measured for the projected stripe state
on $24\times 24$ lattice with $t'/t=-0.2$. 
The electron density is 0.875.
}
\label{stripe-nk}
\end{figure}

\begin{figure}[htbp]
\includegraphics[width=8cm]{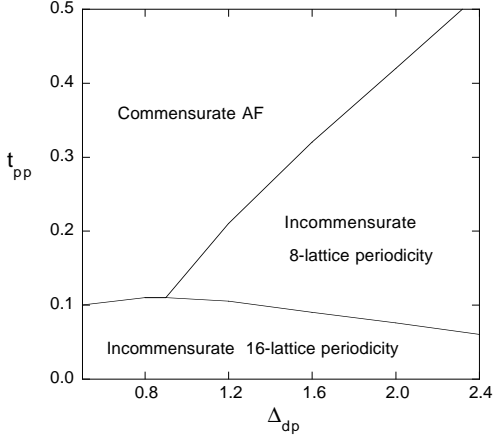}
\caption{
Phase diagram of stable antiferromagnetic state in the plane of
$\Delta_{dp}=\epsilon_p-\epsilon_d$ and $t_{pp}$ obtained for
$16\times 4$ lattice.
}
\label{phase-inc}
\end{figure}

\begin{figure}[htbp]
\includegraphics[width=8cm]{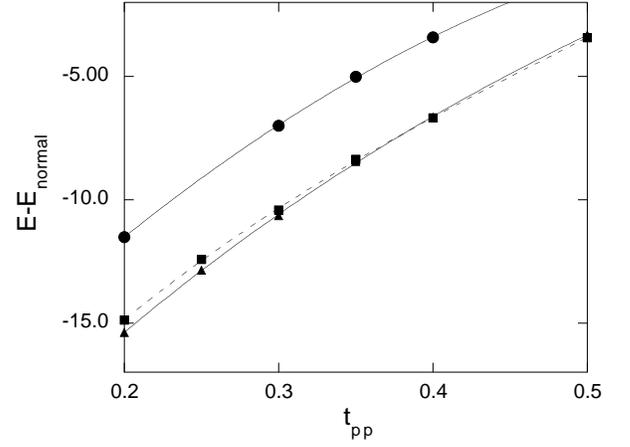}
\caption{
Energy as a function of $t_{pp}$ for 
$16\times 16$ square lattice at $x=1/16$.
Circles, triangles and squares denote the energy for 4-lattice stripes,
8-lattice stripes, and commensurate SDW, respectively, where
$n$-lattice stripe is the incommensurate state with one stripe per $n$
ladders.
The boundary conditions are antiperiodic in $x$-direction and
periodic in $y$-direction.\cite{yan02,yan09}
}
\label{str16}
\end{figure}

\begin{figure}[htbp]
\includegraphics[width=\columnwidth]{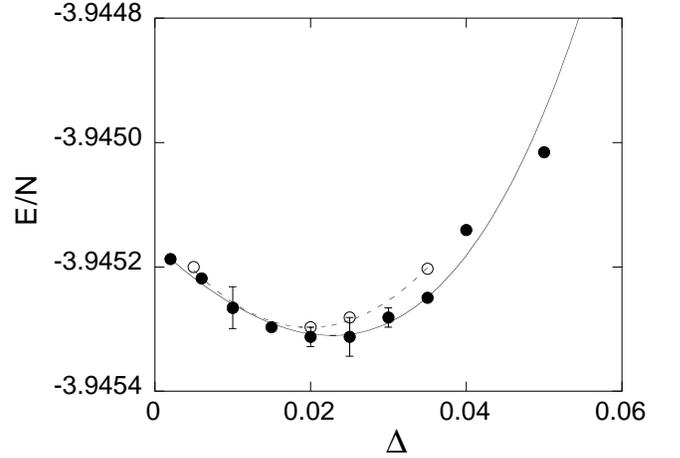}
\caption{
Energy of the coexistent state as a function of the SC order parameter
for $x=0.125$ on $16\times 4$ lattice.
We assume the incommensurate antiferromagnetic order (stripe).
Parameters are $\epsilon_p=0$, $\epsilon_d=-2$, $t_{pp}=0.4$  and $U_d=8$
in $t_{dp}$ units.
For solid circles the SC gap function is taken as
$\Delta_{i,i+\hat{x}}=\Delta{\rm cos}(Q_x(x_i+\hat{x}/2))$ and
$\Delta_{i,i+\hat{y}}=-\Delta{\rm cos}(Q_x(x_i))$, while for the open circles
$\Delta_{i,i+\hat{x}}=\Delta{\rm cos}|(Q_x(x_i+\hat{x}/2))|$ and
$\Delta_{i,i+\hat{y}}=-\Delta|{\rm cos}(Q_x(x_i))|$.
$Q_x=2\pi\delta=\pi/4$.
}
\label{3bcoex}
\end{figure}

\begin{figure}[htbp]
\includegraphics[width=\columnwidth]{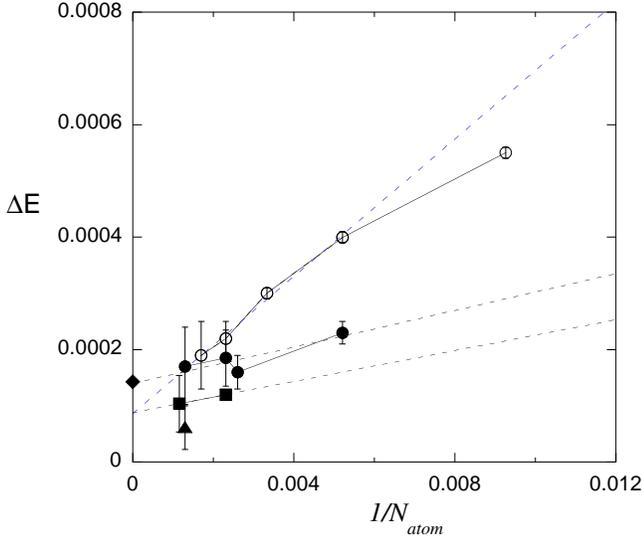}
\caption{
Energy gain due to the SC order parameter as a function of the system
size $N_{atom}=3N_s$.
Parameters are $\epsilon_p=0$, $\epsilon_d=-2$, $t_{pp}=0.4$ and $U_d=8$.
The open circle is for the simple $d$-wave pairing at the hole density $x=0.2$.
The solid symbols indicate the energy gain of the coexistent state;
the solid circle is at $x=0.125$, the solid
square is at $x=0.08333$ and the solid triangle is at $x=0.0625$.
The diamond shows the SC condensation energy obtained on the basis of
specific heat measurements\cite{lor93}.
}
\label{dE-N}
\end{figure}

\begin{figure}[htbp]
\includegraphics[width=7.5cm]{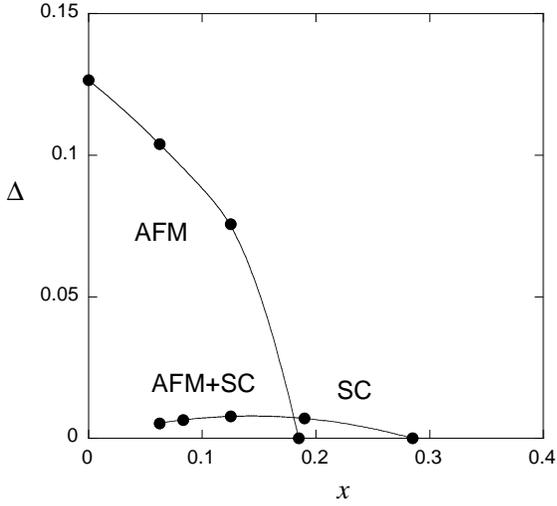}
\caption{
Phase diagram of the d-p model based on the Gutzwiller wave function.\cite{yan09}
}
\label{d-p-phase}
\end{figure}


\begin{figure}[htbp]
\includegraphics[width=8cm]{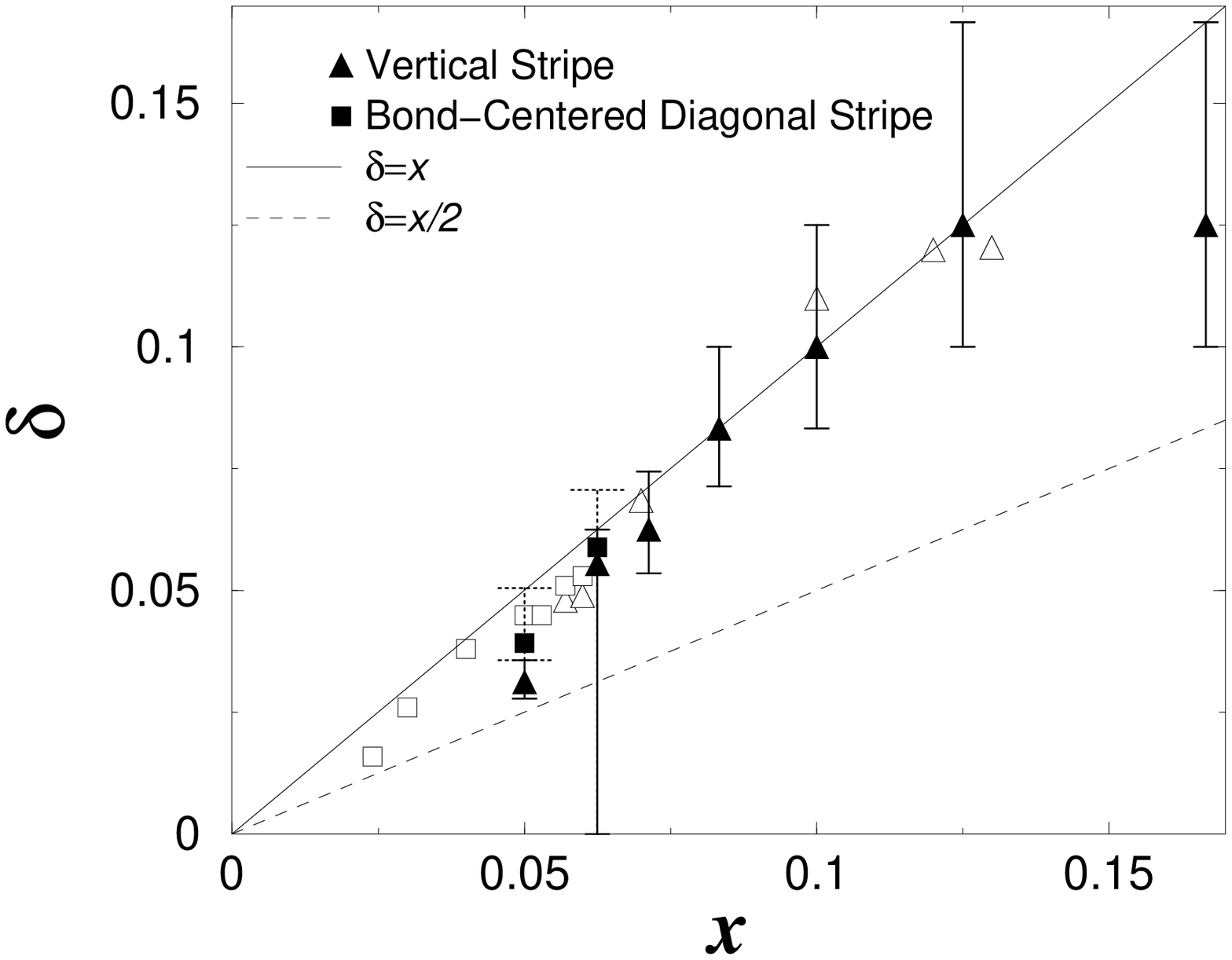}
\caption{
Incommensurability $\delta$ as a function of the hole density $x$ for
$U=8$ and $t'=-0.2$.\cite{miy04}
The numerical results for the vertical and the bond-centered diagonal stripe
state are represented by solid triangles and square symbols, respectively.
Open triangles and squares show the results of the vertical and diagonal
incommensurate SDW order observed from neutron scattering measurements,
respectively.\cite{fuj02}
}
\label{delta-x}
\end{figure}

\subsection{Antiferromagnetic State}

When the density of doped holes is zero or small, the 2D single-band or
three-band Hubbard model takes an antiferromganetic state as its ground
state.  
The magnetic order is destroyed and superconductivity appears
with the increase of doped hole density.
The transition between the $d$-wave SC and the uniform SDW states has been
investigated by computing the energy of the SDW state by using the
variational Monte Carlo method.
The trial SDW wave function is written as
\begin{equation}
\psi_{AF} = P_G\psi_{SDW} ,
\end{equation}
\begin{eqnarray}
\psi_{SDW} &=& \prod_{{\bf k}}(
u_{{\bf k}}c^{\dag}_{{\bf k}\uparrow}+v_{{\bf k}}c^{\dag}_{{\bf k}+Q\uparrow})
\nonumber\\
&\times& \prod_{{\bf k}'}(
u_{{\bf k}'}c^{\dag}_{{\bf k}'\downarrow}-v_{{\bf k}'}c^{\dag}_{{\bf k}'+Q\downarrow})
|0\rangle ,
\label{sdwh}
\end{eqnarray}
\begin{eqnarray}
u_{{\bf k}}&=& [(1-w_{{\bf k}}/(w_{{\bf k}}^2+\Delta_{AF}^2)^{1/2})/2]^{1/2} ,\\
v_{{\bf k}}&=& [(1+w_{{\bf k}}/(w_{{\bf k}}^2+\Delta_{AF}^2)^{1/2})/2]^{1/2} ,\\
w_{{\bf k}}&=& (\epsilon_{{\bf k}}-\epsilon_{{\bf k}+Q})/2 .
\end{eqnarray}
Summation over ${\bf k}$ and ${\bf k}'$ in eq.(\ref{sdwh}) is performed over the
filled ${\bf k}$-points, as in the calculation of the normal state energy.
$Q$ is the SDW wave vector given by $(\pi,\pi)$ and $\Delta_{AF}$ is the
SDW potential amplitude.

As shown in Fig.\ref{fig19}, the energy gain per site in the SDW state rises very sharply
from $n_e\sim 0.84$ for $t'=0$.\cite{yam98}  At $n_e\sim 0.84$ it is slightly larger
than that in the SC state, and at $n_e=0.80$ there is no more stable SDW
state.  Thus the boundary between the SDW and the SC states is given at
$n_e\sim 0.84$.
The results of the bulk limit calculations indicate that the energy gain in the
SC state at $n_e=0.84$ takes the extremely small value and the value at
$n_e=0.80$ vanishes as $N\rightarrow 0$.
Hence the pure $d$-wave SC state possibly exists near the boundary at
$n_e\sim 0.84$, but the region of pure SC state is very restricted.

Let us turn to the three-band model.
We show the antiferromagnetic-paramagnetic boundary for $t_{pp}=0.0$ and
$\epsilon_p-\epsilon_d=2$  
in the plane of $U$ and the hole density in Fig.\ref{fig20} where AF denotes the
antiferromagnetic region.\cite{yan02}
The value $\epsilon_p-\epsilon_d=2$ is taken
from the estimations by cluster calculations.\cite{hyb90,esk89,mcm90}
The phase boundary in the region of small $U$ is drawn from an extrapolation.
For the intermediate values of $U\sim 8-12$, the antiferromagnetic long-range
ordering exists up to about 20 percent doping.
Thus the similar features are observed compared to the single-band Hubbard model.

Since the three-band Hubbard model contains several parameters, we must examine
the parameter dependence of the energy of SDW state.
The energy gain $\Delta E_{SDW}$ in the SDW state is shown in Fig.\ref{fig21} as a
function of doping ratio for several values of 
$\Delta_{dp}\equiv\epsilon_p-\epsilon_d$.  $\Delta E_{SDW}$ increases as
$\Delta_{dp}$ increases as expected.
In Fig.\ref{fig22} $t_{pp}$- and $U_d$-dependencies of $\Delta E_{SDW}$ are
presented. 
The SDW phase extends up to 30 percent doping when $U_d$ is large.
It follows from the calculations that the SDW region will be reduced if 
$\epsilon_p-\epsilon_d$ and $U_d$ decrease.

From the calculations for the SDW wave functions, we should set
$\epsilon_p-\epsilon_d$ and $U_d$ small so that the SDW phase does not
occupy a huge region near half-filling.  In Figs.\ref{phase-6x6} and \ref{dEP} we show
energy gains for both the SDW and SC states for $U_d=8$, $t_{pp}=0$, $0.2$ and
$\epsilon_p-\epsilon_d=2$, where the right hand side and left hand side indicate
the hole-doped and electron-doped case, respectively.
Solid symbols indicate the results for
$8\times 8$ and open symbols for $6\times 6$.
For this set of parameters the SDW region extends up to 20 percent doping
and the pure $d$-wave phase exists outside of the SDW phase.
The $d$-wave phase may be possibly identified with superconducting phase in
the overdoped region in the high-$T_c$ superconductors.

\subsection{Stripes and its Coexistence with Superconductivity}

Incommensurate magnetic and charge peaks have been observed from
the elastic neutron-scattering experiments in the underdoped region of the
Nd-doped La$_{2-x-y}$Nd$_y$Sr$_x$CuO$_4$.\cite{tra96} (Fig.\ref{tranq})
Recent neutron experiments have also revealed the incommensurate
spin structures.\cite{suz98,yama98,ara99,moo00,wak00} 
Rapid decrease of the Hall resistivity in this region suggests that the
electric conduction is approximately one dimensional.\cite{nod99}
The angle-resolved photo-emission spectroscopy measurements also suggested
a formation of two sets of one-dimensional Fermi surface.\cite{zho99}
Then it has been proposed that these results might be understood in the
framework of the stripe state where holes are doped in the domain wall
between the undoped spin-density-wave domains.  This state is a kind of
incommensurate SDW state.
It was also shown that the incommensurability is proportional to the hole
density in the low-doping region in which the hole number per
stripe is half of the site number along one stripe.\cite{tra96,yama98}
A static magnetically ordered phase has been observed by $\mu$SR over a
wide range of SC phase for $0.05<x<0.1$ in La$_{2-x}$Sr$_x$CuO$_4$
(LSCO).\cite{nie98}
Thus the possibility of superconductivity that occurs in the stripe state is
a subject of great interest.\cite{kim99,mats99,mat00,fuj02}
The incommensurate magnetic scattering spots around $(\pi,\pi)$ were observed
in the SC phase in the range of $0.05<x<0.13$ in the elastic and inelastic
neutron-scattering experiments with LSCO.\cite{kim99,mats99,fuj02}
The hole dependence of the incommensurability and the configuration of the
spots around the Bragg spot in the SC phase indicated the vertical stripe.
The neutron-scattering experiments have also revealed that a diagonal
spin modulation occurs across the insulating spin-glass phase in
La$_{2-x}$Sr$_x$CuO$_4$ for $0.02\leq x\leq 0.05$, where a one-dimensional
modulation is rotated by 45 degrees from the stripe in the SC phase.
The incommensurability $\delta$ versus hole density is shown in Fig.\ref{incoms} 
schematically.\cite{mat00,fuj02}
The diagonal stripe changes into the vertical stripe across the boundary
between the insulating and SC phase.

Let us investigate the doped system 
from the point of modulated spin 
structures.\cite{gia91,mac89,poi89,kat90,sch90b,zaa96,ici99,whi98,whi98b,hel99,kob99}
The stripe SDW state has been studied theoretically by using the mean-filed
theory.\cite{mac89,poi89,kat90,sch90b,zaa96}
They found that the stripe state appears when an incommensurate nesting
becomes favorable in the hole-doped 2D Hubbard model.  
When the electron correlation correlation is strong or intermediate, it was
shown that the stripe state is more stable than the commensurate spin-density-wave
state with the wave vector $(\pi,\pi)$ in the ground state of the 2D Hubbard 
model by using the variational Monte Carlo method.\cite{gia91}
It has also been confirmed by the same means that the stripe states are
stabilized in the d-p model.\cite{yan03}
The purpose of this section is to examine whether the superconductivity can
coexist with static stripes in the 2D Hubbard model in a wider doping region
and investigate the doping dependence of the coexisting state.

We consider the 2D Hubbard model on a square lattice.
We calculate the variational energy in the coexistent state that is defined by
\begin{equation}
\psi_{coexist}= P_{N_e}P_G\phi_{coexist}^{MF},
\end{equation}
where $\phi_{coexist}^{MF}$ is a mean-field wave function.
The effective mean-field Hamiltonian for the coexisting state is\cite{miy02} 
represented by
\begin{eqnarray}
H_{MF}=\sum_{ij}(c_{i\uparrow}^{\dag}c_{i\downarrow})
\left(
\begin{array}{cc}
H_{ij\uparrow} & F_{ij} \\
F_{ji}^* & -H_{ji\downarrow} \\
\end{array}
\right)
\left(
\begin{array}{c}
c_{j\uparrow} \\
c_{j\downarrow}^{\dag} \\
\end{array}
\right),
\label{hstripes}
\end{eqnarray}
where the diagonal terms describe the incommensurate spin-density wave
state:
\begin{equation}
H_{ij\sigma}=-t_{ij}-\mu+\frac{U}{2}[n_i+{\rm sign}(\sigma)(-1)^{x_i+y_i}
m_i]\delta_{ij},
\end{equation}
where $\mu$ is the chemical potential.
The vertical stripe state is represented by the charge density $n_i$ and
the spin density $m_i$ that are spatially modulated as
\begin{eqnarray}
n_i&=& 1-\sum_{\ell}\alpha/\cosh\left(\frac{y_i-Y_{\ell}}{\xi_c}\right),\\
m_i&=& m\prod_{\ell}\tanh\left(\frac{y_i-Y_{\ell}}{\xi_c}\right),
\end{eqnarray}
where $Y_{\ell}$ denotes the position of vertical stripes.
The amplitude $\alpha$ is fixed by $\sum_in_i=N_e$.
The off-diagonal terms in eq.(\ref{hstripes}) are defined in terms of the
$d$-wave SC gap as
\begin{equation}
F_{ij}= \sum_{\hat{e}}\Delta_{ij}\delta_{ji+\hat{e}},
\end{equation}
where $\hat{e} =\pm\hat{x}$, $\pm\hat{y}$ (unit vectors).
We consider two types of the spatially inhomogeneous superconductivity:
anti-phase and in-phase defined as
\begin{eqnarray}
\Delta_{i.i+\hat{x}}= \Delta\cos(q_y(y_i-Y)),\\
\Delta_{i,i+\hat{y}}=-\Delta\cos(q_y(y_i-Y+\hat{y}/2)),
\end{eqnarray}
and
\begin{eqnarray}
\Delta_{i.i+\hat{x}}= \Delta|\cos(q_y(y_i-Y))|,\\
\Delta_{i,i+\hat{y}}=-\Delta|\cos(q_y(y_i-Y+\hat{y}/2))|,
\end{eqnarray}
respectively.  Here, ${\bf q}=(0,2\pi\delta)$ and $\delta$ is a
incommensurability given by the stripe's periodicity in the y direction with
regard to the spin.
The anti-phase (in-phase) means that the sign if the superconducting gap is
(is not) changed between nearest domain walls.

The wave function $\psi^0_{coexist}$ is made from the solution of the
Bogoliubov-de Gennes equation represented by
\begin{eqnarray}
\sum_j(H_{ij\uparrow}u^{\lambda}_j +F_{ij}v^{\lambda}_j)&=&E^{\lambda}u^{\lambda}_i ,
\\
\sum_j(F^*_{ji}u^{\lambda}_j -H_{ji\downarrow}v^{\lambda}_j)&=&E^{\lambda}v^{\lambda}_i.
\end{eqnarray}
The Bogoliubov quasiparticle operators are written in the form
\begin{eqnarray}
\alpha_{\lambda}&=& \sum_i(u^{\lambda}_ic_{i\uparrow}
+v^{\lambda}_ic^{\dag}_{i\downarrow})~~(E^{\lambda}>0),\\
\alpha_{\bar{\lambda}}&=& \sum_i(u^{\bar{\lambda}}_ic_{i\uparrow}
+v^{\bar{\lambda}}_ic^{\dag}_{i\downarrow})~~(E^{\bar{\lambda}}<0).
\end{eqnarray}
Then the coexistence wave function is written as\cite{miy02,him02}
\begin{eqnarray}
\psi^0_{coexist}&=& P_{N_e}\prod_{\lambda}\alpha_{\lambda}
\alpha^{\dag}_{\bar{\lambda}}|0\rangle\nonumber\\
&=& C P_{N_e}{\rm exp}\left(-\sum_{ij}(U^{-1}V)_{ij}c^{\dag}_{i\uparrow}
c^{\dag}_{j\downarrow}\right)|0\rangle\nonumber\\
&=& C' \left (
\sum_{ij}(U^{-1}V)_{ij}c^{\dag}_{i\uparrow}c^{\dag}_{j\downarrow}
\right )
^{N_e/2}|0\rangle ,
\end{eqnarray}
for constants $C$ and $C'$.  The calculations are performed for the wave
function $\psi_{coexist}=P_G\psi^0_{coexist}$.  The variational parameters
are $\mu$, $m$, $g$, $\xi_c$ and $\xi_s$.
The system parameters were chosen as $t'=-0.20$ and $U=8$ suitable for
cuprate superconductors.  It has been shown that the "anti-phase" configuration
is more stable than the "in-phase" one.\cite{miy02}

Here, the system parameters are $t'=-0.2$ and $U=8$.  We use the periodic
boundary condition in the $x$-direction and anti-periodic one in the
$y$-direction.  
In Fig.\ref{miyafig1}, we show the total energy of the coexistent state,
$E_{coexist}$, as a function of the SC gap $\Delta$ for the cases of anti-phase
and in-phase.
The SC condensation energy $\Delta E_{coexist}$ is estimated as 0.0008$t$ per 
site at the hole
density 0.125 on the $12\times 8$
lattice with periodic boundary condition in $x$-direction and antiperiodic
one in $y$-direction.  $\Delta E_{coexist}$ in the coexistence state is
defined as the decrease of energy due to finite $\Delta$.
If we use $t\sim 0.5$eV, this is evaluated as $\sim 0.4$meV.
The SC condensation energy per site is shown as a function of hole density
in Fig.\ref{coexi}.  One finds that $\Delta E_{coexist}$ in the stripe state
decreases as the hole density decreases.  This tendency is reasonable since
the SC order is weakened in the domain of the incommensurate SDW because of
the vanishingly small carrier concentration contributing the superconductivity  
in this domain.  This behavior is consistent with the SC condensation energy
estimated from the specific heat measurements.\cite{lor96}

There is a large renormalization of the Fermi surface due to the correlation
effect in the striped state.\cite{miy12}
We considered the next-nearest transfer $t'$ in the trial function as
a variational parameter $\tilde{t}'$.
In Fig.\ref{stripe-t2}, optimized values of $\tilde{t}'/t$ for the striped
state are shown as a function of $U/t$.  The $\tilde{t}'/t$ increases as
$U/t$ increases.  We also mention that the optimized $t''/t$ almost
vanishes.
The renormalized Fermi surface of $\tilde{t}'/t=-0.30$, $-0.40$ and $-0.50$
are plotted in Fig.\ref{stripe-FS}.  
The system is a $16\times 16$ lattice with $t'/t=-0.2$ and the electron
density 0.875.
As $U/t$ is incresed, the Fermi
surface is more deformed. 
We show the the gradient of the momentum distribution function, $|\nabla n_k|$,
calculated in the optimized stripe state in Fig.\ref{stripe-nk}.
The brighter areas coincide with the renormalized Fermi surface with
$\tilde{t}'/t=-0.31$ and $\tilde{t}''/t=0.0$ for $U/t=8$.

The calculations for the three-band Hubbard model has also been done taking
into account the coexistence of stripes and SC.\cite{yan01c,yan09}  
The energy of antiferromagnetic state would be lowered further if we consider
the incommensurate spin correlation in the wave function.
The phase diagram in Fig.\ref{phase-inc} presents the region of stable AF phase
in the plane of $t_{pp}$ and $\Delta_{dp}=\epsilon_p-\epsilon_d$.
For large $\Delta_{dp}=\epsilon_p-\epsilon_d$, we have the region of the AF
state with an eight-lattice periodicity in accordance with the results of
neutron-scattering measurements\cite{tra96,wak00}.
The energy at $x=1/16$ is shown in Fig.\ref{str16} where the 4-lattice 
stripe
state has higher energy than that for 8-lattice stripe for all the values
of $t_{pp}$.

The Bogoliubov-de Gennes
equation is extended to the case of three orbitals $d$, $p_x$ and $p_y$.
$(H_{ij\sigma})$ and $(F_{ij})$ are now $3N\times 3N$ matrices.
The energy of the state with double order parameters $\Delta$ and $m$ is shown
in Fig.\ref{3bcoex} on the $16\times 4$ lattice at the doping rate 1/8.
The SC condensation energy per site is evaluated as $\sim 0.00016t_{dp}$ for
$U_d=8$, $t_{pp}=0.4$ and $\epsilon_p-\epsilon_d=2$. 
If we use $t_{dp}\sim 1.5$eV,\cite{hyb90,esk89,mcm90} we obtain 
$\Delta E_{coexist}\sim 0.24$meV
which is slightly smaller than and close to the value obtained for the 
single-band Hubbard model.
We show the size dependence of the SC condensation energy for $x=0.2$, 0.125
0.08333 and 0.0625 in Fig.\ref{dE-N}.
We set the parameters as $\epsilon_p-\epsilon_d=2$ and $t_{pp}=0.4$ in
$t_{dp}$ units, which is reasonable from the viewpoint of the density of states
and is remarkably in accordance with cluster estimations\cite{hyb90,esk89,mcm90},
and also in the region of eight-lattice periodicity at $x=1/8$.
We have carried out the Monte Carlo calculations up to $16\times 16$ sites
(768 atoms in total).
In the overdoped region in the range of $0.18<x<0.28$,
we have the uniform $d$-wave pairing state as the ground state.
The periodicity of spatial variation  increases as the doping rate $x$
decreases proportional to $1/x$.  In the figure we have the 12-lattice
periodicity at $x=0.08333$ and 16-lattice periodicity at $x=0.0625$.
For $x=0.2$, 0.125 and 0.08333, the results strongly suggest a finite condensation
energy in the bulk limit.
The SC condensation energy obtained on the basis of specific heat
measurements agrees well with the variational Monte Carlo computations\cite{lor93}.
In general, the Monte Carlo statistical errors are much larger than those
for the single-band Hubbard model.  The large number of Monte Carlo
steps (more than 5.0$\times 10^7$) is required to get convergent expectation
values for each set of parameters.

In Fig.\ref{d-p-phase}  the order parameters $\Delta_{AF}$ and $\Delta_{SC}$ were
evaluated using the formula $E_{cond}=(1/2)N(0)\Delta^2$ where
$N(0)$ is the density of states.  Here we have set $N(0)\sim 5/t_{dp}$
since $N(0)$ is estimated as $N(0)\sim 2$ to 3 $(eV)^{-1}$ for
optimally doped YBCO using $N(0)(k_BT_c)^2/2$\cite{and98}.
The phase diagram is consistent with the recently reported phase
diagram for layered cuprates\cite{muk06}.

\subsection{Diagonal Stripe States in the Light-Doping Region}

Here we examine whether the relationship $\delta\sim x$ holds
in the lower doping region or not, and whether the diagonal stripe state is obtained
in the further lower doping region.\cite{miy04}
The elastic neutron scattering experiments of LSCO in the light-doping region,
$0.03<x<0.07$, revealed that the position of incommensurate magnetic peaks 
changed from $(1/2,1/2\pm\delta)$ to $(1/2\pm\delta',1/2\pm\delta')$ as $x$
becomes less than 0.06.\cite{mat00,fuj02}
This means that the stripe direction rotates by 45 degrees to become diagonal
at this transition.  In the diagonal stripe states, the magnetic peaks are
observed to keep a relation $\delta\sim x$ that holds in the vertical
stripe state in the low doping region.

In Fig.\ref{delta-x}, we show the incommensurability of the most stable stripe
state as a function of $x$.  Open squares and triangles are values for diagonal
and vertical incommensurate SDW's obtained in the elastic neutron scattering
experiments on LSCO, respectively.  Solid squares and triangles show our results
for the diagonal and vertical stripes, respectively.
These results are in a good agreement with experimental data. 
We also found that the phase boundary $x_{critical}$ between the diagonal and vertical
stripe states lies at or above 0.0625 in the case of $U=8$ and $t'=-0.2$.
The following factors may give rise to slight changes of the phase boundary 
$x_{critical}$: the diagonal stripe state may be stabilized in the 
low-temperature-orthorhombic (LTO) phase in LSCO.  The diagonal stripe state is
probably stabilized further by forming a line along larger next-nearest hopping
direction due to the anisotropic $t'$ generated by the Cu-O buckling in the
LTO phase.

\subsection{Checkerboard States}

A checkerboard-like density modulation with a roughly $4a\times 4a$ period
($a$ is a lattice constant) has also been observed by scanning tunneling
microscopy (STM) experiments in  Bi$_2$Sr$_2$CaCu$_2$O$_{8+\delta}$,\cite{hof02}
Bi$_2$Sr$_{2-x}$La$_x$CuO$_{6+\delta}$,\cite{wis08} and
Ca$_{2-x}$Na$_{x}$CuO$_2$Cl$_2$ (Na-CCOC)\cite{han04}.
It is important to clarify whether these inhomogeneous states can be 
understood within the framework of strongly correlated electrons.

Possible several electronic checkerboard states have been proposed 
theoretically.\cite{whi04,sei07,kat90}
The charge density $\rho_i$ and spin density $m_i$ are spatially modulated as
\begin{eqnarray}
\rho_i&=& \sum_{\ell}\rho_{\ell}{\rm cos}(Q_{\ell}^c\cdot ({\bf r}_i-{\bf r}_0)),\\
m_i&=& \sum_{\ell}m_{\ell}{\rm cos}(Q_{\ell}^s\cdot ({\bf r}_i-{\bf r}_0)).
\end{eqnarray}
where $\rho_{\ell}$ and $m_{\ell}$ are variational parameters.
The striped incommensurate spin-density wave state (ISDW) is defined by a single Q vector.
On the other hand, the checkerboard ISDW state is described by the double-Q set; for
example, vertical wave vectors $Q_1^s=(\pi,\pi\pm2\pi\delta)$ and
$Q_2^s=(\pi\pm2\pi\delta,\pi)$ describe a spin vertical checkerboard state, where
two diagonal domain walls are orthogonal.
Diagonal wave vectors $Q_1^s=(\pi\pm 2\pi\delta,\pi\pm 2\pi\delta)$ and
$Q_2^s=(\pi\pm 2\pi\delta,\pi\mp 2\pi\delta)$ lead to a spin diagonal checkerboard
state with a $1/\delta$-period.
The hole density forms the charge vertical checkerboard pattern with vertical
wave vectors $Q_1^c=(0,\pm4\pi\delta)$ and $Q_2^c=(2\pi\pm4\pi\delta,2\pi)$ in
which the hole density is maximal on the crossing point of magnetic domain
walls in the spin diagonal checkerboard state.
If $\delta=1/8$ is assumed, the charge modulation pattern is consistent with
the $4a\times 4a$ charge structure observed in STM experiments.

\begin{figure}[htbp]
\includegraphics[width=8cm]{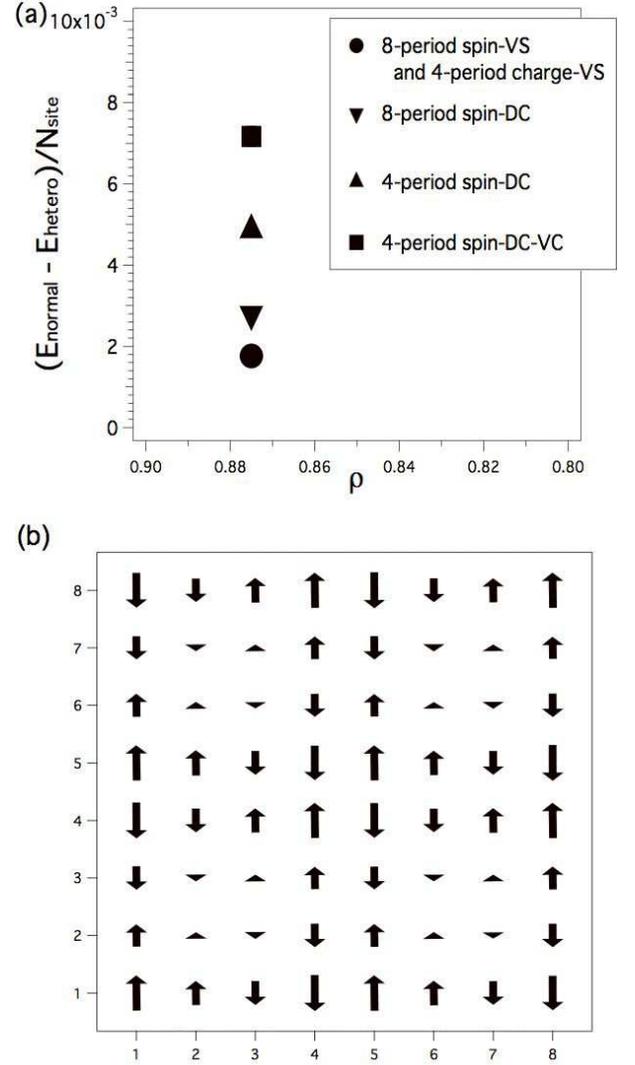}
\caption{
(a) Condensation energies of inhomogeneous states with the bond-centered
magnetic domain wall.  The system is a $16\times 16$ lattice  with
$t'=-0.32$, $t''=0.22$, and $U=8$ for the case of $\rho=0.875$.
The static error bars are smaller than the size of symbols.
(b) Expectation value of the spin density $m_i$ measured in the 
four-period spin-DC-VC solution.  The length of arrows is proportional to
the spin density.
}
\label{checker}
\end{figure}

  We found that the coexistent state of bond-centered four-period diagonal and
vertical spin-checkerboard structure characterized by a multi-Q set is stabilized and
composed of $4\times 4$ period checkerboard spin modulation.\cite{miy09}
In Fig.\ref{checker}(a), we show the condensation energies of some
heterogeneous states, $(E_{normal}-E_{hetero})/N_{site}$, with fixing the
transfer energies $t'=-0.32$ and $t''=0.22$ suitable for Bi-2212.
The system is a $16\times 16$ lattice with the electron-filling 
$\rho=N_e/N_{site}=0.875$.  The energy of the normal state $E_{normal}$ is 
calculated by adopting $m_{\ell}=\rho_{\ell}=0$.
In our calculations, the condensation energies of both bond-centered stripe
and checkerboard states are always larger than those of site-centered stripe
and checkerboard states.  The vertical stripe state is not suitable in this
parameter set since this state is only stabilized with the LSCO-type band.
The four-period spin-diagonal checkerboard (DC) state is significantly
more stable than the eight-period spin-DC state.
We found that the coexistent state of the bond-centered four-period 
spin-DC and four-period spin-vertical checkerboard (VC) with $\rho_{\ell}=0$
is most stable, as shown in Fig.\ref{checker}(a).
The measured expectation value of the spin density is shown in 
Fig.\ref{checker}(b).

\subsection{Improved Gutzwiller Function}

We have presented the results based on the Gutzwiller functions for the
normal state, SDW state and BCS state.  We must consider a method to go
beyond the Gutzwiller function-based Monte Carlo method.
One method to achieve this purpose is to multiply the Jastrow correlation
operators to take into account the intersite correlations.
The simplest possible candidate is an introduction of the diagonal intersite 
correlation factor:\cite{jas55}
\begin{equation}
\psi_{Jastrow}= \prod_{j\ell(\neq 0)}\prod_{\sigma\sigma'}
[1-(1-g(\ell))n_{j\sigma}n_{j+\ell\sigma'}]\psi_n ,
\end{equation}
for the variational parameter $g(\ell)$.
We have investigated the 2D Hubbard model by using the Jastrow-Gutzwiller
function.\cite{yam08}  The ground-state energy is lowered considerably
by considering the intersite correlations such as nearest neighbor and
next nearest neighbor spin and charge correlations.

Here we consider the method to improve the wave functions by an
off-diagonal 
Jastrow correlation operators.\cite{bla81,yan98,oht92} 
The off-diagonal correlation factors are more effective to lower the ground
state energy in 2D systems.
Let us consider the wave functions $\psi^{(m)}$ defined in the following 
way:\cite{yan98}
\begin{eqnarray}
\psi^{(1)}&=& \psi_G= {\rm e}^{-\alpha V}\psi_0 ,\\
\psi^{(2)}&=& {\rm e}^{-\lambda K}{\rm e}^{-\alpha V}\psi_0 ,\\
\psi^{(3)}&=& {\rm e}^{-\lambda' K}{\rm e}^{-\alpha' V}\psi^{(2)},\\ 
 \cdots && \cdots
\end{eqnarray}
and so on, where $K$ denotes the kinetic part of the Hamiltonian
\begin{equation}
K = -t\sum_{\langle ij\rangle\sigma}(c^{\dag}_{i\sigma}c_{j\sigma}+{\rm h.c.}),
\end{equation}
and $V$ denotes the on-site Coulomb interaction.
$\lambda$, $\lambda'$, $\lambda''$,$\alpha$, $\alpha'$ and $\alpha''$ are 
variational 
parameters.
It is considered that $\psi^{(m)}$ approaches the true ground state wave function
as $m$ grows larger since the ground state wave function is written as
\begin{equation}
\psi = {\rm e}^{-\beta H}\psi_0\simeq {\rm e}^{-\epsilon_1K}{\rm e}^{-\epsilon_1V}
\cdots {\rm e}^{-\epsilon_mK}{\rm e}^{-\epsilon_mV}\psi_0 ,
\end{equation}
for large $\beta=\epsilon_1+\cdots+\epsilon_m$ and small $\epsilon_i$ 
($i=1,\cdots,m$).
If we can extrapolate the expectation values from the data for $\psi^{(1)}$, $\psi^{(2)}$,
$\cdots$, we can evaluate correct expectation values.

The computations are performed by using the discrete Hubbard-Stratonovich
transformation as described in Section III.A.
In the evaluation of the expectation values we generate the Monte Carlo samples
by the importance sampling\cite{yan98} with the weight function 
$|w|=|w_{\uparrow}w_{\downarrow}|$ where
\begin{equation}
w_{\sigma}= {\rm det}(\phi^{\sigma\dag}_0{\rm exp}(V^{\sigma}(u,\alpha))
\cdots {\rm exp}(V^{\sigma}(s,\alpha)\phi^{\sigma}_0).
\end{equation}
Since the Monte Carlo samplings are generated with the weight $|w|$, the
expectation values are calculated with the sign of $w$ in the summation over the
generated samples.  In our calculations the negative sign problem has become
less serious due to the variational treatment, although we encounter the
inevitable negative sign problem in the standard projector Monte Carlo 
approaches.\cite{rae85}

\begin{figure}[htbp]
\includegraphics[width=7cm]{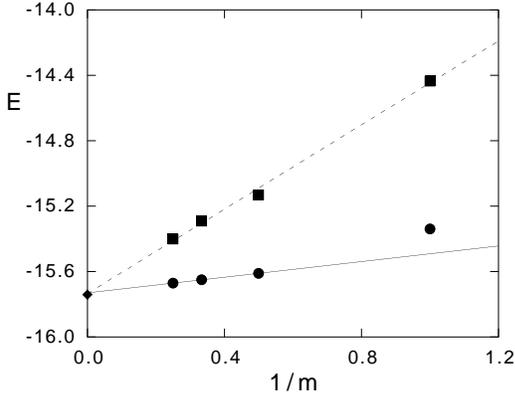}
\caption{
Energy as a function of $1/m$ for the single-band Hubbard model for
$N_e=14$ and $U=4$ on the lattice of $4\times 4$ sites.
For the upper and lower curves, the initial wave function $\psi_0$ is the
Fermi sea and SDW state, respectively.  The diamond indicates the exact
value obtained from the diagonalization.\cite{yan98}
}
\label{Evsm}
\end{figure}

\begin{figure}[htbp]
\includegraphics[width=7cm]{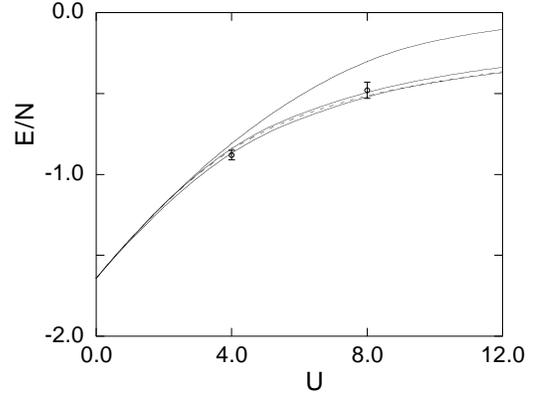}
\caption{
Energy as a function of $U$ for $8\times 8$ lattice at half-filling for the
single-band Hubbard model.
From the top the energies for $\psi_n$, $\psi_{AF}$, $\psi^{(3)}$ and
extrapolated values are shown.  The quantum Monte Carlo results are shown
by open circles.\cite{yan98}
}
\label{EhfH}
\end{figure}

In Fig.\ref{Evsm} the energy is shown as a function of $1/m$ where the SDW and normal
states are chosen as the initial state $\psi_0$.
The extrapolated values for different initial states coincide with each other
within Monte Carlo statistical errors.
The energy expectation values  as a function of $U$ for $8\times 8$ square lattice 
are presented in
Fig.\ref{EhfH} for $\psi_n=\psi^{(1)}$, $\psi_{AF}$, $\psi^{(3)}$.  The extrapolated
curve is shown by the solid curve and the results obtained by the quantum Monte Carlo
simulation (QMC)\cite{hir85} are also shown as a reference. 
A good agreement between two calculations support the method although the
QMC gives slightly higher energy for $U=8$.

\begin{figure}[htbp]
\includegraphics[width=7cm]{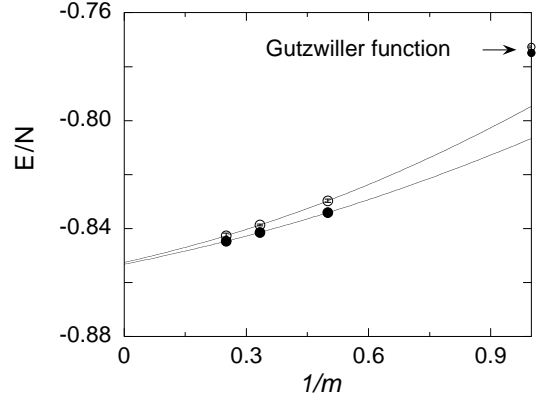}
\caption{
The energy versus $1/m$ for the single-band Hubbard model on the lattice of
$10\times 10$ sites.  Solid and open symbols are obtained for $\psi_m$ with
the normal and $d$-wave state initial functions $\psi_0$ and $\psi_{BCS}$, 
respectively.
Parameters are given by $U=8$, $t'=-0.09$ and $N_e=80$.\cite{yan99}
}
\label{sowf}
\end{figure}

One can formulate an approach to consider the BCS function with
correlation operators.\cite{yan99}
For this end the electron-hole transformation is introduced for the down 
spin\cite{yok88}:
$d_{\bf k}= c^{\dag}_{-{\bf k}\downarrow},~~d^{\dag}_{{\bf k}}=
c_{-{\bf k}\downarrow}$.
The up-spin electrons are unaltered.

We show the energy versus $1/m$ in Fig.\ref{sowf} for $\psi^{(m)}$ and
$\psi^{(m)}_s$.  
From an extrapolation to the limit $m\rightarrow\infty$, both formulations
predict the same limiting value for the energy.
The energy is lowered considerably due to the correlation
operators compared to that for the Gutzwiller function.
The energy in the $d$-wave SC state is always lower than that in the
normal state for each $m$.  The energy gain in the SC state remains the
same order after the multiplication of correlation factors.

\begin{figure}[htbp]
\includegraphics[width=8cm]{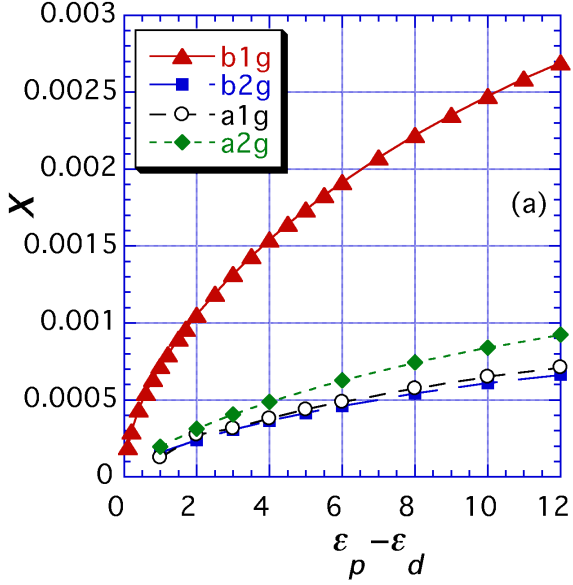}
\caption{
The exponent $x$ (superconductivity strength) as a function of 
$\epsilon_p-\epsilon_d$,
where the level difference $\epsilon_p-\epsilon_d$ is positive.
}
\label{x-ep-ed-a}
\end{figure}

\begin{figure}[htbp]
\includegraphics[width=8.5cm]{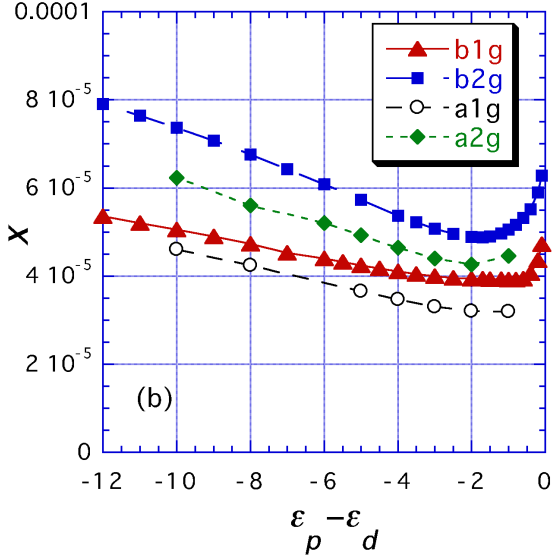}
\caption{
The exponent $x$ as a function of $\epsilon_p-\epsilon_d$,
where the level difference $\epsilon_p-\epsilon_d$ is negative.
}
\label{x-ep-ed-b}
\end{figure}

\subsection{$T_c$ and $\epsilon_p-\epsilon_d$}

Relationships between $T_c$ and structural features in cuprate 
high-temperature superconductors are very interesting.
Torrance and Metzger found the first such relationship between $T_c$
and the Madelung potential difference $\Delta V_M$.\cite{tor89}
Here $\Delta V_M$ is the potential difference between Cu and O sites
in the CuO$_2$ plane.
$T_c$ was found to increase with decreasing $\Delta V_M$.
There is an interesting tendency of increasing $T_c$ with increasing
relative ratio of hole density at oxygen site against that at copper 
site.\cite{zhe95}
 
Here we show the results obtained by using the perturbation
theory\cite{kon01,hlu99,koi01,koi06,yan08}. 
There have been many similar works by making some kind of approximation
such as Random phase approximation (RPA)\cite{sca86,shi88,mon91}, 
fluctuation-exchange
approximation (FLEX)\cite{bic89,pao94,mon94,yama99},
effective spin-fluctuations\cite{miy86,mor90,mor00},
and perturbation theory in terms of $U$\cite{hot93,juj99,nom00}.
An application was made for Sr$_2$RuO$_4$ where we need to consider
the multi-band structure including $\alpha$ and $\beta$
orbitals\cite{koi03}, and also to the three-dimensional d-p model\cite{koi10}.
In our formulation the gap function is
written as
\begin{equation}
\Delta = \exp\left( -\frac{2}{xU_d^2} \right).
\end{equation}
The exponent $x$ indicates the strength of superconductivity.
The results are in Figs. \ref{x-ep-ed-a} and \ref{x-ep-ed-b}.\cite{yam13}
As shown in the figure, for positive $\epsilon_p-\epsilon_d$, 
with increase of $\epsilon_p-\epsilon_d$ the exponent $x$ increases 
monotonously.  This means the increase of superconducting gap and 
so of $T_c$, and is consistent with the wide-range tendency of the 
variational Monte Carlo calculation.\cite{yan01,koik03}
This tendency can be understood in terms of
\begin{equation}
U_{eff}= U_d\langle u_{{\bf k}}^0\rangle^2= \frac{U_d}{4}\left(
1+\frac{1}{\sqrt{1+16t_{dp}^2/(\epsilon_p-\epsilon_d)^2}}\right)^2,
\end{equation}
where $u_{{\bf k}}^0$ is the weight ofd electrons.
This clearly indicates that increase of $\epsilon_p-\epsilon_d$ leads
to the increase of $U_{eff}$ and subsequently of $x$.
In the case of $\epsilon_p-\epsilon_d<0$, we take account of finite Coulomb 
repulsion $U_p$ on oxygen sites.
The effective interaction coming from $U_p$ is similarly given by the
susceptibility with the weight of $p$ electrons.
The results of $x$ with $U_p/U_d=6/8$ indicates that all four types of
even parity ($b_{1g}$, $b_{2g}$, $a_{1g}$ and $a_{2g}$) SC strength
values increase, so that $T_c$ is raised, as the absolute value
$|\epsilon_p-\epsilon_d|$ increases in this region.
This result shows that $U_p$ also plays an important role as well.

Let us give a discussion on this result.
Increase of $|\epsilon_p-\epsilon_d|$ in the region of 
$\epsilon_p-\epsilon_d<0$
means decrease of $\Delta V_M=|e|(V_M^O-V_M^{Cu})$ since
$\epsilon_p-\epsilon_d=(A_2^O-I_3^{Cu})+|e|(V_M^O-V_M^{Cu})$, where $A_2^O$
is the second electron affinity of oxygen atom and $I_3^{Cu}$ is the
third electron ionicity of copper atom and $e$ is the charge of electron.
Therefore, this relation is consistent with the systematics reported in
\cite{tor89}.
With increase of the distance of the apex oxygen away from the CuO$_2$
plane, cuprate superconductors are known to increase $T_c$.\cite{pav01}
The accompanying raise of $\epsilon_d$ should tend to increase $T_c$.

The Coulomb interaction between p electrons on oxygen atom will raise
the level of p electrons effectively.
This leads to the lowering of p hole level $\epsilon_p$ or the
raise of $\epsilon_d$ relatively.
This indicates that $T_c$ will be increased by the Coulomb interaction
between p electrons.

\section{Quantum Monte Carlo Studies}

\subsection{Quantum Monte Carlo Method}

The Quantum Monte Carlo (QMC) method is a numerical method that is employed to
simulate the behavior of correlated electron systems.
We outline the QMC method in this section.
The Hamiltonian is the Hubbard model that contains the on-site Coulomb
repulsion and is written as

\begin{eqnarray}
H&=&-\sum_{ij\sigma}t_{ij}(c^{\dag}_{i\sigma}c_{j\sigma}+h.c.)
+ U\sum_jn_{j\uparrow}n_{j\downarrow},
\label{hamil}
\end{eqnarray}
where $c^{\dag}_{j\sigma}$ ($c_{j\sigma}$) is the creation
(annihilation) operator of an electron with spin $\sigma$
at the $j$-th site and $n_{j\sigma}=c^{\dag}_{j\sigma}c_{j\sigma}$.
$t_{ij}$ is the transfer energy between the sites $i$ and $j$.
$t_{ij}=t$ for the nearest-neighbor bonds. For all other cases $t_{ij}=0$.
$U$ is the on-site Coulomb energy.
The number of sites is $N$ and
the linear dimension of the system is denoted as $L$.
The energy unit is given by $t$ and
the number of electrons is denoted as $N_e$.

In a Quantum Monte Carlo simulation, the ground state wave function is
\begin{equation}
\psi= {\rm e}^{-\tau H}\psi_0 ,
\label{psi}
\end{equation}
where $\psi_0$ is the initial one-particle state represented by a Slater
determinant.
For large $\tau$, ${\rm e}^{-\tau H}$ will project out the ground state from
$\psi_0$.
We write the Hamiltonian as $H=K+V$ where
K and V are the kinetic and interaction terms of the Hamiltonian in
Eq.(\ref{hamil}), respectively.
The wave function in Eq.(\ref{psi}) is written as
\begin{equation}
\psi= ({\rm e}^{-\Delta\tau (K+V)})^M \psi_0 \approx
({\rm e}^{-\Delta\tau K}{\rm e}^{-\Delta\tau V})^M \psi_0 ,
\end{equation}
for $\tau=\Delta\tau\cdot M$.
Using the Hubbard-Stratonovich transformation\cite{hir83,bla81}, we have
\begin{eqnarray}
{\rm exp}(-\Delta\tau Un_{i\uparrow}n_{i\downarrow})
&=& \frac{1}{2}\sum_{s_i=\pm 1}{\rm exp}( 2as_i(n_{i\uparrow}
-n_{i\downarrow})\nonumber\\
&-&\frac{1}{2}U\Delta\tau(n_{i\uparrow}+n_{i\downarrow}) ),
\end{eqnarray}
for $({\rm tanh}a)^2={\rm tanh}(\Delta\tau U/4)$ or
${\rm cosh}(2a)={\rm e}^{\Delta\tau U/2}$. 
The wave function is expressed as a summation of the one-particle Slater
determinants
over all the configurations of the auxiliary fields
$s_j=\pm 1$. 
The exponential operator is expressed as
\begin{eqnarray}
({\rm e}^{-\Delta\tau K}{\rm e}^{-\Delta\tau V})^M
&=& \frac{1}{2^{NM}}\sum_{\{s_i(\ell)\}}\prod_{\sigma}B_M^{\sigma}(s_i(M))
\nonumber\\ 
&\times& B_{M-1}^{\sigma}(s_i(M-1))\cdots B_1^{\sigma}(s_i(1)),\nonumber\\
\end{eqnarray}
where we have defined
\begin{equation}
B_{\ell}^{\sigma}(\{s_i(\ell)\})={\rm e}^{-\Delta\tau K_{\sigma}}
{\rm e}^{-V_{\sigma}(\{s_i(\ell)\})},
\label{bmat}
\end{equation}
for
\begin{eqnarray}
V_{\sigma}(\{s_i\})&=& 2a\sigma\sum_i s_in_{i\sigma}-\frac{1}{2}
U\Delta\tau \sum_in_{i\sigma},\\
K_{\sigma}&=&-\sum_{ij}t_{ij}(c_{i\sigma}^{\dag}c_{j\sigma}+h.c.).
\end{eqnarray}
The ground-state wave function is
\begin{equation}
\psi= \sum_mc_m\phi_m ,
\label{wf}
\end{equation}
where $\phi_m$ is a Slater determinant corresponding to a configuration
$m=\{s_i(\ell)\}$ ($i=1,\cdots,N; \ell=1,\cdots,M$)
of the auxiliary fields:
\begin{eqnarray}
\phi_m&=& \prod_{\sigma}B_M^{\sigma}(s_i(M))\cdots B_1^{\sigma}(s_i(1))\psi_0
\nonumber\\
&\equiv& \phi_m^{\uparrow}\phi_m^{\downarrow}.
\end{eqnarray}
The coefficients $c_m$ are constant real numbers: $c_1=c_2=\cdots$.
The initial state $\psi_0$ is a one-particle state.
If electrons occupy the wave numbers $k_1$, $k_2$, $\cdots$, $k_{N_{\sigma}}$
for each spin $\sigma$, $\psi_0$ is given by the product
$\psi_0^{\uparrow}\psi_0^{\downarrow}$ where $\psi_0^{\sigma}$ is the matrix
represented as\cite{ima89}
\begin{equation}
\left( \begin{array}{ccccc}
{\rm e}^{ik_1\cdot r_1} & {\rm e}^{ik_2\cdot r_1} & \cdots & \cdots &
{\rm e}^{ik_{N_{\sigma}}\cdot r_1}  \\
{\rm e}^{ik_1\cdot r_2} & {\rm e}^{ik_2\cdot r_2} & \cdots & \cdots &\cdots\\
\cdot & \cdot & \cdot & \cdot & \cdot \\
{\rm e}^{ik_1\cdot r_N} & {\rm e}^{ik_2\cdot r_N} & \cdots & \cdots &
\end{array} \right).
\end{equation}
$N_{\sigma}$ is the number of electrons for spin $\sigma$.
In actual calculations we can use a real representation where the matrix
elements are cos$(k_i\cdot r_j)$ or sin$(k_i\cdot r_j)$.
In the real-space representation, the matrix of $V_{\sigma}(\{s_i\})$ is a
diagonal matrix given as
\begin{equation}
V_{\sigma}(\{s_i\})={\rm diag}(2a\sigma s_1-U\Delta\tau/2,\cdots,
2a\sigma s_N-U\Delta\tau/2).
\end{equation}
The matrix elements of $K_{\sigma}$ are
\begin{eqnarray}
(K_{\sigma})_{ij}&=& -t~~~i,j~ {\rm are~ nearest~ neighbors}\nonumber\\
&=& 0~~~{\rm otherwise}.
\end{eqnarray}
$\phi_m^{\sigma}$ is an $N\times N_{\sigma}$ matrix given by the product
of the matrices ${\rm e}^{-\Delta\tau K_{\sigma}}$, ${\rm e}^{V_{\sigma}}$
and $\psi_0^{\sigma}$.
The inner product is thereby calculated as a determinant\cite{zha97},
\begin{equation}
\langle\phi_m^{\sigma}\phi_n^{\sigma}\rangle=
{\rm det}(\phi_m^{\sigma\dag}\phi_n^{\sigma}).
\end{equation}
The expectation value of the quantity $Q$ is evaluated as
\begin{equation}
\langle Q\rangle = \frac{\sum_{mn}\langle\phi_m Q\phi_n\rangle}
{\sum_{mn}\langle\phi_m\phi_n\rangle}.
\label{qexpe}
\end{equation}
If $Q$ is a bilinear operator $Q_{\sigma}$ for spin $\sigma$, we have
\begin{eqnarray}
\langle Q_{\sigma}\rangle &=& \frac{
\sum_{mn}\langle\phi_m^{\sigma}Q_{\sigma}\phi_n^{\sigma}\rangle
\langle\phi_m^{-\sigma}\phi_n^{-\sigma}\rangle}
{\sum_{mn}\langle\phi_m^{\sigma}\phi_n^{\sigma}\rangle
\langle\phi_m^{-\sigma}\phi_n^{-\sigma}\rangle}
\nonumber\\
&=& \frac{\sum_{mn}\langle\phi_m^{\sigma}Q_{\sigma}\phi_n^{\sigma}\rangle
{\rm det}(\phi_m^{-\sigma\dag}\phi_n^{-\sigma})}
{\sum_{mn}{\rm det}(\phi_m^{\sigma\dag}\phi_n^{\sigma})
{\rm det}(\phi_m^{-\sigma\dag}\phi_n^{-\sigma})}
\nonumber\\
&=& \sum_{mn}\frac{{\rm det}(\phi_m^{\sigma\dag}\phi_n^{\sigma})
{\rm det}(\phi_m^{-\sigma\dag}\phi_n^{-\sigma})}
{\sum_{m'n'}{\rm det}(\phi_{m'}^{\sigma\dag}\phi_{'n}^{\sigma})
{\rm det}(\phi_{m'}^{-\sigma\dag}\phi_{n'}^{-\sigma})}\nonumber\\
&\times& \frac{\langle\phi_m^{\sigma}Q_{\sigma}\phi_n^{\sigma}\rangle}
{\langle\phi_m^{\sigma}\phi_n^{\sigma}\rangle}.
\end{eqnarray}
The expectation value with respect to the Slater determinants
$\langle\phi_m^{\sigma}Q_{\sigma}\phi_n^{\sigma}\rangle$ is evaluated
using the single-particle Green's function\cite{ima89,zha97},
\begin{equation}
\frac{\langle\phi_m^{\sigma}c_{i\sigma}c_{j\sigma}^{\dag}\phi_n^{\sigma}
\rangle}
{\langle\phi_m^{\sigma}\phi_n^{\sigma}\rangle}
=\delta_{ij}-(\phi_n^{\sigma}
(\phi_m^{\sigma\dag}\phi_n^{\sigma})^{-1}\phi_m^{\sigma\dag})_{ij}.
\end{equation}
In the above expression,
\begin{equation}
P_{mn}\equiv {\rm det}(\phi_m^{\sigma}\phi_n^{\sigma})
{\rm det}(\phi_m^{-\sigma}\phi_n^{-\sigma})
\end{equation}
can be regarded as the weighting factor to obtain the Monte Carlo samples.
Since this quantity is not necessarily positive definite, the weighting factor
should be $|P_{mn}|$; the resulting relationship is,
\begin{eqnarray} 
\langle Q_{\sigma}\rangle&=& \sum_{mn}P_{mn}\langle Q_{\sigma}\rangle_{mn}
/\sum_{mn}P_{mn}\nonumber\\
&=& \sum_{mn}|P_{mn}|sign(P_{mn})\langle Q_{\sigma}\rangle_{mn}
/\sum_{mn}|P_{mn}|sign(P_{mn})\nonumber\\
\end{eqnarray}
where
$sign(a)=a/|a|$ and
\begin{equation}
\langle Q_{\sigma}\rangle_{mn}=\frac{\langle\phi_m^{\sigma}Q_{\sigma}
\phi_n^{\sigma}\rangle}{\langle\phi_m^{\sigma}\phi_n^{\sigma}\rangle}.
\end{equation} 
This relation can be evaluated using a Monte Carlo procedure if an
appropriate algorithm, such as the Metropolis or heat bath method, is
employed\cite{bla81}.
The summation can be evaluated using appropriately defined Monte Carlo
samples,
\begin{equation}
\langle Q_{\sigma}\rangle= \frac{ \frac{1}{n_{MC}}\sum_{mn}sign(P_{mn})
\langle Q_{\sigma}\rangle_{mn}}{\frac{1}{n_{MC}}\sum_{mn}sign(P_{mn})},
\label{qqmc}
\end{equation}
where $n_{MC}$ is the number of samples.
The sign problem is an issue if the summation of $sign(P_{mn})$
vanishes within statistical errors.  In this case it is indeed impossible
to obtain definite expectation values.

\subsection{Quantum Monte Carlo Diagonalization}

\subsubsection{Basic Method and Optimization}

Quantum Monte Carlo diagonalization (QMD) is a method for the evaluation of
$\langle Q_{\sigma}\rangle$ without {\em the negative sign problem}.\cite{yan07}
A bosonic version of this method was developed before in Ref.\cite{miz86}.
The configuration space of the probability $\|P_{mn}\|$ in Eq.(\ref{qqmc})
is generally very strongly peaked.
The sign problem lies in the distribution of $P_{mn}$ in the configuration
space.
It is important to note that the distribution of the basis functions $\phi_m$
($m=1,2,\cdots$) is uniform since $c_m$ are constant numbers: $c_1=c_2=\cdots$.
In the subspace $\{\phi_m\}$, selected from all configurations of auxiliary
fields, the right-hand side of Eq.(\ref{qexpe}) can be determined.
However, the large number of basis states required to obtain accurate
expectation values is beyond the current storage capacity of computers.
Thus we use the variational principle to obtain the expectation values.

From the variational principle,
\begin{equation}
\langle Q\rangle = \frac{\sum_{mn}c_mc_n\langle\phi_m Q\phi_n\rangle}
{\sum_{mn}c_mc_n\langle\phi_m\phi_n\rangle},
\end{equation}
where $c_m$ ($m=1,2,\cdots$) are variational parameters.
In order to minimize the energy
\begin{equation}
E = \frac{\sum_{mn}c_mc_n\langle\phi_m H\phi_n\rangle}
{\sum_{mn}c_mc_n\langle\phi_m\phi_n\rangle},
\end{equation}
the equation $\partial E/\partial c_n=0$ ($n=1,2,\cdots$) is solved for,
\begin{equation}
\sum_m c_m\langle\phi_n H\phi_m\rangle-E\sum_m c_m\langle\phi_n\phi_m\rangle
=0.
\end{equation}
If we set
\begin{eqnarray}
H_{mn}&=&\langle\phi_m H\phi_n\rangle,\\
A_{mn}&=&\langle\phi_m\phi_n\rangle,
\end{eqnarray}
the eigen-equation is
\begin{equation}
Hu=EAu,
\end{equation}
for $u=(c_1,c_2,\cdots)^t$.
Since $\phi_m$ ($m=1,2,\cdots$) are not necessarily orthogonal,
$A$ is not a diagonal matrix.
We diagonalize the Hamiltonian $A^{-1}H$, and then
calculate the expectation values of correlation functions
with the ground state eigenvector.

  In Quantum Monte Carlo simulations an extrapolation is performed to obtain
the expectation values for the ground-state wave function.
If $M$ is large enough, the wave function in Eq.(\ref{wf}) will approach
the exact ground-state wave function, $\psi_{exact}$, as the number of basis
functions, $N_{states}$, is increased.
If the number of basis functions is large enough, the wave function
will approach, $\psi_{exact}$, as $M$ is increased.
In either case the method  employed for the reliable
extrapolation of the wave function is a key issue in calculating the
expectation values.
The variance method was recently proposed in
variational and Quantum Monte Carlo simulations, where the extrapolation is
performed as a function of the energy variance.
We can expect linearity in some cases\cite{sor01}:
\begin{equation}
\langle Q\rangle-Q_{exact}\propto v,
\end{equation}
where $v$ denotes the variance defined as 
\begin{equation}
v= \frac{\langle (H-\langle H\rangle)^2\rangle}{\langle H\rangle^2}
\label{vari}
\end{equation}
and $Q_{exact}$ is the expected exact value of the quantity $Q$.

The simplest procedure for optimizing the ground-state wave function is to
increase the number of basis states $\{\phi_m\}$ by random sampling.
First, we set $\tau$ and $M$, for example, $\tau=0.1$, 0.2, $\cdots$, and
$M=20$, 30, $\cdots$.
We denote the number of basis functions as $N_{states}$.
We start with $N_{states}=100\sim 300$ and then increase up to 10000.
This procedure can be outlined as follows:\\
\\
A1. Generate the auxiliary fields $s_i$ ($i=1,\cdots,N$) in
$B_{\ell}^{\sigma}(\{s_i\}))$  randomly for $\ell=1,\cdots,M$ for
$\phi_m$ ($m=1,\cdots,N_{states}$), and
generate $N_{states}$ basis wave function $\{\phi_m\}$.\\
\\
A2. Evaluate the matrices $H_{mn}=\langle\phi_m H\phi_n\rangle$ and
$A_{mn}=\langle\phi_m\phi_n\rangle$, and diagonalize the matrix
$A^{-1}H$ to obtain $\psi=\sum_m c_m\phi_m$.
Then calculate the expectation values and the energy
variance.\\
\\
A3. Repeat the procedure from A1 after increasing the number of basis
functions.\\
\\
For small systems this random method produces reliable energy  results.
The diagonalization plays an importance producing fast convergence.
In order to lower the ground-state energy
efficiently, we can employ a genetic algorithm\cite{gol89}  to generate 
the basis set from the initial basis set.
One idea is to replace some parts of $\{s_i(\ell)\}$
($i=1,\cdots,N; \ell=1,\cdots,M$) in $\phi_n$ that has the large weight
$\left|c_n\right|^2$ to generate a new basis function $\phi'_n$.
The new basis function $\phi'_n$ obtained in this way is expected to also
have a
large weight and contribute to $\psi$.
The details of the method are shown in Ref.\cite{yan07}.

\subsubsection{Ground State Energy and Correlation Functions}

\begin{figure}[htbp]
\includegraphics[width=8cm]{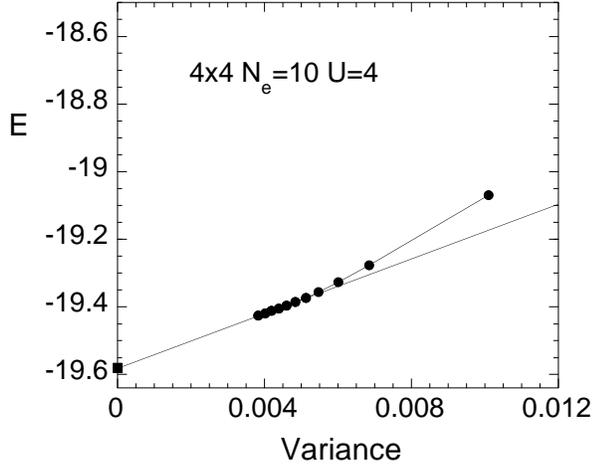}
\caption{
Energy as a function of the variance for $4\times 4$, $U=4$ and $N_e=10$.
The square is the exact result.
The data fit using a straight line using the least-square method
as the variance is reduced.
We started with $N_{states}=100$ (first solid circle) and then increase
up to 2000.
}
\label{4x4E}
\end{figure}

\begin{figure}[htbp]
\includegraphics[width=8cm]{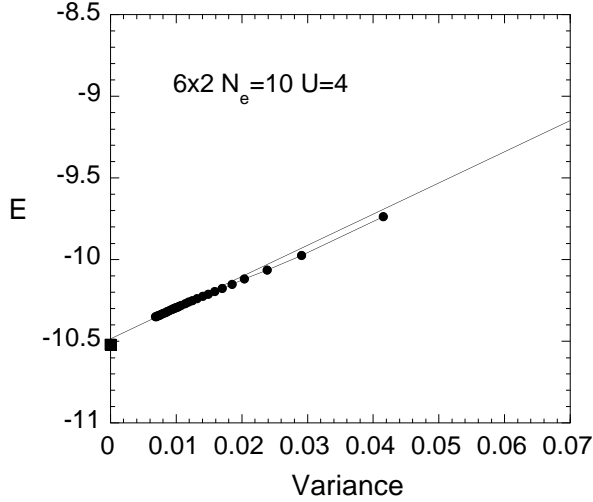}
\caption{
Energy as a function of the variance  for $6\times 2$
$N_e=10$ and $U=4$.
The square is the exact value obtained using exact diagonalization.
}
\label{6x2E}
\end{figure}

\begin{figure}[htbp]
\includegraphics[width=10cm]{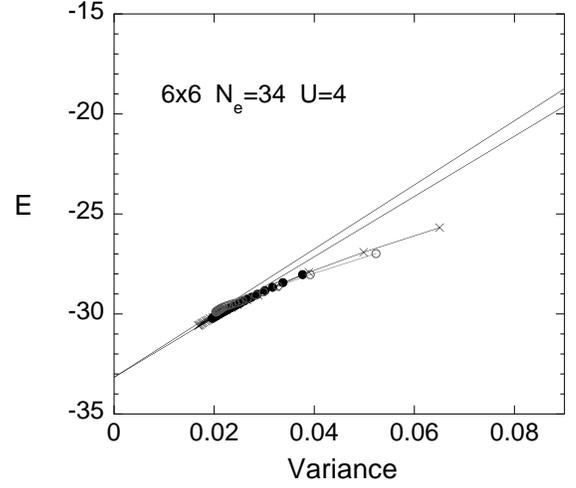}
\caption{
Energy as a function of the variance $v$ for $6\times 6$. 
with the periodic boundary conditions.
Solid circles and crosses are data obtained from the QMD method for
two different initial configurations of the auxiliary fields.
Gray open circles show results obtained from a simple $1/2^N$-method
with 300 basis wave functions\cite{yan07}.
}
\label{6x6-34E}
\end{figure}

The energy as a function of the variance is presented in Figs.\ref{4x4E},
\ref{6x2E}  and \ref{6x6-34E}
for $4\times 4$, $6\times 2$ and $6\times 6$, respectively.
To obtain these results the genetic algorithm was employed to produce the
basis functions except the open symbols in Fig.\ref{6x2E}.
The $4\times 4$ where $N_e=10$ in Fig.\ref{4x4E} is the energy for the closed shell
case up to 2000 basis states.
The other two figures are for open shell cases, where evaluations were
performed up to 3000 states.
We show the results for the $4\times 4$, $6\times 2$ and $6\times 6$ systems
in Table II.

\begin{figure}[htbp]
\includegraphics[width=9cm]{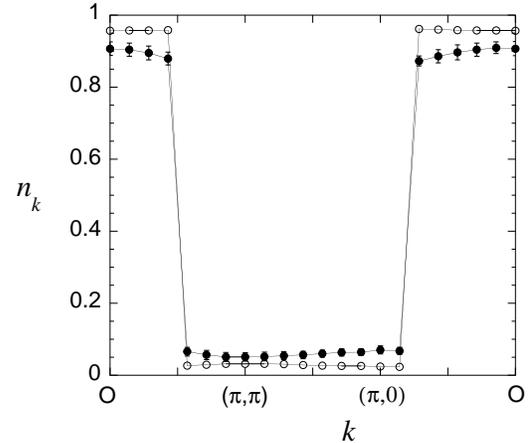}  
\caption{
Momentum distribution function for the $14\times 14$ lattice.
Parameters are $U=4$ and $N_e=146$.  The boundary conditions are periodic
in both directions.
The results for the Gutzwiller function (open circle) are also provided.
}
\label{2Dnk}
\end{figure}

The Fig. \ref{2Dnk} is the momentum distribution function $n({\bf k})$,
\begin{equation}
n({\bf k})= \frac{1}{2}\sum_{\sigma}\langle c_{{\bf k}\sigma}^{\dag}
c_{{\bf k}\sigma}\rangle,
\end{equation}
for $14\times 14$ sites where the results for the Gutzwiller VMC and
the QMD are indicated.
The Gutzwiller function gives the results that $n(k)$ increases as $k$
approaches $k_F$ from above the Fermi surface.
This is clearly unphysical. 
This flaw of the Gutzwiller function near the Fermi surface is not
observed for the QMD result. 

\subsubsection{Spin Gap in the Hubbard Ladder}
\begin{figure}[htbp]
\includegraphics[width=9.5cm]{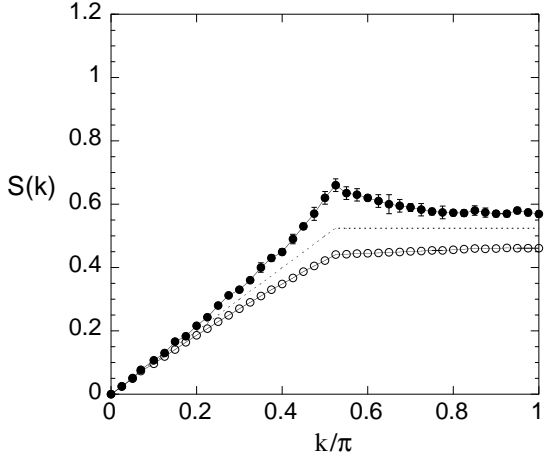}
\caption{
Spin (solid circle) and charge (open circle) correlation functions
obtained from the QMD method for the one-dimensional
Hubbard model with $80$ sites.  The number of electrons is $66$.
We set $U=4$ and use the
periodic boundary condition.
}
\label{1dsk}
\end{figure}

Here we show the results for one-dimensional models.
The ground state of the 1D Hubbard model is no longer Fermi liquid for $U>0$.
The ground state is insulating at half-filling and metallic for less than
half-filling.
The Fig. \ref{1dsk} is the spin and charge correlation
functions, $S(k)$ and $C(k)$,
as a function of the wave number, for the 1D Hubbard model where $N=80$.
The $2k_F$ singularity can be clearly identified where the dotted line is
for $U=0$.
The spin correlation is enhanced and the charge correlation function is
suppressed slightly  because of the Coulomb interaction.

\begin{figure}[htbp]
\includegraphics[width=9cm]{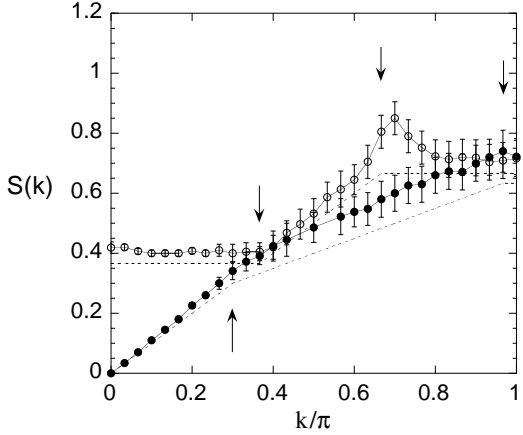}
\caption{
Spin correlation function obtained from the QMD method for the ladder
Hubbard model for $60\times 2$ sites with periodic boundary condition.
The number of electrons is $80$ and $U=4$.
The upper line is for the upper band and the lower line is for the lower band.
Singularities are at $k_{F1}-k_{F2}$, $2k_{F2}$, $k_{F1}+k_{F2}$ and
$2k_{F1}$ from left.
The dotted lines are for $U=0$.
}
\label{ldsk}
\end{figure}

The spin correlation function $S({\bf k})$ for the Hubbard ladder is presented in
Fig.\ref{ldsk},
where $U=4$ and $t_d=1$.
$S({\bf k})$ is defined as
\begin{equation}
S({\bf k})= \frac{1}{N}\sum_{i\ell,j\ell'}{\rm e}^{i{\bf k}\cdot({\bf R}_{i\ell}
-{\bf R}_{j\ell'})}\langle(n_{\ell i\uparrow}-n_{\ell i\downarrow})
(n_{\ell' j\uparrow}-n_{\ell' j\downarrow})\rangle,
\end{equation}
where ${\bf R}_{i\ell}$ denotes the site $(i,\ell)$ ($\ell$=1,2).
We use the convention that ${\bf k}=(k,k_y)$ where $k_y=0$ and $\pi$
indicate the lower band and upper band, respectively.
There are four singularities at $2k_{F1}$, $2k_{F2}$, $k_{F1}-k_{F2}$,
and $k_{F1}+k_{F2}$ for the Hubbard ladder, where $k_{F1}$ and
$k_{F2}$ are the
Fermi wave numbers
of the lower and upper band, respectively.

\begin{figure}[htbp]
\includegraphics[width=7cm]{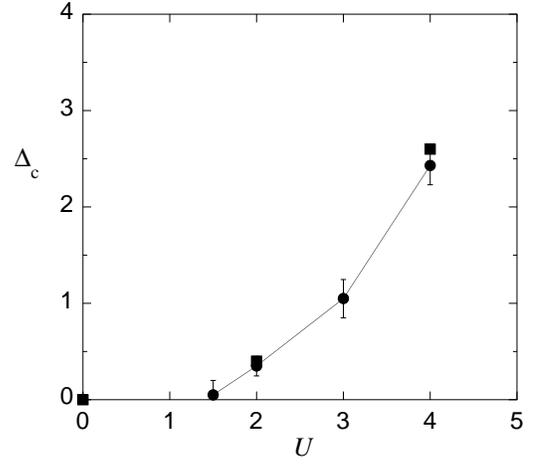}
\caption{
Charge gap as a function of $U$ for $t_d=1$ (circles).
The DMRG results (squares) are provided for comparison\cite{dau00}.
}
\label{ldDc}
\end{figure}

It has been expected that the charge gap opens up as $U$ turns on at half-filling
for the Hubbard ladder model.
In Fig.\ref{ldDc} the charge gap at half-filling is shown as a function of $U$.
The charge gap is defined as
\begin{equation}
\Delta_c= E(N_e+2)+E(N_e-2)-2E(N_e),
\end{equation}
where $E(N_e)$ is the ground state energy for the $N_e$ electrons.
The charge gap in Fig.\ref{ldDc} was estimated using the extrapolation to
the infinite
system from the data for the $20\times 2$, $30\times 2$, and $40\times 2$
systems.
The data suggest the exponentially small
charge gap for small $U$ or the existence of the
critical value $U_c$ in the range
of $0\leq U_c <1.5$, below which the charge gap vanishes.

\begin{table}
\caption{Ground state energy per site from the Hubbard model.
The boundary conditions are periodic in both directions.
The current results are presented under the column labeled QMD.
The constrained
path Monte Carlo (CPMC) results
are from Refs.\cite{zha97}.
The column VMC is the results obtained for the optimized variational
wave function $\psi_{\lambda}^{(2)}$ except for $6\times 2$ for which
$\psi_{\lambda}^{(1)}$ is employed.
The QMC results are from Ref.\cite{fur92}.
Exact results are obtained using
diagonalization\cite{par89}.}
\begin{tabular}{cccccccc}
\colrule
Size & $N_e$ & $U$ &  QMD & VMC & CPMC &  QMC & Exact \\
\colrule
$4\times 4$ & 10 & 4  & -1.2237   & -1.221(1) & -1.2238 &      & -1.2238 \\
$4\times 4$ & 14 & 4  & -0.9836   & -0.977(1) & -0.9831 &   & -0.9840 \\
$4\times 4$ & 14 & 8  & -0.732(2) & -0.727(1) & -0.7281 &   & -0.7418 \\
$4\times 4$ & 14 & 10 & -0.656(2) & -0.650(1) &         &   & -0.6754 \\
$4\times 4$ & 14 & 12 & -0.610(4) & -0.607(2) & -0.606  &   & -0.6282 \\
$6\times 2$ & 10 & 2  & -1.058(1) & -1.040(1) &         &   & -1.05807\\
$6\times 2$ & 10 & 4  & -0.873(1) & -0.846(1) &         &   & -0.8767 \\
$6\times 6$ & 34 & 4  & -0.921(1) & -0.910(2) &         & -0.925 & \\
$6\times 6$ & 36 & 4  & -0.859(2) & -0.844(2) &         & -0.8608 & \\
\colrule
\end{tabular}
\end{table}

\begin{figure}[htbp]
\includegraphics[width=8cm]{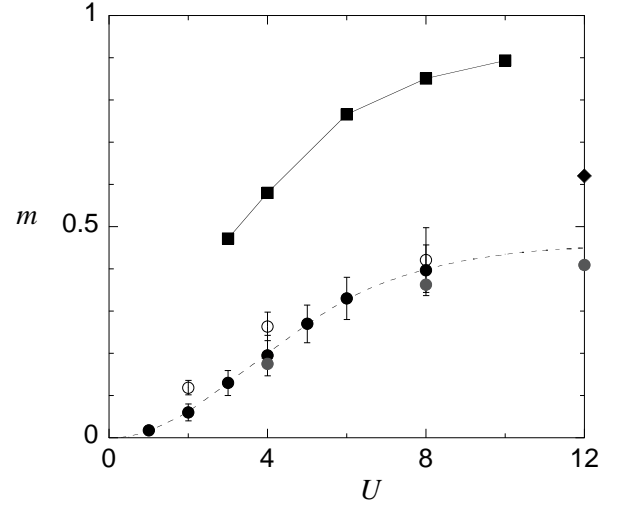} 
\caption{
Magnetization as a function of $U$ for the half-filled Hubbard model
after extrapolation at the limit of large $N$.
Solid circles are the QMD results, and open circles are results obtained
from the QMC method\cite{hir85}.  The squares are the Gutzwiller-VMC
results\cite{yok87} and gray solid
circles show the 3rd $\lambda$-function ($\psi_{\lambda}^{(3)}$) VMC results
carried out on the $8\times 8$ lattice\cite{yan98}.
The diamond symbol is the value from the two-dimensional Heisenberg model
where $m=0.615$\cite{rie89,cal98}.
}
\label{m-U}
\end{figure}

\begin{figure}[htbp]
\includegraphics[width=8cm]{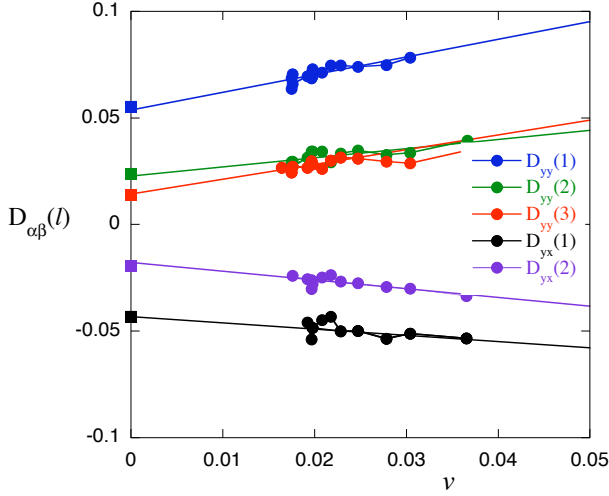}
\caption{ 
Pair correlation function $D_{yy}(\ell)$ and $D_{yx}(\ell)$
for $4\times 3$, $U=4$ and $N_e=10$
obtained by the diagonalization quantum Monte Carlo method. 
The square are the exact results obtained by the exact diagonalization method.
The data fit using a straight line using the least-square method
as the variance is reduced.
We started with $N_{states}=100$ (first solid circles) and then increase
up to 2000.
}
\label{4x3U4qmd}
\end{figure}

\begin{figure}[htbp]
\includegraphics[width=8cm]{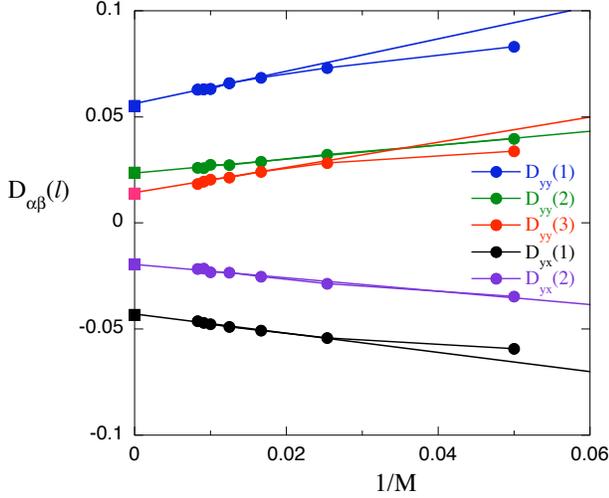}
\caption{ 
Pair correlation function $D_{yy}(\ell)$ and $D_{yx}(\ell)$
for $4\times 3$, $U=4$ and $N_e=10$
obtained by the Metropolis quantum Monte Carlo method. 
The square are the exact results obtained by the exact diagonalization method.
An extrapolation is performed as a function of $1/M$.
}
\label{4x3U4qmc}
\end{figure}

\begin{figure}[htbp]
\includegraphics[width=8cm]{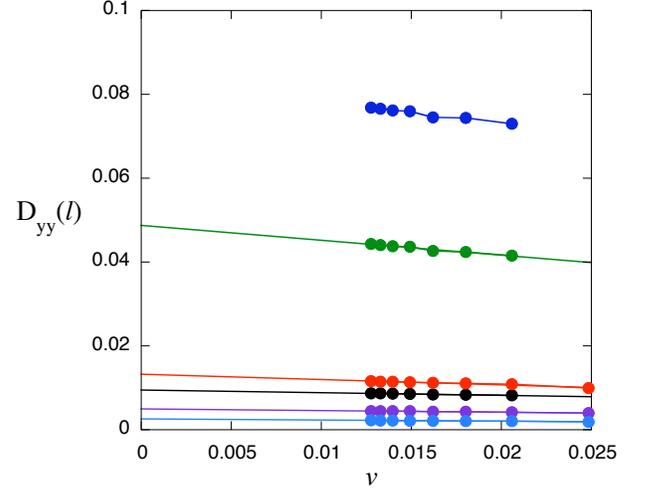}
\caption{ 
Pair correlation function $D_{yy}(\ell)$ as a function of the energy
variance $v$  for $30\times 2$, $U=4$ and $N_e=48$ obtained by
the diagonalization quantum Monte Carlo method.
}
\label{30x2U4qmd}
\end{figure}

\begin{figure}[htbp]
\includegraphics[width=8cm]{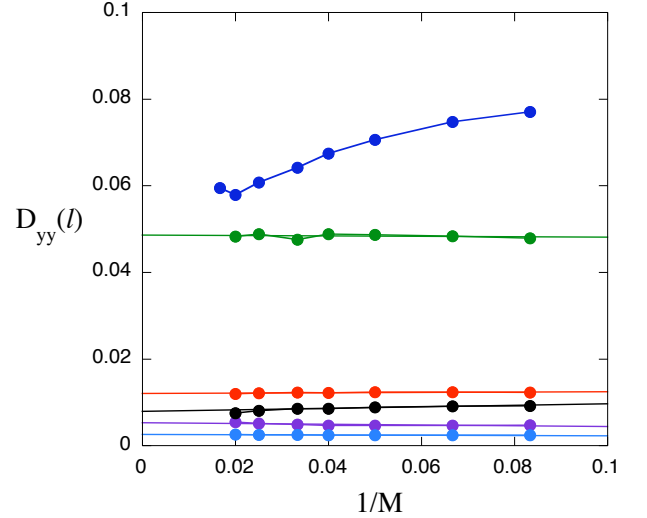}
\caption{ 
Pair correlation function $D_{yy}(\ell)$ as a function of $1/M$
for $30\times 2$, $U=4$ and $N_e=48$ obtained by
the Metropolis quantum Monte Carlo method.
}
\label{30x2U4qmc}
\end{figure}

\begin{figure}[htbp]
\includegraphics[width=7cm]{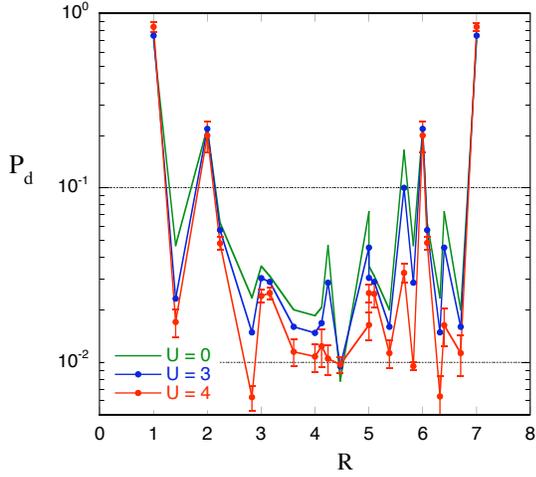}
\caption{ 
Pair correlation function $P_{d}$ as a function of the distance $R=|\ell|$
on $8\times 8$ lattice for the half-filled case $N_e=64$.
We set $t'=0.0$ and $U=0$, 3 and 4.
To lift the degeneracy of electron configurations at the Fermi energy
in the half-filled case,
we included a small staggered magnetization $\sim 10^{-4}$ in the
initial wave function $\psi_0$.
}
\label{64-64-scr}
\end{figure}

\begin{figure}[htbp]
\includegraphics[width=7cm]{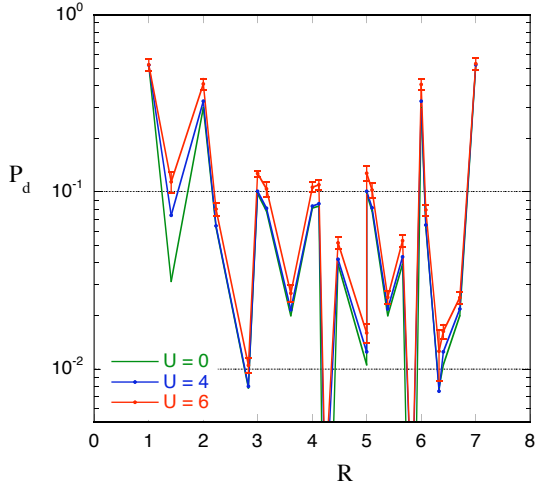}
\caption{ 
Pair correlation function $P_{d}$ as a function of the distance $R=|\ell|$
on $8\times 8$ lattice for $N_e=54$.
We set $t'=-0.2$ and $U=0$, 4 and 6.
}
\label{64-54-scr}
\end{figure}

\begin{figure}[htbp]
\begin{center}
\includegraphics[width=7cm]{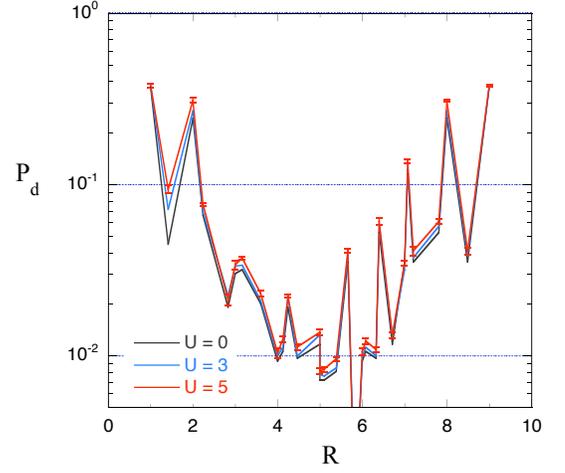}
\caption{
Pair correlation function $P_{d}$ as a function of the distance $R=|\ell|$
on $10\times 10$ lattice for $N_e=82$ and $t'=-0.2$.
The strength of the Coulomb interaction is $U=0$, 3 and 5.
}
\label{100-82sc-r}
\end{center}
\end{figure}

\begin{figure}[htbp]
\begin{center}
\includegraphics[width=7cm]{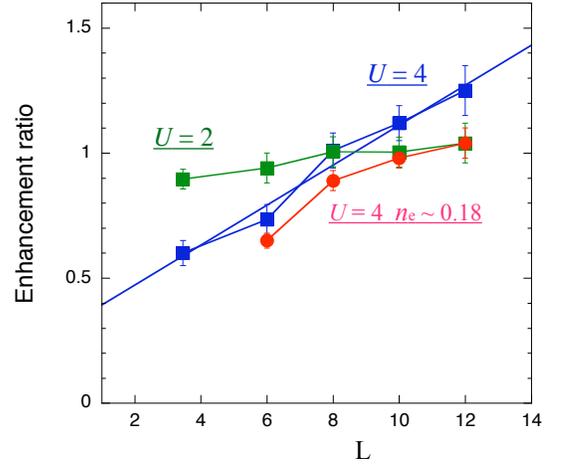}
\caption{
Enhancement ratio of pair correlation function $P_d|_U/P_d|_{U=0}$
as a function of the linear system size $L$ for $U=4$ and $U=2$.
The electron density $n_e$ is about 0.8: $n_e\sim 0.8$ for squares.
The data for $U=4$ and $n_e\sim 0.18$ are also shown by circles.
}
\label{enhance-r}
\end{center}
\end{figure}

\begin{figure}[htbp]
\begin{center}
\includegraphics[width=7cm]{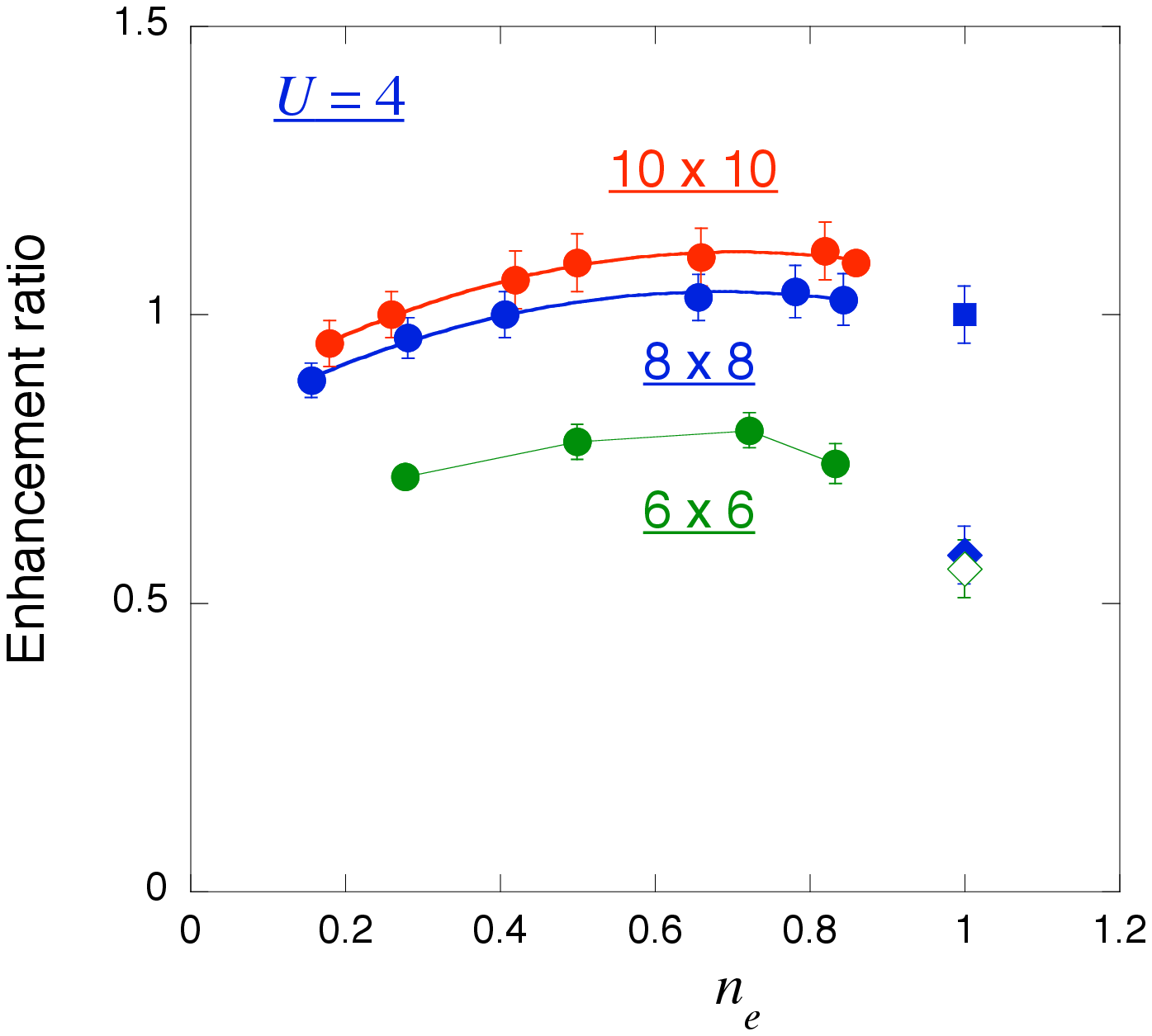}
\caption{
Enhancement ratio of pair correlation function $P_d|_U/P_d|_{U=0}$
as a function of the electron density $n_e$.
We adopt $t'=-0.2$ and $U=4$.  For the half-filled case, the diamonds show
that for $t'=0$ on $8\times 8$ lattice (solid diamond) and $6\times 6$
lattice (open diamond).
The square is for $t'=-0.2$ on $8\times 8$ and $10\times 10$ where
there is no enhancement.
}
\label{enhance-n}
\end{center}
\end{figure}

\subsubsection{Magnetization in 2D Hubbard Model}

The ground state of the 2D Hubbard model at half-filling is antiferromagnetic
for $U>0$ because of the nesting
due to the commensurate vector $Q=(\pi,\pi)$.
The Gutzwiller function predicts that the magnetization
\begin{equation}
m=\left|\frac{1}{N}\sum_j(n_{j\uparrow}-n_{j\downarrow})
{\rm e}^{iQ\cdot R_j}\right|
\end{equation}
increases rapidly as $U$ increases and approaches $m=1$ for large $U$.
In Fig.\ref{m-U} the QMD results are presented for $m$ as a function of $U$.
The previous results obtained using the QMC method are plotted as open circles.
The gray circles are for the $\lambda$-function VMC method and squares
are the Gutzwiller VMC data.  Clearly, the magnetization is reduced considerably
because of the fluctuations, and is smaller than the Gutzwiller VMC method by
about 50 percent.

\begin{figure}[htbp]
\includegraphics[width=7cm]{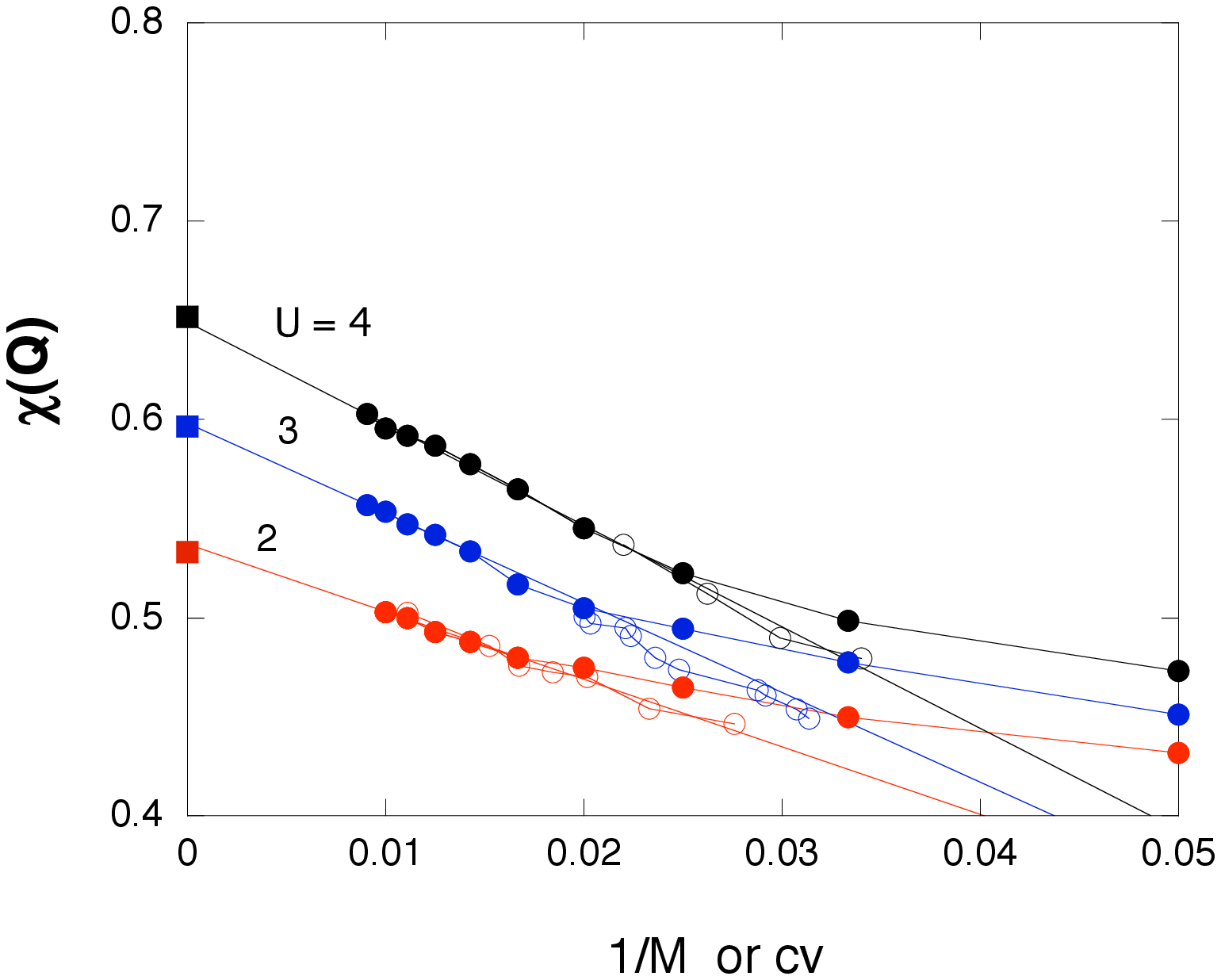}
\caption{ 
Spin susceptibility $\chi (Q)$ as a function of $1/M$ or the variance $v$
for a $6\times 2$ lattice with the  periodic boundary condition.
The number of electrons is 10.
We set $\Delta\tau=0.01$.
The solid circles and open circles are obtained by using the QMC method and
the QMD method, respectively.
The squares indicate exact values.
The variance $v$ is multiplied by a numerical constant.
We set $U=2$, 3, and 4 in units of $t$.
}
\label{2x6-xQ}
\end{figure}

\begin{figure}[htbp]
\includegraphics[width=\columnwidth]{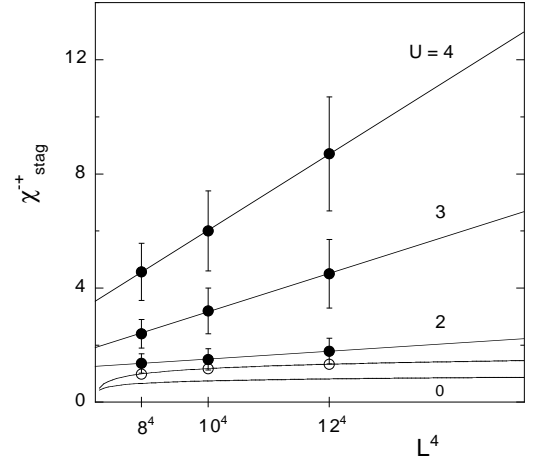}
\caption{
Staggered spin susceptibility $\chi^{-+}_{stag}$ as a function of $L^4$ 
at half-filling
with $t'=0$ for $U=2$, 3, and 4.
We use the periodic and antiperiodic boundary conditions in the $x$ and
$y$ directions, respectively.
The lowest line is for $U=0$, which is fitted by a logarithmic curve.
The open circles show the results for the Gutzwiller function with $U=4$,
which exhibits a logarithmic dependence.
}
\label{xQ-L}
\end{figure}

\begin{figure}[htbp]
\begin{center}
\includegraphics[width=7cm]{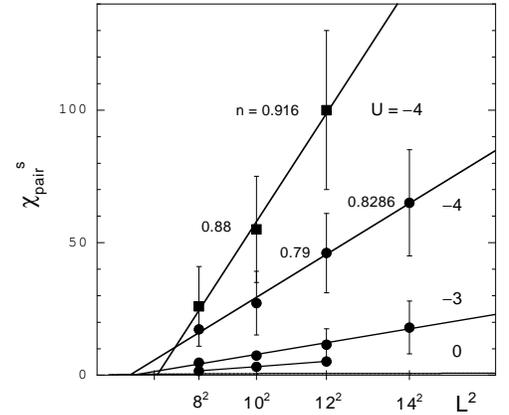}
\caption{
Isotropic $s$-wave susceptibility $\chi^s_{pair}$ as a function of $L^2$
for the negative-$U$ Hubbard model with $U=-2$, -3, and -4, and $t'=0$.
The circles indicate the results for $n_e\sim 0.8$, where
we use the periodic boundary condition in both the $x$ and $y$ directions, 
and the chemical potential is set at the center of the level spacing between
adjacent energy levels.
The lowest dotted line is for $U=0$ ($n_e\sim 0.75$), which is fitted by a 
logarithmic
curve, that is, 
$\chi^s_{pair}\sim {\rm log}(L)$. 
We show $\chi^s_{pair}$ for $n_e\sim 0.9$ and $U=-4$ by squares, where the
boundary condition is antiperiodic in one direction and periodic in the other
direction.
}
\label{xpair-s-N}
\end{center}
\end{figure}

\begin{figure}[htbp]
\begin{center}
\includegraphics[width=6.5cm]{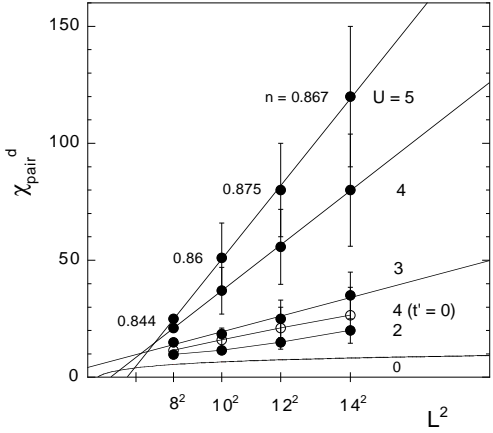}
\caption{
The $d$-wave susceptibility $\chi^d_{pair}$ as a function of $L^2$
for the repulsive-$U$ Hubbard model with $U=2$, 3, 4, and 5.
We use the periodic boundary condition in both the $x$ and $y$ directions.
The solid circles present the results with $t'=-0.2$ and $n_e\sim 0.87$
for $U=2$, 3, 4 and 5.
For the solid squares the parameters are $t'=-0.1$ and $n_e\sim 0.82$ with $U=4$.
The open squares are for $t'=-0.2$, $n_e\sim 1$ (near half-filling) and
$U=4$.
The open circles indicate the results for $t'=0$, $n_e\sim 0.85$ and $U=4$.
The lowest line for $U=0$ is fitted by a logarithmic curve:
$\chi^d_{pair}\sim {\rm log}(L)$.
}
\label{Xpair-N}
\end{center}
\end{figure}

\subsection{Pair Correlation Function}

The pair correlation function $D_{\alpha\beta}$ is defined by
\begin{equation}
D_{\alpha\beta}(\ell)= \langle \Delta_{\alpha}^{\dag}(i+\ell)
\Delta_{\beta}(i)\rangle,
\end{equation}
where $\Delta_{\alpha}(i)$, $\alpha=x,y$, denote the annihilation operators
of the singlet electron pairs for the nearest-neighbor sites:
\begin{equation}
\Delta_{\alpha}(i)= c_{i\downarrow}c_{i+\hat{\alpha}\uparrow}
-c_{i\uparrow}c_{i+\hat{\alpha}\downarrow}.
\end{equation}
Here $\hat{\alpha}$ is a unit vector in the $\alpha(=x,y)$-direction.
We consider the correlation function of d-wave pairing:
\begin{equation}
P_d(\ell)= \langle \Delta_d(i+\ell)^{\dag}\Delta_d(i)\rangle,
\end{equation}
where
\begin{equation}
\Delta_d(i)= \Delta_x(i)+\Delta_{-x}(i)-\Delta_y(i)-\Delta_{-y}(i).
\end{equation}
$i$ and $i+\ell$ denote sites on the lattice.

We show how the pair correlation function is evaluated in quantum
Monte Carlo methods.
We show the pair correlation functions $D_{yy}$ and $D_{yx}$ on the lattice
$4\times 3$ in Figs.\ref{4x3U4qmd} and \ref{4x3U4qmc}.
The boundary condition is open in the 4-site direction and is periodic
in the other direction.
An extrapolation is performed as a function of $1/M$ in the QMC
method with Metropolis algorithm and as a function of the energy
variance $v$ in the QMD method with diagonalization.
We keep $\Delta\tau$ a small constant $\simeq 0.02\sim 0.05$ and
and increase $\tau=\Delta\tau\cdot M$, where $M$ is the division number.
In the Metropolis QMC method, we calculated averages over $5\times 10^5$
Monte Carlo steps.
The exact values were obtained by using the exact diagonalization method.
Two methods give consistent results as shown in figures.
All the $D_{yy}(\ell)$ and
$D_{yx}(\ell)$ are suppressed on $4\times 3$ as $U$ is increased.
In general, the pair correlation functions are suppressed in small
systems.
In Figs.\ref{30x2U4qmd} and \ref{30x2U4qmc}, we show the inter-chain 
pair correlation function
for the ladder model $30\times 2$.
We use the open boundary condition.
The number of electrons is $N_e=48$, and the strength of the Coulomb interaction
is $U=4$.
$\Delta_y(i)$ indicates the electron pair along the rung, and $D_{yy}(\ell)$
is the expectation value of the parallel movement of the pair along the
ladder.
The results obtained by two methods are in good agreement except
$\ell=(1,0)$ (nearest-neighbor correlation).

We first consider the half-filled case with $t'=0$; in this case the
antiferromagnetic correlation is dominant over the superconductive
pairing correlation and thus the pairing correlation function is
suppressed as the Coulomb repulsion $U$ is increased.
The Fig.\ref{64-64-scr} exhibits the d-wave pairing correlation function
$P_d$ on $8\times 8$ lattice as a function of the distance.
The $P_d$ is suppressed due to the on-site Coulomb interaction,
as expected.  Its reduction is, however, not so considerably large
compared to previous QMC studies \cite{zha97b} where the pairing
correlation is almost annihilated for $U=4$.
We then turn to the case of less than half-filling.
We show the results on $8\times 8$ with electron number
$N_e=54$.
We show $P_d$ as a function of the distance in Fig.\ref{64-54-scr} ($N_e=54$).
In the scale of this figure, $P_d$ for $U>0$ is almost the same
as that of the non-interacting case, and is enhanced slightly for
large $U$.
Our results indicate that the pairing correlation is not suppressed
and is indeed enhanced by the Coulomb interaction $U$, and its
enhancement is very small.

The Fig.\ref{100-82sc-r} shows $P_d$ on $10\times 10$ lattice.
This also indicates that the pairing correlation function is
enhanced for $U>0$.
There is a tendency that $P_d$ is easily suppressed as the system
size becomes small.
We estimated the enhancement ratio compared to the non-interacting
case $P_d(\ell)|_U/P_d(\ell)|_{U=0}$ at $|\ell|\sim L/2$
for $n_e\sim 0.8$ as shown in Fig.\ref{enhance-r}.
This ratio increases as the system size is increased.
To compute the enhancement, we picked the sites, for example on
$8\times 8$ lattice, $\ell=(3,2)$, (4,0), (4,1), (3,3), (4,2), (4,3), (5,0),
(5,1) with $|\ell|\sim 4-5$ and evaluate the mean value.
In our computations, the ratio increases almost linearly
indicating a possibility of superconductivity.
This indicates $P_d(\ell)\sim LP_d(\ell)\sim \ell P_d(\ell)$ for $\ell\sim L$.
Because $P_d(\ell)|_{U=0}\sim 1/|\ell|^3$, we obtain
$P_d(\ell)\sim \ell P_d(\ell)\sim 1/|\ell|^2$ for $|\ell|\sim L$.
This indicates that the exponent of the power law is 2.
When $U=2$, the enhancement is small and is almost independent of $L$.
In the low density case, the enhancement is also suppressed being equal
to 1.
In Fig.\ref{enhance-n}, the enhancement ratio is shown as a function of 
the electron
density $n_e$ for $U=4$.  A dome structure emerges even in small systems.
The square in Fig.\ref{enhance-n} indicates the result for the half-filled
case with $t'=-0.2$ on $8\times 8$ lattice.
This is the open shell case and causes a difficulty in computations
as a result of the degeneracy due to partially occupied electrons.
The inclusion of $t'<0$ enhances $P_d$ compared to the case with
$t'=0$ on $8\times 8$ lattice.  $P_d$ is, however, not enhanced
over the non-interacting case at half-filling.
This also holds for $10\times 10$ lattice where the enhancement ratio
$\sim 1$.
This indicates the absence of superconductivity at half-filling.

\subsection{Spin Susceptibility}

We proposde a method to compute the magnetic susceptibility at absolute zero
($T=0$).\cite{yan10}  We add the source term $H_1$ to the Hamiltonian as follows
\begin{equation}
H_1= g\sum_j c_{j\uparrow}^{\dag}c_{j\downarrow}e^{iq\cdot R_j}+\mbox{h.c.}
=g (S_{-{\bf q}}^+ +h.c.),
\end{equation}
where $g$ is a small real number of the order $10^{-3}$ or $10^{-4}$.
We calculate $-\langle S_{{\bf q}}^-\rangle/g$ in the ground state,
which is, as shown by the linear response theory, the magnetic  
susceptibility
\begin{equation}
-\langle S_{{\bf q}}^-\rangle/g= \int_{-\infty}^{\infty}dtG_{ret}
(t,{\bf q})\Bigr|_{T\rightarrow 0}  
= \chi^{-+}({\bf q},\omega=0)\Bigr|_{T\rightarrow 0},
\end{equation}
in the limit of small $g$,
where $G_{ret}$ is the retarded Green function and $\chi^{-+}({\bf q},\omega)$
is the dynamical susceptibility,
\begin{equation}
\chi^{-+}({\bf q},\omega)= i\int_0^{\infty}dt e^{i\omega t}
\langle[S_{{\bf q}}^-(t),S_{-{\bf q}}^+(0)]\rangle.
\end{equation}
Indeed, the above formula gives the correct spin susceptibility
$\chi^{-+}({\bf q},\omega=0)$ on the finite lattice for the
noninteracting case, which is given by
$\sum_k(f(\xi_{k+q})-f(\xi_k))/(\xi_k-\xi_{k+q})$ with the Fermi
distribution function $f(\xi)$.
We calculate $-\langle S_{{\bf q}}^-\rangle/g$ by using the quantum Monte
Carlo method to obtain $\chi^{-+}({\bf q},\omega=0)$.

We examine the results obtained for the susceptibilities.
Figure \ref{2x6-xQ} shows
the spin susceptibility $\chi(Q)=\chi^{-+}(Q,\omega=0)$ for $Q=(\pi,\pi)$
on a  $6\times 2$ lattice as a
function of $1/m$ or the energy variance $v$.  The number of electrons is 10.
The expectation values agree well with exact values given by
the exact diagonalization method.

We now compute the staggered susceptibility $\chi^{-+}_{stag}$
by adding the source term
$H_1=g\sum_j(c_{j\uparrow}^{\dag}c_{j\downarrow}(-1)^{j_x+j_y}+\mbox{h.c.})$ to
the Hamiltonian, where $j=(j_x,j_y)$.
Here we set the
periodic and antiperiodic boundary conditions in the $x$ and $y$ directions,
respectively, to avoid a numerical difficulty caused by the degeneracy between
states ${\bf k}$ and ${\bf k}+Q$ where $Q=(\pi,\pi)$.
It has been shown that a long-range spin correlation exists in the ground
state of the half-filled Hubbard model with $t'=0$ for
$U>0$\cite{kas01,yan07,rie89,cal98}.
In the case $U=0$, $\chi^{-+}_{stag}$ exhibits a  double logarithmic behavior
$({\rm log}(L))^2$.
$\chi^{-+}_{stag}$ is shown as a function of $L$ in Fig.\ref{xQ-L} for $U=2$, 3, and 4.
The obtained values are well fitted by $L^4$ and $\chi^{-+}_{stag}$ diverges
in the limit of a large system size $L$:
\begin{equation}
\chi^{-+}_{stag}\sim L^4 .
\end{equation}
This result is consistent with the existence of the long-range spin correlation
for $U>0$\cite{rie89,cal98}.
The degree of divergence of $\chi^{-+}_{stag}$
is beyond the criterion of the Kosterlitz-Thouless transition,
and thus the long-range order represented by $\chi^{-+}_{stag}$ belongs to a
different category.
The $L^4$ behavior of $\chi_{stag}$ is consistent with the predictions of
perturbation theory in the 2D non-linear sigma model at low temperatures\cite{has93}.

\subsection{Pair Susceptibility}

In this section we consider a method to evaluate the pair susceptibility
$\chi_{pair}$ at $T=0$.
In order to compute the pair susceptibility, we use an electron-hole
transformation for the down spin, $c_{i\downarrow}= d_i^{\dag}$,
whereas the up-spin electrons are unaltered, $c_{i\uparrow}=c_i$.
For the on-site $s$-wave pairing, the source term is
given by the following expression
\begin{equation}
H_1^s= g\sum_i(c_i^{\dag}d_i+\mbox{h.c.}).
\end{equation}
For the anisotropic $d$-wave pairing, we add
\begin{eqnarray}H_1^d&=& g\sum_{i,\mu(=\pm x,\pm y)}(a_{\mu}c_{i+\mu\uparrow}^{\dag}
c_{i\downarrow}^{\dag}+h.c.)\nonumber\\
&=& g\sum_{i,\mu(=\pm x,\pm y)}(a_{\mu}c_{i+\mu}^{\dag}d_i+\mbox{h.c.}),
\end{eqnarray}
where $a_{\pm x}=1$ and $a_{\pm y}=-1$.
The $s$-wave and $d$-wave pair susceptibility are respectively:
\begin{eqnarray} 
\chi_{pair}^s&=& -\frac{1}{g}\frac{1}{N} 
\sum_{i}(\langle c_{i}^{\dag}d_i\rangle+\mbox{h.c.}),\nonumber\\
\chi_{pair}^d&=& -\frac{1}{g}\frac{1}{N} 
\sum_{i,\mu(=\pm x,\pm y)}(a_{\mu}\langle c_{i+\mu}^{\dag}d_i\rangle+\mbox{h.c.}).
\end{eqnarray}
Using the Fourier transformation, the source term for the pair potential is
written as follows
$H_1^a= g\sum_k z_k(c_{-k\downarrow}^{\dag}c_{k\uparrow}^{\dag}+\mbox{h.c.})$
for $a=s$ or $d$ with the $k$-dependence factor $z_k$.
If we define $\Delta_k=\langle c_{k\uparrow}c_{-k\downarrow}\rangle$, then
for a small value of $g$, we have the following
\begin{equation}
\Delta_k/g= -\sum_{k'}z_{k'}\int_{-\infty}^{\infty}dt'
G_{ret}(t-t';k,k'),
\end{equation}
where
\begin{equation}
G_{ret}(t-t';k,k')= i\theta(t-t')\langle[b_k(t),b_{k'}^{\dag}(t')]\rangle,
\end{equation}
for $b_k= c_{k\uparrow}c_{-k\downarrow}$ and
$b_k(t)= e^{iHt}b_{k}e^{-iHt}$.
On the basis of analytic continuation, using the thermal Green function,
$\Delta_k$ is written as
\begin{equation}
\Delta_k/g= -\sum_{k'}\int_0^{\beta}d\tau e^{i\omega_n\tau}
\langle T_{\tau}b_k(\tau)b_{k'}^{\dag}(0)\rangle_{i\omega_n\rightarrow 0}.
\end{equation}
In the noninteracting system, this formula exhibits logarithmic divergence
on the finite lattice $L\times L$:
$\chi_{pair}= A\langle |z_k|^2\rangle {\rm log}(cL)$ with constants $A$ and $c$,
which can be confirmed by numerical estimations on finite systems.

In the Kosterlitz-Thouless theory, the susceptibility is scaled
as follows\cite{kos73,cha02}:
$\chi\sim \xi^{2-\eta}$,
where $\xi$ is the coherence length.  $\xi$ is of order $L$
on a lattice $L\times L$ if long-range coherence exists.
The exponent $\eta$ is expected to be 0 at absolute zero.  Thus $\chi$
scales as $\chi\sim L^2$ in the ground state if the Kosterlitz-Thouless
transition occurs at some temperature.

First, we investigate $\chi^s_{pair}$ for the attractive Coulomb interaction
$U<0$.  For this model, the existence of a Kosterlitz-Thouless transition
has been predicted on the basis of quantum Monte Carlo methods\cite{mor91,cha02}.
The results in Fig.\ref{xpair-s-N} show that the size dependence for $t'=0$ and
$n_e\sim 0.8$ is
\begin{equation} 
\chi^s_{pair}\sim L^2,
\end{equation}
which is consistent with previous studies,
 and shows the existence of a
Kosterlitz-Thouless transition for the attractive interaction.
At near half-filling, $\chi_{pair}^s$ is more enhanced than that at $n_e\sim 0.8$.
Second, let us investigate the $d$-wave pair susceptibility $\chi^d_{pair}$ for the
repulsive Coulomb interaction.
Pair susceptibilities are sensitively dependent on the band structure, particularly
the energy of the van Hove singularity, as a characteristic of
two-dimensional systems.
We compute $\chi^d_{pair}$ at an electron density $n_e\sim 0.87$, a value
near that of optimally doped high-temperature cuprates.
We set $t'=-0.2$.
Figure \ref{Xpair-N} shows $\chi^d_{pair}$ as a function of $L^2$ for
$U=2$, 3, 4, and 5 with $t'=-0.2$ and $n_e\sim 0.87$.
This shows that
\begin{equation}
\chi^d_{pair}\sim L^2,
\end{equation}
if $U$ is moderately large.  This result shows that a $d$-wave superconducting
Kosterlitz-Thouless transition may
exist for the repulsive interaction if we adjust the band parameters
in the region of optimal doping.

\section{Summary}

We have investigated the superconductivity of electronic origin on the basis of
the (single-band and three-band) two-dimensional Hubbard model.
First, we employ the variational Monte Carlo method to clarify the phase
diagram of the ground state of the Hubbard model. 
The superconducting condensation energy per site obtained by the Gutzwiller ansatz
is reasonably close to  experimental value 0.17$\sim 0.26$meV/site.
We have examined the stability of striped and checkerboard states in the under-doped
region.  The relation of the incommensurability and hole density, $\delta\sim x$,
is satisfied in the lower doped region.  We have found that the $4\times 4$
period checkerboard spin modulation is stabilized in the two-dimensional Hubbard
model with the Bi-2212 type band structure.

We have further performed investigation by using the quantum Monte Carlo method that
is an exact unbiased method.
We have presented an algorithm of the quantum Monte Carlo diagonalization to avoid
the negative sign problem in quantum simulations of many-fermion systems.
We have computed d-wave pair correlation functions.
In the half-filled case $P_d$ is suppressed for the repulsive $U>0$, and
when doped away from half-filling $N_e<N$, $P_d$ is enhanced slightly
for $U>0$.
It is noteworthy that the correlation function $P_d$ is
indeed enhanced and is increased as the system size increases in the 2D
Hubbard model.
The enhancement ratio increases almost linearly $\propto L$ as the
system size is increased,
which is an indication of the existence of superconductivity.
Our criterion is that when the enhancement ratio as a function of the
system size $L$ is proportional to a certain power of $L$, superconductivity
will be developed.
This ratio depends on $U$ and is reduced as $U$ is decreased.
The dependence on the band filling shows a dome structure as a function
of the electron density.
In the $10\times 10$ system, the ratio is greater than 1 in the range
$0.3 < n_e < 0.9$. 
Let us also mention on superconductivity at half-filling.
Our result indicates the absence of superconductivity in the
half-filling case because there is no enhancement of pair
correlation functions.

\begin{center}
{\bf Acknowledgment}
\end{center}
We thank I. Hase, S. Koikegami, S. Koike and
J. Kondo for stimulating discussions.
This work was supported by a Grant-in Aid for Scientific Research from the
Ministry of Education, Culture, Sports, Science and Technology of Japan, and
CREST Program of Japan Science and Technology Agency.
A part of numerical calculations was performed at the facilities in the
Supercomputer Center of the Institute for Solid State Physics of the
University of Tokyo.


\begin{thebibliography}{}

\bibitem{bed86}J. G. Bednorz and K. A. M\"{u}ller, Z. Phys. B{\bf 64}, 189 (1986).
\bibitem{dag94}E. Dagotto, Rev. Mod. Phys. {\bf 66}, 763 (1994).
\bibitem{and97}P. W. Anderson, {\em The Theory of Superconductivity in
the High-T$_c$ Cuprates} (Princeton University Press, Princeton, 1997).
\bibitem{mor00}T. Moriya and K. Ueda, Adv. Phys. {\bf 49}, 555 (2000).
\bibitem{ben03}{\em The Physics of Superconductor} Vol.II,
edited by K. H. Bennemann and J. B. Ketterson (Springer, Berlin, 2003).
\bibitem{ste84}G. R. Stewart, Rev. Mod. Phys. {\bf 56}, 755 (1984).
\bibitem{lee86}P. A. Lee, T. M. Rice, J. W. Serene, L. J. Sham and J. W.
Wilkins, Comments Cond. Matter Phys. {\bf 12}, 99 (1986).
\bibitem{ott87}H. R. Ott, Prog. Low Temp. Phys. {\bf 11}, 215 (1987).
\bibitem{map00}M. B. Maple, {\em Handbook on the Physics and Chemistry of
Rare Earths} Vol. 30 (North-Holland, Elsevier, Amsterdam, 2000).
\bibitem{ish98}T. Ishiguro, K. Yamaji and G. Saito, {\em Organic
Superconductors} (Springer-Verlag, Berlin, 1998).
\bibitem{tsu94}C. C. Tsuei et al., Phys. Rev. Lett. {\bf 73}, 593 (1994).
\bibitem{wol95}D. A. Wollman et al., Phys. Rev. Lett. {\bf 74}, 797 (1995).
\bibitem{tsu00}C. C. Tsuei and J. R. Kirtlry, Phys. Rev. Lett. {\bf 85}, 182 (2000).
\bibitem{sat01}T. Sato, T. Kamiyama, T. Takahashi, K. Kurahashi, and 
K. Yamada, Science {\bf 291}, 1517 (2001).
\bibitem{yan01c}T. Yanagisawa, S. Koikegami, H. Shibata, S. Kimura, S. Kashiwaya,
A. Sawa, N. Matsubara and K. Takita, J. Phys. Soc. Jpn. {\bf 70}, 2833 (2001).
\bibitem{shi87}G. Shirane, Y. Endoh, R. Birgeneau, M. A. Kastner, Y. Hidaka,
M. Oka, M. Suzuki, and T. Murakami, Phys. Rev. Lett. {\bf 59}, 1613 (1987).
\bibitem{lyo88}K. B. Lyons, P. A. Fleury, L. F. Schncemmeyer, and
J. V. Waszczak, Phys. Rev. Lett. {\bf 60}, 732 (1988).
\bibitem{man89}E. Manousakis and R. Salvadoe, Phys. Rev. Lett. {\bf 62}, 1310
(1989).
\bibitem{pre88}P. Prelovsek, Phys. Lett. A{\bf 126}, 287 (1988).
\bibitem{inu88}M. Inui and S. Doniach, Phys. Rev. B{\bf 38}, 6631 (1988).
\bibitem{yan92}T. Yanagisawa, Phys. Rev. Lett. {\bf 68}, 1026 (1992).
\bibitem{yan93}T. Yanagisawa and Y. Shimoi, Phys. Rev. B{\bf 48}, 6104 (1993).
\bibitem{yan96}T. Yanagisawa and Y. Shimoi, Int. J. Mod. Phys. B10, 3383
(1996) (http://staff.aist.go.jp/t-yanagisawa/paper/sprevc.pdf).
\bibitem{yan01}T. Yanagisawa, S. Koike, and K. Yamaji, Phys. Rev. B{\bf 64},
184509 (2001).
\bibitem{tra95}J. M. Tranquada, B. J. Sternlieb, J. D. Axe, Y. Nakamura,
and S. Uchida, Nature {\bf 375}, 561 (1995).
\bibitem{hub63}J. Hubbard, Proc. Roy. Soc. A{\bf 276}, 238 (1963); A{\bf 277}, 
237 (1964);
A{\bf 281}, 401(1964).
\bibitem{hir83}J. E. Hirsch, Phys. Rev. Lett. {\bf 51}, 1900 (1983).
\bibitem{hir85}J. E. Hirsch: Phys. Rev. B{\bf 31}, 4403 (1985).
\bibitem{sor88}S. Sorella, E. Tosatti, S. Baroni, R. Car and M. Parrinell,
Int. J. Mod. Phys. B{\bf 2}, 993 (1988).
\bibitem{whi89}S. R. White, D. J. Scalapino, R. L. Sugar, E. Y. Loh,
J. E. Gubernatis, and R. T. Scalettar, Phys. Rev. B{\bf 40}, 506 (1989).
\bibitem{ima89}M. Imada and Y. Hatsugai, J. Phys. Soc. Jpn. {\bf 58}, 3752 (1989).
\bibitem{sor89}S. Sorella, S. Baroni, R. Car and M. Parrinello, Europhys.
Lett. {\bf 8}, 663 (1989).
\bibitem{loh90}E. Y. Loh, J. E. Gubernatis, R. T. Scalettar, S. R. White,
D. J. Scalapino, and R. L. Sugar, Phys. Rev. B{\bf 41}, 9301 (1990).
\bibitem{mor91}A. Moreo, D. J. Scalapino, and E. Dagotto, Phys. Rev. B{\bf 56},
11442 (1991).
\bibitem{fur92}N. Furukawa and M. Imada, J. Phys. Soc. Jpn. {\bf 61}, 3331 (1992).
\bibitem{mor92}A. Moreo, Phys. Rev. B{\bf 45}, 5059 (1992).
\bibitem{fah91}S. Fahy and D. R. Hamann, Phys. Rev. B{\bf 43}, 765 (1991).
\bibitem{zha97}S. Zhang, J. Carlson and J. E. Gubernatis, Phys. Rev. B{\bf 55},
7464 (1997).
\bibitem{zha97b}S. Zhang, J. Carlson and J. E. Gubernatis, Phys. Rev. Lett. {\bf 78},
4486 (1997).
\bibitem{kas01}T. Kashima and M. Imada, J. Phys. Soc. Jpn. {\bf 70}, 2287 (2001).
\bibitem{yan07}T. Yanagisawa, Phys. Rev. B{\bf 75}, 224503 (2007)
(arXiv: 0707.1929).
\bibitem{yan13}T. Yanagisawa, New J. Phys. 15, 033012 (2013).
\bibitem{yok87}H. Yokoyama and H. Shiba, J. Phys. Soc. Jpn. {\bf 56}, 1490 (1987);
ibid. {\bf 56}, 3582 (1987).
\bibitem{gro87}C. Gros, R. Joynt, and T. M. Rice, Phys. Rev. B{\bf 36}, 381 (1987).
\bibitem{nak97}T. Nakanishi, K. Yamaji and T. Yanagisawa, J. Phys. Soc. Jpn. {\bf 66}, 
294 (1997).
\bibitem{yam98}K. Yamaji, T. Yanagisawa, T. Nakanishi and S. Koike,
 Physica C{\bf 304}, 225 (1998).
\bibitem{yan02}T. Yanagisawa, S. Koike and K. Yamaji, J. Phys.: Condens. 
Matter {\bf 14}, 21 (2002).
\bibitem{yan03}T. Yanagisawa, S. Koike, S. Koikegami and K. Yamaji, Phys. Rev. 
B{\bf 67}, 132400 (2003). 
\bibitem{yan05}T. Yanagisawa, M. Miyazaki and K. Yamaji, J. Phys. Soc. Jpn. {\bf 74}, 
835 (2005).
\bibitem{miy04}M. Miyazaki, K. Yamaji and T. Yanagisawa,
J. Phys. Soc. Jpn. {\bf 73}, 1643 (2004).
\bibitem{yam94}K. Yamaji and Y. Shimoi, Physica C{\bf 222}, 349 (1994).
\bibitem{yam94b}K. Yamaji, Y. Shimoi and T. Yanagisawa, Physica C{\bf 235}-{\bf 240},
2221 (1994).
\bibitem{koi99}S. Koike, K. Yamaji and T. Yanagisawa, J. Phys. Soc. Jpn. {\bf 68}, 
1657  (1999).
\bibitem{koi00}S. Koike, K. Yamaji and T. Yanagisawa, J. Phys. Soc. Jpn. {\bf 69}, 
2199 (2000). 
\bibitem{noa96}R. M. Noack, S. R. White, and D. J. Scalapino, Physica C{\bf 270}, 
281 (1996).
\bibitem{noa97}R. M. Noack, N. Bulut, D. J. Scalapino, and M. G. Zacher, 
Phys. Rev. B{\bf 56}, 7162 (1997).
\bibitem{kur96}K. Kuroki, T. Kimura and H. Aoki, Phys. Rev. B{\bf 54}, 15641 (1996),
\bibitem{dau00}S. Daul and D. J. Scalapino, Phys. Rev. B{\bf 62}, 8658 (2000).
\bibitem{san05}K. Sano, Y. Ono, and Y. Yamada, J. Phys. Soc. Jpn. {\bf 74}, 2885
(2005).
\bibitem{bul02}N. Bulut, Adv. in Phys. {\bf 51}, 1587 (2002).
\bibitem{aim07}T. Aimi and M. Imada, J. Phys. Soc. Jpn. {\bf 76}, 113708 (2007).
\bibitem{kon01}J. Kondo, J. Phys. Soc. Jpn. {\bf 70}, 808 (2001).
\bibitem{hlu99}R. Hlubina, Phys. Rev. B{\bf 59}, 9600 (1999).
\bibitem{koi01}S. Koikegami and T. Yanagisawa, J. Phys. Soc. Jpn. {\bf 70}, 3499 
(2001); {\bf 71}, 761 (2002) (E).
\bibitem{koi06}S. Koikegami and T. Yanagisawa, J. Phys. Soc. Jpn. {\bf 75}, 
034715 (2006).
\bibitem{yan08}T. Yanagisawa, New J. Phys. {\bf 10}, 023014 (2008).
\bibitem{har67}A. B. Harris and R. V. Lange, Phys. Rev. {\bf 157}, 295 (1967).
\bibitem{gut63}M. C. Gutzwiller, Phys. Rev. Lett. {\bf 10}, 159 (1963).
\bibitem{kan63}J. Kanamori, Prog. Theor. Phys. {\bf 30}, 275 (1963).
\bibitem{bet31}H. Bethe, Z. Phys. {\bf 71}, 205 (1931).
\bibitem{yan67}C. N. Yang, Phys. Rev. Lett. {\bf 19}, 1312 (1967).
\bibitem{lie68}E. H. Lieb and F. Y. Wu, Phys. Rev. Lett. {\bf 20}, 1445 (1968).
\bibitem{sch90}H. J. Schulz, Phys. Rev. Lett. {\bf 64}, 2831 (1990).
\bibitem{fra90}H. Frahm and V. E. Korepin, Phys. Rev. B{\bf 42}, 10552 (1990).
\bibitem{kaw90}N. Kawakami and S. K. Yang, Phys. Lett. A{\bf 148}, 359 (1990).
\bibitem{hal81}F. D. M. Haldane, J. Phys. C{\bf 14}, 2585 (1981).
\bibitem{eme87}V.J. Emery: Phys. Rev. Lett. {\bf 58}, 2794 (1987).
\bibitem{tje89}L.H. Tjeng, H. Eskes, and G.A. Sawatzky, 
{\em Strong Correlation and Superconductivity}  edited by H. Fukuyama,
S. Maekawa and A.P. Malozemoff (Springer, Berlin Heidelberg, 1989). p.33.
\bibitem{ste89}W.H. Stephan, W. Linden and P. Horsch, Phys. Rev. B{\bf 39},
2924 (1989).
\bibitem{hir89}J.E. Hirsch, E.Y. Loh, D.J. Scalapino and S. Tang, Phys. Rev.
B{\bf 39}, 243 (1989).
\bibitem{sca91}R.T. Scalettar, D.J. Scalapino, R.L. Sugar and S.R. White,
Phys. Rev. B{\bf 44}, 770 (1991).
\bibitem{dop90}G. Dopf, A. Muramatsu and W. Hanke, Phys. Rev. B{\bf 41}, 9264
(1990).
\bibitem{dop92}G. Dopf, A. Muramatsu and W. Hanke, Phys. Rev. Lett. {\bf 68}, 
353 (1992).
\bibitem{asa96}T. Asahata, A. Oguri and S. Maekawa, J. Phys. Soc. Jpn.
{\bf 65}, 365 (1996).
\bibitem{kur96b}K. Kuroki and H. Aoki, Phys. Rev. Lett. {\bf 76}, 4400 (1996).
\bibitem{tak97}T. Takimoto and T. Moriya, J. Phys. Soc. Jpn. {\bf 66},
2459 (1997).
\bibitem{gue98}M. Guerrero, J.E. Gubernatis and S. Zhang, Phys. Rev.
B{\bf 57}, 11980 (1998).
\bibitem{koi00b}S. Koikegami and K. Yamada, J. Phys. Soc. Jpn. {\bf 69}, 768
(2000).
\bibitem{hyb90}M.S. Hybertsen, E.B. Stechel, M. Schl\"{u}ter and D.R.
Jennison, Phys. Rev. B{\bf 41}, 11068 (1990).
\bibitem{esk89}H. Eskes, G.A. Sawatzky and L.F. Feiner, Physica C{\bf 160},
424 (1989).
\bibitem{mcm90}A.K. McMahan, J.F. Annett and R.M. Martin, Phys. Rev. 
B{\bf 42}, 6268 (1990).
\bibitem{cep77}D. Ceperley, G.V. Chester and K.H. Kalos, Phys. Rev. 
B{\bf 16}, 3081 (1977).
\bibitem{umr88}C.J. Umrigar, K.G. Wilson and J.W. Wilkins, Phys. Rev. Lett.
{\bf 60}, 1719 (1988).
\bibitem{bla81}R. Blankenbecler , D.J. Scalapino and R.L. Sugar, Phys. Rev.
D{\bf 24}, 2278 (1981).
\bibitem{yan98}T. Yanagisawa, S. Koike and K. Yamaji, J. Phys. Soc. Jpn.
{\bf 67}, 3867 (1998). 
\bibitem{yan99}T. Yanagisawa, S. Koike and K. Yamaji, J. Phys. Soc. Jpn.
{\bf 68}, 3608 (1999).
\bibitem{yam00b}K. Yamaji, T. Yanagisawa and S. Koike, Physica B{\bf 284}-{\bf 288},
415 (2000).
\bibitem{fei96}L.F. Feiner, J.H. Jefferson and R. Raimondi, Phys. Rev. B{\bf 53},
8751 (1996).
\bibitem{lor93}J.W. Loram, K.A. Mirza, J.R. Cooper and W.Y. Liang, Phys.
Rev. Lett. {\bf 71}, 1470 (1993).
\bibitem{and98}P.W. Anderson, Science {\bf 279}, 1196 (1998).
\bibitem{hao91}Z. Hao, J.R. Clem, M.W. McElfresh, L. Civale, A.P. Malozemoff
and F. Holtzberg, Phys. Rev. B{\bf 43}, 2844 (1991).
\bibitem{yok96}H. Yokoyama and M. Ogata, J. Phys. Soc. Jpn. {\bf 65}, 3615 (1996).
\bibitem{mat04}T. Matsuzaki, N. Momono, M. Oda and M. Ido, J. Phys. Soc. Jpn.
{\bf 73}, 2232 (2004).
\bibitem{toh00}T. Tohyama and S. Maekawa, Supercond. Sci. Technol. {\bf 13},
17 (2000).
\bibitem{miy02}M. Miyazaki, K. Yamaji and T. Yanagisawa, J. Phys. Chem. Solids 
{\bf 63}, 1403 (2002).
\bibitem{yan09}T. Yanagisawa, M. Miyazaki and K. Yamaji, 
J. Phys. Soc. Jpn. {\bf 78}, 013706 (2009).
\bibitem{sin92}D. J. Singh and W. E. Pickett, Physica C{\bf 203}, 193 (1992).
\bibitem{hus03}N. E. Hussey et al., Nature {\bf 425}, 814 (2003).
\bibitem{pla05}M. Plate  et al., Phys. Rev. Lett. {\bf 95}, 07001 (2005).
\bibitem{lee06}W. S. Lee et al., arXiv: cond-mat/0600347 (2006).
\bibitem{yam08}K. Yamaji, T. Yanagisawa, M. Miyazaki and R. Kadono, Physica
C{\bf 468}, 1125 (2008).
\bibitem{yam11}K. Yamaji, T. Yanagisawa, M. Miyazaki and R. Kadono,
J. Phys. Soc. Jpn. {\bf 80}, 083702 (2011).
\bibitem{fab92}M. Fabrizio, A. Parola and T. Tosatti, Phys. Rev. B{\bf 46},
3159 (1992).
\bibitem{fab93}M. Fabrizio, Phys. Rev. B{\bf 48}, 15838 (1993).
\bibitem{yan95}T. Yanagisawa, Y. Shimoi and K. Yamaji, Phys. Rev. B{\bf 52},
3860 (1995).
\bibitem{bal96}L. Balents and M.P.A. Fisher, Phys. Rev. B{\bf 53}, 12133 (1996).
\bibitem{zhe95}G.-q. Zheng, Y. Kitaoka, K. Ishida and K. Asayama,
J. Phys. Soc. Jpn. {\bf 64}, 2524 (1995).
\bibitem{tra96}J. Tranquada, J.D. Axe, N. Ichikawa, Y. Nakamura, S. Uchida
and B. Nachumi, Phys. Rev. B{\bf 54}, 7489 (1996).
\bibitem{suz98}T. Suzuki, T. Goto, K. Chiba, T. Fukase, H. Kimura, K. Yamada,
M. Ohashi and Y. Yamaguchi, Phys. Rev. B{\bf 57}, 3229 (1998).
\bibitem{yama98}K. Yamada, C.H. Lee, K. Kurahashi, J. Wada, S. Wakimoto,
S. Ueki, H. Kimura and Y. Endoh, Phys. Rev. B{\bf 57}, 6165 (1998).
\bibitem{ara99}M. Arai, T. Nishijima, Y. Endoh, T. Egami, S. Tajima,
K. Tomimoto, Y. Shiohara, M. Takahashi, A. Garrett and S.M. Bennington, 
Phys. Rev. Lett. {\bf 83}, 608 (1999).
\bibitem{moo00}H.A. Mook, D. Pengcheng, F. Dogan and R.D. Hunt, Nature
{\bf 404}, 729 (2000).
\bibitem{wak00}S. Wakimoto, R.J. Birgeneau, M.A. Kastner, Y.S. Lee, R. Erwin,
P.M. Gehring, S.H. Lee, M. Fujita, K. Yamada, Y. Endoh, K. Hirota and
G. Shirane, Phys. Rev. B{\bf 61}, 3699 (2000).
\bibitem{nod99}T. Noda, H. Eisaki and S. Uchida, Science {\bf 286}, 265
(1999).
\bibitem{zho99}X.J. Zhou, P. Bogdanov, S.A. Keller, T. Noda, H. Eisaki,
S. Uchida, Z. Hussain and Z.X. Shen, Science {\bf 286}, 268 (1999).
\bibitem{nie98}C. Niedermayer, C. Bernhard, T. Blasius, A. Golnik,
A. Moodenbauch and J.I. Budnick, Phys. Rev. Lett. {\bf 80}, 3843 (1998).
\bibitem{kim99}H. Kimura {\em et al}., Phys. Rev. B{\bf 59}, 6517 (1999).
\bibitem{mats99}H. Matsushita, H. Kimura, M. Fujita, K. Yamada, K. Hirota
and Y. Endoh, J. Phys. Chem. Solids {\bf 60}, 1079 (1999).
\bibitem{mat00}M. Matsuda, M. Fujita, K. Yamada, R.J. Birgeneau, M.A. Kastner,
H. Hirai, Y. Endoh, S. Wakimoto and G. Shirane, Phys. Rev. B{\bf 62},
9148 (2000).
\bibitem{fuj02}M. Fujita, K. Yamada, H. Hiraka, P. M. Gehring, S. H. Leem
S. Wakimoto and G. Shigane, Phys. Rev. B{\bf 65}, 064505 (2002).
\bibitem{gia91}T. Giamarchi and C. Lhuillier, Phys. Rev. B{\bf 43}, 12943 
(1991).
\bibitem{mac89}K. Machida, Physica C{\bf 158}, 192 (1989).
\bibitem{poi89}D. Poilblanc and T.M. Rice, Phys. Rev. B{\bf 39}, 9749 (1989).
\bibitem{kat90}M. Kato, K. Machida, H. Nakanishi and M. Fujita,
J. Phys. Soc. Jpn. {\bf 59}, 1047 (1990).
\bibitem{sch90b}H. Schulz, Phys. Rev. Lett. {\bf 64}, 1445 (1990).
\bibitem{zaa96}J. Zaanen and A.M. Oles, Annalen der Physik {\bf 508}, 224 (1996).
\bibitem{ici99}M. Ichioka and K. Machida, J. Phys. Soc. Jpn. {\bf 68}, 4020 
(1999).
\bibitem{whi98}S. White and D.J. Scalapino, Phys. Rev. Lett. {\bf 80}, 1272
(1998).
\bibitem{whi98b}S. White and D.J. Scalapino, Phys. Rev. Lett. {\bf 81}, 3227
(1998).
\bibitem{hel99}C. S. Hellberg and E. Manousakis, Phys. Rev. Lett. {\bf 83},
132 (1999).
\bibitem{kob99}K. Kobayashi and H. Yokoyama, Physica B{\bf 259}-{\bf 261},
506 (1999).
\bibitem{him02}A. Himeda, T. Kato and M. Ogata, Phys. Rev. Lett. {\bf 88},
117001 (2002).
\bibitem{lor96}.W. Loram, K.A. Mirza, J.R. Cooper, N. Athanassopoulou and
W. Liang, {\em Proc. 10th Ann. HTS Workshop} (World Scientific, Singapore,
1996) p.341.
\bibitem{miy12}M. Miyazaki, K. Yamaji, T. Yanagisawa and K. Yonemitsu,
Physcs Procedia {\bf 27}, 64 (2012).
\bibitem{muk06}H. Mukuda et al., Phys. Rev. Lett. {\bf 96}, 087001 (2006).
\bibitem{hof02}J. E. Hoffman et al., Science {\bf 295}, 466 (2002).
\bibitem{wis08}W. D. Wise et al., Nature Phys. {\bf 4}, 696 (2008).
\bibitem{han04}T. Hanaguri et al., Nature {\bf 430}, 1001 (2004).
\bibitem{whi04}S. R. White and D. J. Scalapino, Phys. Rev. B{\bf 70},
220506 (2004).
\bibitem{sei07}G. Seibold, J. Lorenzana and M. Grilli, Phys. Rev. B{\bf 75},
100505 (2007).
\bibitem{miy09}M. Miyazaki, K. Yamaji, T. Yanagisawa and R. Kadono,
J. Phys. Soc. Jpn. {\bf 78}, 043706 (2009).

\bibitem{jas55}R. Jastrow, Phys. Rev. {\bf 55}, 1479 (1955).
\bibitem{oht92}H. Ohtsuka, J. Phys. Soc. Jpn. {\bf 61}, 1645 (1992).
\bibitem{rae85}H. De Raedt, Phys. Rep. {\bf 127}, 233 (1985).
\bibitem{yok88}H. Yokoyama and H. Shiba, J. Phys. Soc. Jpn. {\bf 57}, 2482
(1988).
\bibitem{tor89}J. B. Torrance and R. M. Metzger, Phys. Rev. Lett. {\bf 63},
1515 (1989). 


\bibitem{sca86}D.J. Scalapino, E. Loh and J.E. Hirsch, Phys. Rev. B{\bf 34},
8190 (1986).
\bibitem{shi88}H. Shimahara and S. Takada, J. Phys. Soc. Jpn. {\bf 57}, 1044 (1988).
\bibitem{mon91}P. Monthoux, A.V. Balatsky and D. Pines, Phys. Rev. Lett.
{\bf 67}, 3448 (1991).
\bibitem{bic89}N.E. Bickers, D.J. Scalapino and S.R. White, Phys. Rev. Lett. {\bf 62},
961 (1989).
\bibitem{pao94}C.-H. Pao and N.E. Bickers, Phys. Rev. B{\bf 49}, 1586 (1994);
Phys. Rev. Lett. {\bf 72}, 1870 (1994).
\bibitem{mon94}P. Monthoux and D.J. Scalapino, Phys. Rev. Lett. {\bf 72}, 1874 (1994).
\bibitem{yama99}Y. Yanase and K. Yamada, J. Phys. Soc. Jpn. {\bf 68}, 2999 (1999).
\bibitem{miy86}K. Miyake, S. Schmidt-Rink and C.M. Varma, Phys. Rev.
B{\bf 34}, 6554 (1986).
\bibitem{mor90}T. Moriya, Y. Takahashi and K. Ueda, J. Phys. Soc. Jpn.
{\bf 59}, 2905 (1990).
\bibitem{hot93}T. Hotta, J. Phys. Soc. Jpn. {\bf 62}, 4414 (1993).
\bibitem{juj99}T. Jujo, S. Koikegami and K. Yamada, J. Phys. Soc. Jpn. {\bf 68},
1331 (1999).
\bibitem{nom00}T. Nomura and K. Yamada, J. Phys. Soc. Jpn. {\bf 69}, 3678 (2000).
\bibitem{koi03}S. Koikegami, Y. Yoshida and T. Yanagisawa, Phys. Rev. B{\bf 67}, 
134517 (2003).
\bibitem{koi10}S. Koikegami and T. Yanagisawa, J. Phys. Soc. Jpn. {\bf 79},
064701 (2010).
\bibitem{yam13}K. Yamaji and T. Yanagisawa, Physica C, (2013) in press
(http://dx.doi.org/10.1016/j.physc.2013.04.017).
\bibitem{koik03}S. Koike et al., Physica C{\bf 388}-{\bf 389}, 65 (2003).
\bibitem{pav01}E. Pavarini et al.,
Phys. Rev. Lett. {\bf 87}, 047003 (2001).

\bibitem{miz86}T. Mizusaki, M. Honma and T. Otsuka, Phys. Rev. C53, 2786 (1986).
\bibitem{sor01}S. Sorella, Phys. Rev. B{\bf 64}, 024512 (2001).
\bibitem{gol89}D. E. Goldberg, {\em Genetic Algorithms in Search, Optimization
and Machine Learning} (Addison-Wesley, Boston, 1989).
\bibitem{par89}A. Parola, S. Sorella, S. Baroni, R Car, M. Parrinello and
E. Tosatti, Physica C{\bf 162}-{\bf 164}, 771 (1989).
\bibitem{yan10}T. Yanagisawa, J. Phys. Soc. Jpn. {\bf 79}, 063708 (2010).
\bibitem{rie89}J. A. Riera and A. P. Young, Phys. Rev. B{\bf 39}, 9697 (1989).
\bibitem{cal98}M. Calandra Buonaura and S. Sorella, Phys. Rev. B{\bf 57}, 11446
(1998).
\bibitem{has93}P. Hasenfratz and F. Niedernayer, Z. Physik B{\bf 92} (1993) 91.
\bibitem{kos73}J. M. Kosterlitz and D. J. Thouless, J. Phys. C{\bf 6} (1973) 1181.
\bibitem{cha02}S. Chandrasekharan and J. C. Osborn, Phys. Rev. B{\bf 66} (2002) 
045113.




\end{thebibliography}
\end{document}